\documentclass[lettersize,journal]{IEEEtran}
\usepackage{amsmath,amsfonts}
\usepackage{algorithm}
\usepackage{array}
\usepackage[caption=false,font=normalsize,labelfont=sf,textfont=sf]{subfig}
\usepackage{textcomp}
\usepackage{stfloats}
\usepackage{url}
\usepackage{verbatim}
\usepackage{graphicx}
\usepackage{cite}

\usepackage{algpseudocode}

\usepackage{psfrag,amsmath, amssymb, amscd, amsfonts, latexsym, epsf, graphicx,amscd}

\def\bbbr{{\mathbb R}} 
\def\bbbz{{\mathbb Z}} 

\def\cont{\scriptsize\mbox{cont}}
\def\disc{\scriptsize\mbox{disc}}
\def\refref{\scriptsize\mbox{ref}}
\def\inin{\scriptsize\mbox{in}}
\def\outout{\scriptsize\mbox{out}}
\def\dB{\scriptsize\mbox{dB}}
\def\Gabor{\scriptsize\mbox{Gabor}}
\def\timecaus{\scriptsize\mbox{time-caus}}

\def\argmax{\operatorname{argmax}}

\begin{document}

\title{{\scriptsize This article has been accepted for publication in IEEE Transactions on Information Theory. Citation information:
DOI 10.1109/TIT.2024.3507879} \\
A time-causal and time-recursive analogue of the Gabor transform}

\author{Tony Lindeberg%
\thanks{The support from the Swedish Research Council 
  (contract 2022-02969) is gratefully acknowledged.}\\$\,$\\
Computational Brain Science Lab, Division of Computational Science and Technology,\\ KTH Royal Institute of Technology, Stockholm, Sweden}

\markboth{Lindeberg: Time-causal and time-recursive analogue of the Gabor transform}{}

\maketitle

\begin{abstract}
This paper presents a time-causal analogue of the Gabor filter, as
well as a both time-causal and time-recursive analogue of the Gabor
transform, where the proposed time-causal representations obey both
temporal scale covariance and a cascade property over temporal
scales. The motivation behind these constructions is to enable
theoretically well-founded time-frequency analysis over multiple
temporal scales for real-time situations, or for physical or
biological modelling situations, when the future cannot be accessed,
and the non-causal access to the future in Gabor filtering is therefore
not viable for a time-frequency analysis of the system.

We develop a principled axiomatically determined theory for
formulating these time-causal time-frequency representations, obtained
by replacing the Gaussian kernel in the Gabor filtering with a
time-causal kernel, referred to as the time-causal limit kernel, and
which guarantees simplification properties from finer to coarser
levels of scales in a time-causal situation, similar as the Gaussian
kernel can be shown to guarantee over a non-causal temporal domain. We
do also develop an axiomatically determined theory for implementing a
discrete analogue of the proposed time-causal frequency analysis
method on discrete data, based on first-order recursive filters
coupled in cascade, with provable variation-diminishing properties
that strongly suppress the influence from local perturbations and
noise, and with specially chosen time constants to achieve
self-similarity over scales and temporal scale covariance.

In these ways, the proposed time-frequency representations guarantee
well-founded treatment over multiple temporal scales, in situations when the
characteristic scales in the signals, or physical or biological
phenomena, to be analyzed may vary substantially, and additionally all
steps in the time-frequency analysis have to be fully time-causal.
\end{abstract}


\section{Introduction}

The Gabor filter, proposed by Gabor \cite{Gab46}, defines a time-frequency
transform corresponding to a windowed Fourier transform, with the
Gaussian kernel used as a temporal window function.
By varying the temporal duration of that Gaussian window
function, time-frequency decompositions can be obtained at different
temporal scales. Specifically, when using this type of time-frequency
decomposition, the time-frequency representations at any coarser level of
temporal scale will be related to the time-frequency representations at
any finer temporal scales, by both a temporal scale covariance property and a
cascade smoothing property over temporal scales.

The Gabor function is, however, not a time-causal kernel, in the sense
that the Gaussian window function makes use of information from the future in
relation to any time moment. This property, thus, prevents Gabor filtering from being
used in real-time situations, in which the future cannot be
accessed.

Instead, for real-time processing of {\em e.g.\/}\ auditory
signals, approaches such as the Gammatone filter have been
developed and used by Johannesma \cite{Joh72-HearTheory}, 
Patterson {\em et al.\/} \cite{PatNimHolRic87-GammaTone}, \cite{PatAllGig95-JASA},
and Hewitt and Meddis \cite{HewMed94-JASA}.
The Gammatone filter does, however, not obey similar theoretical
properties over a time-causal temporal domain as the Gabor filter
family obeys over a non-causal temporal domain,
regarding temporal scale covariance and
cascade smoothing properties.
A related form of time-frequency analysis, based on time-causal and
time-recursive temporal filtering operations,
has been proposed by Lindeberg and Friberg \cite{LinFri15-PONE},
however, not reaching
full temporal scale covariance, as in the work to be presented here.
For more general overviews of methods for time-frequency analysis, we
refer the reader to the treatments by Feichtinger and Strohmer \cite{FeiStr98-book},
Qian and Chen \cite{QiaChe99-SignProcMag},
Gr{\"o}chenig \cite{Gro01-book}, Flandrin \cite{Fla18-book}
and the references therein.

The subject of this article is to describe a time-causal analogue of
Gabor filtering, that makes it possible to define time-frequency
representations for real-time situations, and which also obeys
temporal scale covariance and cascade smoothing property over temporal
scales, in such a way that time-frequency representations at different
temporal scales can be theoretically related. To achieve this property,
we shall replace the non-causal Gaussian kernel in the Gabor filter
with a time-causal kernel, referred to as the time-causal limit
kernel, and which ensures theoretically consistent treatment of
structures in temporal signals over
multiple temporal scales, see Lindeberg \cite{Lin23-BICY}.
By a slight modification, the resulting time-frequency representation
can also be made fully time-recursive, implying that it can
be computed in terms of a set of first-order
integrators coupled in cascade, with especially selected time
constants to achieve temporal scale covariance. For discrete
implementation, the first-order integrators can, in turn, be
approximated by a small set of first-order recursive filters coupled in
cascade, making real-time implementations both straightforward and
computationally very efficient on regular
signal processing hardware.

In this way, the presented theory provides both a principled theoretical
framework for expressing time-frequency representations for modelling
and analyzing physical and biological processing for which the future
cannot be accessed, as well as a framework for processing and analyzing
temporal signals in real time.

The presented theoretical framework is intended for applications where
it is critical to perform time-frequency analysis in real time,
specifically where time-delayed analysis of pre-recorded temporal data
is not applicable or else regarded as an unrealistic simplification
of the actual physical situation. Such constraints arise in time-critical technical
systems, where real-time time-frequency analysis may be an integrated part of a
control loop mechanism, or a real-time analysis or monitoring system,
as well as when
modelling physical or biological processes, for example the brain and the nervous
system, (such as the analysis of sound stimuli in auditory perception),
where it constitutes a strong evolutionary advantage both for an
individual and a species to be
able to react fast, {\em e.g.\/}, in fight or flight scenarios.

Due to the presented computationally very efficient discrete implementation of
the proposed time-causal and time-recursive analogue of the Gabor
transform, this time-frequency transform could also, beyond real-time
applications, be beneficial for offline analysis of larger datasets,
since it can be very well
approximated by applying a low number (here 4 to 8) of recursive
filters coupled in cascade.

\subsection{Structure of this article}

In Section~\ref{sec-theor-prop-gab-filt}, we begin by stating two important
properties of the regular Gabor filters and the regular Gabor transform under temporal
scaling transformations, as well as in relation to time-frequency
analysis over multiple temporal scales, that we will then generalize
to a non-causal temporal domain.

Section~\ref{sec-time-caus-anal-gab-filt} then defines time-causal
analogues of Gabor filters and the Gabor transform, obtained by
replacing the Gaussian kernel in these time-frequency representations with a special
time-causal kernel, referred to as the time-causal limit kernel.
The theoretical background to the time-causal limit kernel in terms of
temporal scale-space kernels is described, as well as basic properties
of this kernel that we will build upon.

Section~\ref{sec-theor-prop-time-caus-gabor-filt} then shows how the
desirable properties in terms of temporal scale covariance and a
temporal cascade smoothing property hold for the proposed time-causal
analogues of Gabor filters and the Gabor transform.
Section~\ref{sec-disc-impl} outlines how the time-causal analogue 
of the Gabor transform can be
implemented in practice on discrete signals, in terms of fully
time-causal and time-recursive operations, also suitable for real-time
implementation.

Section~\ref{sed-theor-props-time-caus-gabor} gives a theoretical
analysis of properties of the proposed time-frequency analysis
concept, regarding temporal delays and frequency selectivity characteristics.
Section~\ref{sec-exp} presents experiments of applying the proposed
concepts to multi-scale time-frequency analysis of an audio signal
over multiple temporal scales, 
for four main use cases regarding the parameter settings.
For comparison, we also show the result of computing truncated and
time-delayed Gabor transforms for the same signal, based on using a
similar temporal delay for the truncated and time-delayed Gabor
transforms as for the time-causal and time-recursive analogue of the
Gabor transform, while setting all contributions, that would have implied
forbidden access to the future, to zero.

Section~\ref{sec-theor-est-accuracy-freq-est} then derives a set of
theoretical estimates, to characterize basic properties of the proposed
time-causal and time-recursive analogue of the Gabor transform, in
terms of frequency selectivity,
and how those properties differ from the regular non-causal Gabor
transform, based on fully continuous responses to ideal sine waves.
Section~\ref{sec-exp-robustness-to-noise} then complements that
theoretical analysis will numerical simulation experiments, to characterize
how robust the resulting time-causal frequency estimates will be to
noise, and then also including the effects of discretizations over time and
in the frequency domain, that are necessary for a discrete implementation.

Section~\ref{sec-summ-cov-props} summarizes the basic
covariance properties under transformations of the signal domain, that
the proposed time-causal and time-recursive analogue of the Gabor
transform obeys under basic transformations of the signal domain, and
which thereby imply a certain degree of robustness to these classes of
transformations of the input signal.

Finally, Section~\ref{sec-summ-disc} concludes with a summary of some
of the main conceptual results.

\subsection{Extensive appendix sections}

This paper also comprises an extensive appendix with complementary
theory and technical details, which the treatment in the main part of
the paper builds upon, but where many technical details have been put in the appendix,
to keep the main presentation of the paper as conceptual as possible,
and as a simplification for the first-time reader of the paper.

Appendix~\ref{sec-arg–why-time-caus-anal} gives an extensive treatment
of theoretical motivations why the proposed time-causal time-frequency
analysis concept can be regarded as a time-causal analogue of the
Gabor transform. A set of theoretical arguments, in terms of
information-reducing properties is developed, by which
the Gaussian kernel can be singled out as a canonical choice of
temporal window function when defining a non-causal time-frequency
transform, and which in this way uniquely singles out the Gabor
transform as a canonical time-frequency transform over a non-causal
temporal domain. By then stating as similar as possible requirements
regarding the formulation of a time-causal frequency analysis concept, as can
be stated given the constraints arising from the fact that the future
cannot be accessed in a time-causal scenario,
we present a set of theoretical arguments, that
lead to using the time-causal limit kernel as a canonical temporal
window function for a time-causal time-frequency analysis.

Appendix~\ref{app-class-temp-scsp-kern} then addresses the problem of
formulating an as theoretically well-founded approach for defining a
discrete analogue of the proposed time-causal analogue of the Gabor
transform. Based on a classification of which discrete temporal kernels
guarantee that the number of local extrema, or equivalently the number
of zero-crossings, must not increase when filtering from finer to coarser levels of
temporal scales, we first arrive at a finite set of possible candidate
time-causal kernels, out of which the single choice of first-order
discrete integrators arises as the unique choice, that also constitutes
a true numerical approximation of the continuous first-order
integrators coupled in cascade, and which describe the temporal smoothing
effect of the time-causal limit kernel used for defining the proposed
time-causal analogue of the Gabor transform. Specifically, a detailed
outline of an algorithm for computing the proposed discrete
analogue of the time-causal analogue of the Gabor transform is
given in Appendix~\ref{app-discr-impl}, with a complementary
description for real-time implementations in Appendix~\ref{sec-app-real-time-alg}.

Appendix~\ref{app-disc-anal-gabor-transf} describes a discrete
analogue of the continuous Gabor transform, that arises out of the
classification of the discrete kernels that guarantee non-creation of new
local extrema from finer to coarser levels of scale.

Appendix~\ref{sec-app-inv-transf-time-caus-gabor-transf} outlines how
inverse transforms of the time-causal analogue of the Gabor transform
can be defined. Notably, those inverse transforms are, however, not
time-causal---it is explicitly explained why. Therefore, those inverse
transforms may not be directly applicable for time-critical real-time
applications, why this contribution should then mainly be intended for
temporal processing that is not time-critical, such as offline
analysis. With respect to the main target context of this paper,
concerning real-time computations of time-frequency transforms, the
treatment in Appendix~\ref{sec-app-inv-transf-time-caus-gabor-transf}
should therefore be regarded
as mainly theoretical, in the respect that it demonstrates
that the proposed new time-causal time-frequency
analysis concept constitutes a true time-frequency transform.

\begin{figure*}[hbtp]
  \begin{center}
    \begin{tabular}{ccc}
      $g(t;\; \tau)$ for $\sigma = \sqrt{\tau} = 1$
      & $g(t;\; \tau)$ for $\sigma = \sqrt{\tau} = 2$ \\
       \includegraphics[width=0.25\textwidth]{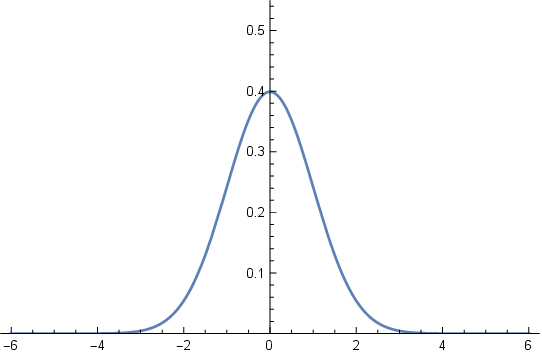}
       &  \includegraphics[width=0.25\textwidth]{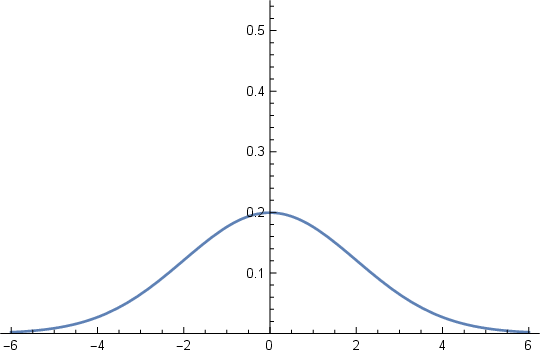} \\
      $\,$ \\
       $\Psi(t;\; \tau, c)$ for $\sigma = \sqrt{\tau} = 1$ and $c = \sqrt{2}$
       & $\Psi(t;\; \tau, c)$ for $\sigma = \sqrt{\tau} = 2$ and $c = \sqrt{2}$ \\
      \includegraphics[width=0.25\textwidth]{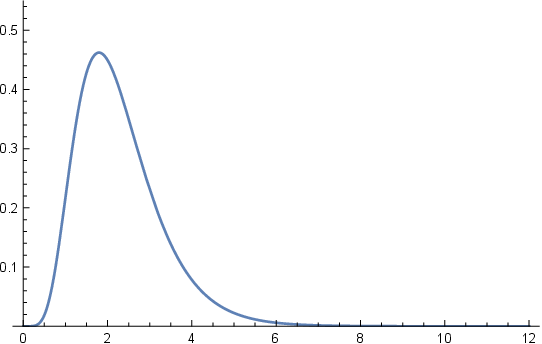}
      & \includegraphics[width=0.25\textwidth]{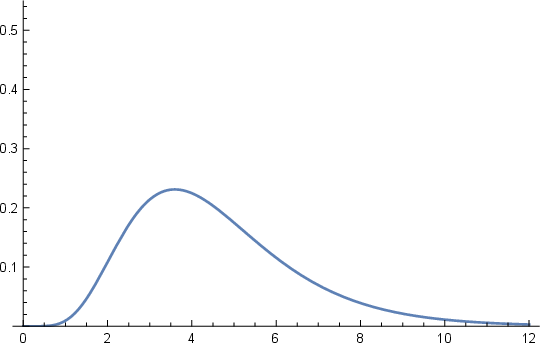} \\
      $\,$ \\
      $\Psi(t;\; \tau, c)$ for $\sigma = \sqrt{\tau} = 1$ and $c = 2$
       & $\Psi(t;\; \tau, c)$ for $\sigma = \sqrt{\tau} = 2$ and $c = 2$ \\
        \includegraphics[width=0.25\textwidth]{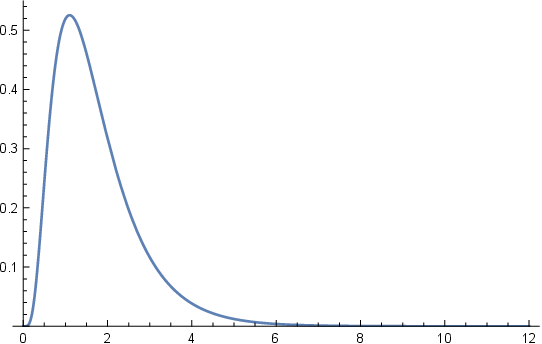}
      & \includegraphics[width=0.25\textwidth]{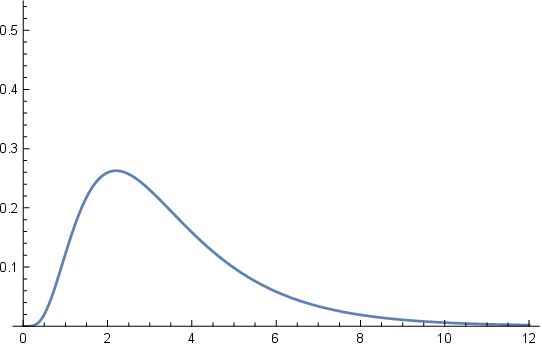} \\
    \end{tabular}
   \end{center}
  \caption{(top row) Gaussian kernels at temporal scale levels
    $\sigma = \sqrt{\tau} = 1$ and $2$.
    (middle row) The time-causal limit kernel at temporal scale levels
    $\sigma = \sqrt{\tau} = 1$ and $2$ for $c = \sqrt{2}$.
    (bottom row) The time-causal limit kernel at temporal scale levels
    $\sigma = \sqrt{\tau} = 1$ and $2$ for $c = 2$.}
  \label{fig-gauss-limit-kern-graphs}
\end{figure*}

\begin{figure*}[hbtp]
  \begin{center}
    \begin{tabular}{ccc}
      $\operatorname{Re} G(t, \omega;\; \tau)$ for $\sigma =
      \sqrt{\tau} = 1$ and $\omega = 10$
      & $\operatorname{Im} G(t, \omega;\; \tau)$ for $\sigma =
          \sqrt{\tau} = 1$ and $\omega = 10$\\
       \includegraphics[width=0.26\textwidth]{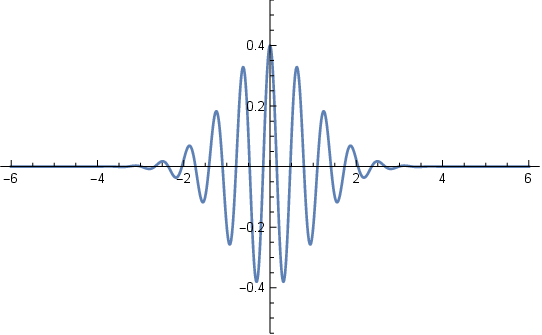}
       & \includegraphics[width=0.26\textwidth]{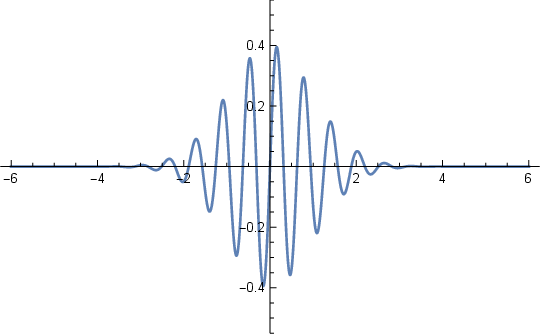}\\
      $\,$ \\
     $\operatorname{Re} G(t,\omega;\; \tau)$ for $\sigma =
      \sqrt{\tau} = 2$ and $\omega = 10$
      & $\operatorname{Im} G(t, \omega;\; \tau)$ for $\sigma =
          \sqrt{\tau} = 2$ and $\omega = 10$\\
        \includegraphics[width=0.26\textwidth]{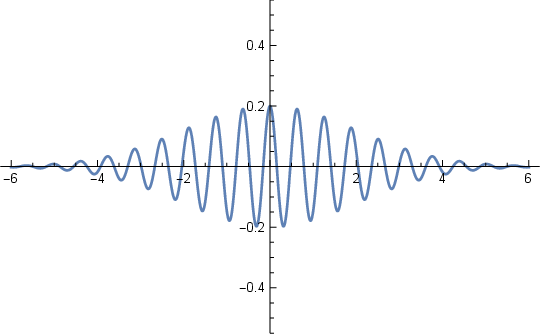}
       & \includegraphics[width=0.26\textwidth]{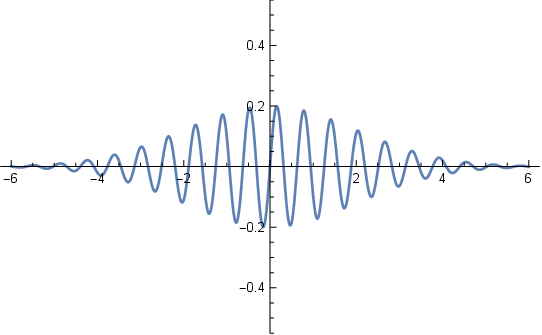}\\
       $\,$ \\
    $\operatorname{Re} \chi(t, \omega;\; \tau, c)$ for $\sigma =
      \sqrt{\tau} = 1$, $\omega = 10$ and $c = \sqrt{2}$
     & $\operatorname{Im} \chi(t, \omega;\; \tau, c)$ for $\sigma =
         \sqrt{\tau} = 1$, $\omega = 10$ and $c = \sqrt{2}$\\
       \includegraphics[width=0.26\textwidth]{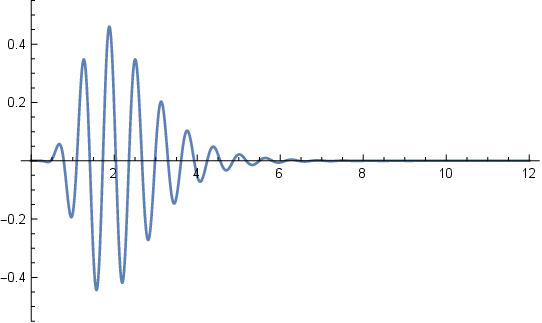}
       & \includegraphics[width=0.26\textwidth]{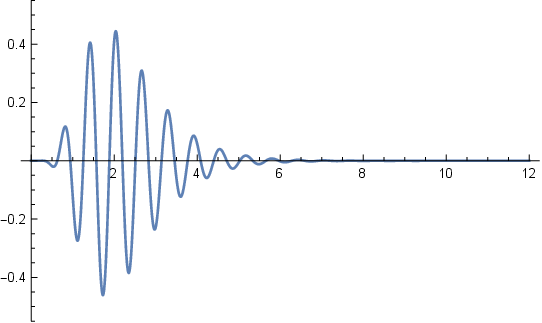}\\    
       $\,$ \\
    $\operatorname{Re} \chi(t, \omega;\; \tau, c)$ for $\sigma =
      \sqrt{\tau} = 2$, $\omega = 10$ and $c = \sqrt{2}$
     & $\operatorname{Im} \chi(t, \omega;\; \tau, c)$ for $\sigma =
         \sqrt{\tau} = 2$, $\omega = 10$ and $c = \sqrt{2}$\\
        \includegraphics[width=0.26\textwidth]{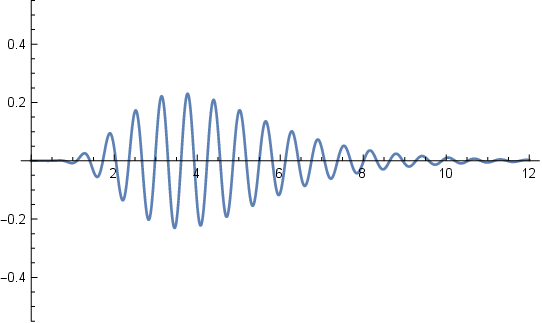}
       & \includegraphics[width=0.26\textwidth]{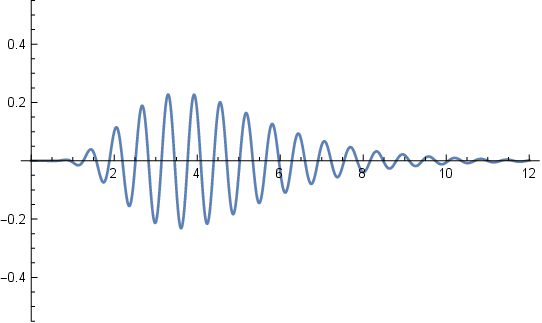} \\
      $\,$ \\
     $\operatorname{Re} \chi(t, \omega;\; \tau, c)$ for $\sigma =
      \sqrt{\tau} = 1$, $\omega = 10$ and $c = 2$
     & $\operatorname{Im} \chi(t, \omega;\; \tau, c)$ for $\sigma =
         \sqrt{\tau} = 1$, $\omega = 10$ and $c = 2$\\
        \includegraphics[width=0.26\textwidth]{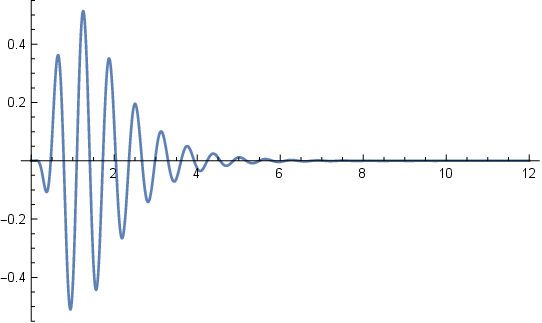}
       & \includegraphics[width=0.26\textwidth]{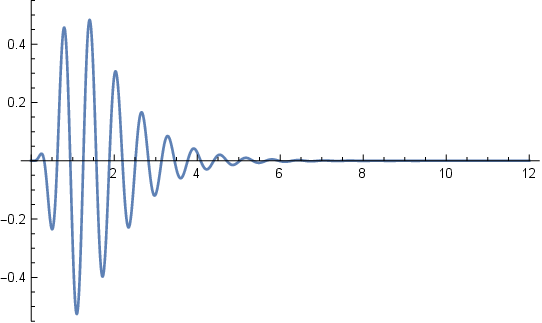}\\    
      $\,$ \\
     $\operatorname{Re} \chi(t, \omega;\; \tau, c)$ for $\sigma =
      \sqrt{\tau} = 2$, $\omega = 10$ and $c = 2$
     & $\operatorname{Im} \chi(t, \omega;\; \tau, c)$ for $\sigma =
         \sqrt{\tau} = 2$, $\omega = 10$ and $c = 2$\\
        \includegraphics[width=0.26\textwidth]{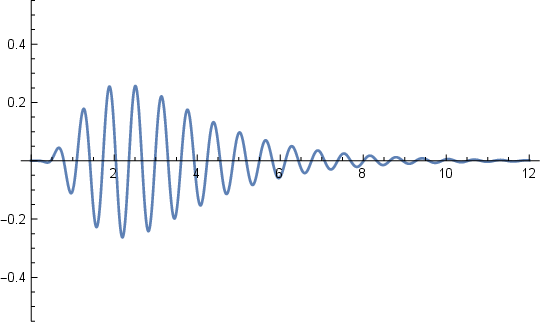}
       & \includegraphics[width=0.26\textwidth]{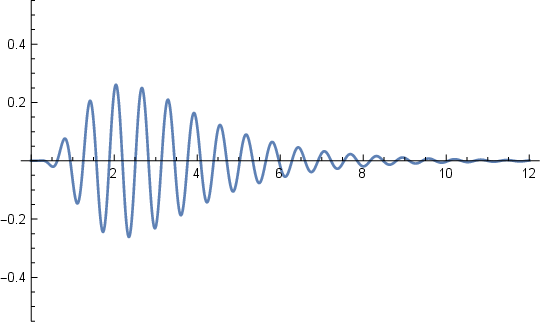} \\
   \end{tabular}
   \end{center}
   \caption{(top two rows) The real and imaginary parts of Gabor kernels at temporal scale levels
    $\sigma = \sqrt{\tau} = 1$ and $2$ for $\omega = 10$.
    (middle two rows) The real and imaginary parts of the
    complex-valued extension of the time-causal limit kernel at temporal scale levels
    $\sigma = \sqrt{\tau}  = 1$ and $2$ for $\omega = 10$ and $c = \sqrt{2}$. 
    (bottom two rows) The real and imaginary parts of the
    complex-valued extension of the time-causal limit kernel at temporal scale levels
    $\sigma = \sqrt{\tau}  = 1$ and $2$ for $\omega = 10$ and $c = 2$.}
  \label{fig-gabor-compl-limit-kern-graphs}
\end{figure*}

\section{Theoretical properties of Gabor filtering}
\label{sec-theor-prop-gab-filt}

With $\tau = \sigma^2$ denoting the variance of the Gaussian window
function $g(t;\; \tau)$, the Gabor filter is defined as
\begin{equation}
  \label{eq-gab-filt}
  G(t, \omega;\; \tau)
  = g(t;\; \tau) \, e^{i \omega t}
  = \frac{1}{\sqrt{2 \pi \tau}} \, e^{-t^2/2\tau} \, e^{i \omega t},
\end{equation}
and the Gabor transform%
\footnote{In this work, we normalize the Gaussian window function to
  having unit $L_1$-norm, because it simplifies both the following
  analysis and the relations to the time-causal limit kernel, which we
will use for defining a time-causal analogue of the Gabor
transform. There are, however, alternative definitions of the Gabor
transform, that do not perform such a normalization, and instead
normalize the Gaussian kernel to having its peak value equal to 1.}
of a function $f(t)$ as
\begin{equation}
  \label{eq-gab-transf}
  ({\cal G} f)(t, \omega;\; \tau) =
  \int_{u=-\infty}^{\infty}
     f(u) \,
    \frac{1}{\sqrt{2 \pi \tau}} \,  e^{-(t-u)^2/2\tau} \, e^{-i \omega  u} \, du.
\end{equation}
Convolution of a signal $f(t)$ with a Gabor function $G(t, \omega;\; \tau)$
is therefore related to the Gabor transform of $f(t)$ according to
\begin{align}
  \begin{split}
     &(G(\cdot, \omega;\; \tau) * f(\cdot))(t, \omega;\; \tau)
  \end{split}\nonumber\\
  \begin{split}
     & = \int_{u=-\infty}^{\infty} G(t-u, \omega;\; \tau) \, f(u) \, du
     \label{eq-rel-gabor-filt-gabor-transf}
    = e^{i \omega t} \,  ({\cal G} f)(t, \omega;\; \tau).
  \end{split}
\end{align}
The top row in Figure~\ref{fig-gauss-limit-kern-graphs} shows Gaussian kernels at a few levels of
scale, and the top row in
Figure~\ref{fig-gabor-compl-limit-kern-graphs} shows the real and imaginary parts
of corresponding Gabor functions, obtained by multiplying the Gaussian
kernel with a complex sine wave.

By varying the temporal scale parameter $\tau$ in the Gabor transform, different trade-offs
can be obtained between the resolution in the temporal domain {\em vs.\/}\
the resolution in the frequency domain. A long temporal scale will
imply a high
resolution in the frequency domain, at the cost of a lower resolution in
the temporal domain, whereas a short temporal scale will increase the
resolution in the temporal domain, at the cost of a lower resolution in
the frequency domain. By using multiple values of the temporal scale
parameter in parallel, it is hence possible to obtain multiple
trade-offs between these conflicting requirements.
When constructing such time-frequency representations at multiple
scales, it is, however, essential that the resulting
representations are mutually compatible, which we will henceforth
show that they are, in terms of temporal scale covariance and temporal
cascade smoothing properties.

\subsection{Temporal scale covariance}

Under a scaling transformation of the temporal domain,
\begin{equation}
  t' = S \, t
\end{equation}
for some temporal scaling factor $S > 0$, the Gaussian kernel
transforms as
\begin{equation}
  \label{eq-sc-prop-gauss-function}
  g(t';\; \tau') = \frac{1}{S} \, g(t;\; \tau),
\end{equation}
provided that the temporal scale parameter $\tau'$ in the transformed
domain is related to the temporal scale parameter $\tau$ in the
original domain according to
\begin{equation}
  \tau' = S^2 \, \tau.
\end{equation}
This property implies that the Gabor function transforms
according to
\begin{equation}
  \label{eq-sc-prop-gabor-function}
  G(t', \omega';\; \tau') = \frac{1}{S} \, G(t, \omega;\; \tau),
\end{equation}
provided that the angular frequency $\omega'$ in the transformed
domain is related to the angular frequency $\omega$ in the original
domain according to
\begin{equation}
  \omega' = \frac{\omega}{S}.
\end{equation}
This transformation property does, in turn, mean that the Gabor transform
of a rescaled temporal signal
\begin{equation}
  f'(t') = f(t)  \quad\quad \mbox{for} \quad\quad t' = S \, t
\end{equation}
is related to the Gabor transform of the original signal according to
\begin{equation}
  ({\cal G} f')(t', \omega';\; \tau') = ({\cal G} f)(t, \omega;\; \tau), 
\end{equation}
with
\begin{multline}
  ({\cal G} f')(t', \omega';\; \tau') = \\ =
  \int_{u'=-\infty}^{\infty}
     f'(u') \,
    \frac{1}{\sqrt{2 \pi \tau'}} \,  e^{-(t'-u')^2/2\tau'} \, e^{-i \omega'  u'} \, du'.
\end{multline}
In other words, this temporal scale covariance property implies that
under a rescaling of the temporal domain, it is possible to perfectly
match the time-frequency representation of the rescaled signal to the
time-frequency representation of the original signal, provided that
the values of the angular frequency parameter $\omega$ and temporal scale
parameter $\tau$ are transformed accordingly.

\begin{figure}[h]
  \[
    \begin{CD}
       ({\cal G} f)(t, \omega;\; \tau)
       @>{\scriptsize\begin{array}{c} \scriptsize t' = S t  \\
                         \scriptsize \omega' = \omega/S \\
                         \scriptsize \tau' = S^2 \tau \end{array}}>> ({\cal G} f')(t', \omega';\; \tau') \\
       \Big\uparrow\vcenter{\rlap{$\scriptstyle{{*T(t;\; 
               \tau)}}$}} & &
       \Big\uparrow\vcenter{\rlap{$\scriptstyle{{*T(t';\; 
               \tau')}}$}} \\
       f(t) @>{\scriptsize t' = S t}>> f'(t')
    \end{CD}
 \]
\caption{Commutative diagram for the regular Gabor transform under
  temporal scaling transformations, for temporal scaling factors $S > 0$.
  (This commutative diagram should be
   read from the lower left corner to the upper right corner, and means
   that irrespective of whether we first rescale the input signal $f(t)$
   to a rescaled signal $f'(t')$ and then compute the Gabor transform, or
   first compute the Gabor transform and then rescale it, we get the same
   result, provided that the values of the angular frequency parameter
   and the temporal scale parameter are matched according to
$\omega' = \omega/S$ and $\tau' = S^2 \tau$.)}
\label{fig-comm-diag-gabor-transf}
\end{figure}

\subsection{Cascade smoothing property with a simplifying temporal
  kernel over temporal scales}

Due to the semi-group property of the Gaussian kernel
\begin{equation}
   g(\cdot;\; \tau_1) * g(\cdot;\; \tau_2) = g(\cdot;\; \tau_1 + \tau_2),
\end{equation}
it follows that the Gaussian kernel at a coarse temporal scale
$\tau_2$ is related to the Gaussian kernel at a finer temporal scale
\begin{equation}
   g(\cdot;\; \tau_2) =  g(\cdot;\; \tau_2- \tau_1) * g(\cdot;\; \tau_1).
\end{equation}
This property does, hence, imply that the Gabor transform
$({\cal G} f)(t, \omega;\; \tau_2)$  at a coarse temporal scale
$\tau_2$ is related to the Gabor transform
$({\cal G} f)(t, \omega;\; \tau_1)$  at any finer temporal scale
$\tau_1 < \tau_2$ according to
\begin{equation}
  \label{eq-casc-smooth-prop-gab-transf}
  ({\cal G} f)(\cdot, \omega;\; \tau_2)
  = g(\cdot;\; \tau_2- \tau_1) * ({\cal G} f)(\cdot, \omega;\; \tau_1).
\end{equation}
In other words, provided that the Gaussian kernel can be regarded as a
simplifying kernel from finer to coarser levels of temporal scales,
which it indeed is, due to the theory of continuous temporal
scale-space kernels%
\footnote{According to the theory of temporal scale-space kernels, the
  only non-trivial temporal smoothing kernels that guarantee that the number of local
  extrema in a signal cannot increase from finer to coarser levels of
  scales are Gaussian kernels, truncated exponential kernels and such
  kernels coupled in cascade (see Lindeberg \cite{Lin23-BICY} Section~2.1).}
in Lindeberg \cite{Lin23-BICY} Section~2.1,
this property implies that
the Gabor transform at any coarse temporal scale $\tau_2$  can be
regarded as a simplification of the corresponding Gabor transform
at any finer temporal scale $\tau_1 < \tau_2$.
This property constitutes a strong support for using the Gabor
transform over multiple temporal scales, to formulate a multi-scale
time-frequency analysis.

\subsection{Temporal shifting property}

Under a temporal shift
\begin{equation}
  f'(t') = f(t) \quad\quad \mbox{for} \quad\quad t' = t + \Delta t,
\end{equation}
the Gabor transform transforms according to
\begin{equation}
  ({\cal G} f')(t', \omega;\; \tau)
  =  ({\cal G} f)(t, \omega;\; \tau) \, e^{-i \omega \Delta t},
\end{equation}
corresponding to a temporal translation covariance property.
The Gabor transform is therefore closed under a shift of the temporal axis.

\section{Time-causal analogue of Gabor filtering}
\label{sec-time-caus-anal-gab-filt}

A fundamental problem when trying to apply the above Gabor filtering
theory to real-world temporal signals, however, is that the temporal
window function, {\em i.e.\/}\ the Gaussian kernel, is not
time-causal. The Gaussian kernel accesses values of the signal from the
future in relation to any temporal moment, and can therefore not be
used for processing real-time signals, for which the future cannot be
accessed.

In the following, we will describe a way to define a time-causal
analogue of Gabor filtering, by replacing the Gaussian kernel with a
specially chosen kernel, referred to as the time-causal limit kernel,
and which obeys both the properties of temporal scale covariance and a
cascade property over temporal scales, to ensure a well-founded
definition of a time-frequency analysis over multiple temporal scales,
also in the case of a truly time-causal temporal domain.
Specifically, for the resulting time-causal analogue of the Gabor
transform, the computations are also fully time-recursive, and
can be implemented as the composed convolution with a set of truncated
exponential kernels coupled in cascade, in turn, equivalent to a set of first-order integrators
coupled in cascade.

\subsection{Theoretical background concerning temporal scale-space kernels}

When to formulate a time-causal analogue of the Gaussian window
function, one may ask: ``Would any time-causal function do?''.
That is, however, not the case. Of crucial importance with respect to
a multi-scale analysis is that the transformation from a finer to a
coarser temporal scale does not introduce new temporal structures at
coarser levels of scale, not present at finer levels of scale.

In the area of temporal scale-space theory, developed by
Koenderink \cite{Koe88-BC},
Lindeberg and Fagerstr{\"o}m \cite{LF96-ECCV} and 
Lindeberg \cite{Lin16-JMIV}, \cite{Lin23-BICY}, this topic has been extensively
treated for the domain of regular temporal or spatio-temporal signals;
in other words, signals without a complementary explicit treatment of
local frequencies in the signal.

Specifically, the notion of a temporal scale-space kernel has been
defined as a kernel that guarantees that the number of local extrema
(or equivalently the number of zero-crossings) in a temporally
smoothed signal must not exceed the number of local extrema in the
original signal. Summarizing the treatment in
Lindeberg \cite{Lin23-BICY} Section~2.1,
it can, based on classical results by Schoenberg \cite{Sch50}, be
shown that a (not necessarily time-causal) 1-D continuous kernel is scale-space kernel if
and only if it can be decomposed into convolutions with a Gaussian kernel%
\footnote{Compare with the established definition of Gabor
  filtering, which precisely uses a Gaussian window function.}
and/or a set of truncated exponential kernels coupled in cascade.

By adding the complementary requirement of temporal causality, it
additionally follows that truncated exponential kernels
\begin{equation}
  \label{eq-trunc-exp-kernel}
    h_{\exp}(t;\; \mu_k) 
    = \left\{
        \begin{array}{ll}
          \frac{1}{\mu_k} \, e^{-t/\mu_k} & t \geq 0, \\
          0         & t < 0,
        \end{array}
      \right.
\end{equation}
coupled in cascade are the only possible temporal scale-space kernels,
that respect forbidden access to the future,
and thus the natural candidate window functions to be used, when to formulate a
time-causal analogue of Gabor filtering.

Each such primitive truncated exponential kernel does, in turn,
correspond to a first-order integrator of the form
\begin{equation}
  \label{eq-first-ord-int}
   (\partial_t f_{\outout})(t) 
   = \frac{1}{\mu_k} \left( f_{\inin}(t) - f_{\outout}(t)\right),
\end{equation}
where $f_{\inin}(t)$ denotes the input signal and $f_{\outout}(t)$ the output
signal. In this way, the temporal smoothing process, resulting from a
temporal scale-space kernel, will follow the input signal, although
not fully, and also with a certain amount of temporal delay, as
determined by the time constant $\mu_k$.

\subsection{The time-causal limit kernel}

Out of the multitude of possible ways of combining truncated
exponential kernels with different time constants in cascade, it can
be shown that a particular way of choosing the time constants does
additionally allow for simultaneous temporal scale covariance and a
cascade property over temporal scales.
Consider the temporal kernel defined as the infinite convolution of a
set of truncated exponential kernels in cascade, having a composed
Fourier transform of the form (see Lindeberg \cite[Equation~(25)]{Lin23-BICY})
\begin{align}
  \begin{split}
     \label{eq-FT-comp-kern-log-distr-limit}
     \hat{\Psi}(\omega;\; \tau, c) 
     & = \prod_{k=1}^{\infty} \frac{1}{1 + i \, c^{-k} \sqrt{c^2-1} \sqrt{\tau} \, \omega},
  \end{split}
\end{align}
where $c > 1$ is a distribution constant that determines the temporal
scale levels according to a geometric distribution.
Over the temporal domain, this function corresponds to an infinite
convolution of truncated exponential kernels in cascade%
\footnote{For actual implementation, as will be addressed later in
                Section~\ref{sec-disc-impl}, the infinite convolution
                can, however,
                because of the rapid convergence of the underlying
                geometric distribution of the time constants,
                for common choices of the distribution parameter $c = \sqrt{2}$
                or $2$, be truncated after the first 4 to 8 convolution stages.
                For discrete implementation, the
                truncated exponential kernels can furthermore be replaced by
                first-order recursive filters, which is
                computationally more efficient compared to explicit
                convolution, and also has excellent
                information-reducing properties.}%
\footnote{Unfortunately, it is hard to express a more explicit
  expression for the time-causal limit kernel over the temporal
  domain. By a partial fraction expansion of the corresponding Laplace
  transform of the kernel, it is, in principle, possible to rewrite the
  explicit convolution of truncated exponential kernels in cascade as
  a linear combination of truncated exponential kernels,
  as further detailed in Lindeberg \cite{Lin23-BICY} Section~6.1.2.
  Using such an explicit partical fraction expansion as the basis for
  actual implementations would, however, not be advisable,
  since the linear combination with a mixture of positive
  and negative weights would lead to a loss of accuracy, if the
  individual components in the linear combination are represented with finite precision.
  For actual implementations, digital or analog, the recommendation
  is instead to then start from the mathematically equivalent
  formulation as a set of first-order integrators of the form
  (\ref{eq-first-ord-int}) coupled in cascade. By the formulation of
  these first-order integrators as explicit damping feedback processes,
  the implementation that they will be rise to will be much more
  well-conditioned with respect to finite precision in the
  representations of the data.
  
  For certain types of approximate theoretical analysis, the time-causal limit kernel
  can, however, be analytically approximated by the scale-time kernel,
  earlier derived by Koenderink \cite{Koe88-BC}, as described in
  Lindeberg \cite{Lin23-BICY} Section~3.3, and which has a much more
  explicit closed-form expression for the dependency on the time
  variable $t$.}
\begin{equation}
  \Psi(t;\; \tau, c) = *_{k = 1}^{\infty} \frac{1}{\mu_k} \, e^{-t/\mu_k}
\end{equation}
where
\begin{equation}
  \mu_k = c^{-k} \sqrt{c^2-1} \sqrt{\tau}.
\end{equation}
Under convolution with the temporal smoothing kernel defined in this
way, a smaller value of the distribution parameter $c$ will lead to a
denser sampling of the temporal
scale levels, at the cost of longer temporal delays and and increased
amount of computations. A larger value of $c$ will, on the other hand,
lead to more
rapid temporal response properties (shorter temporal delays) as well
as less computations, however, at the cost of a more sparse sampling of
the temporal scales, which may affect the results, if a multi-scale
analysis also aims at tracking features across temporal scales.

When approximating the time-causal kernel using a finite number $K$ of
temporal scale levels, the temporal scale levels are instead determined according to
(see Lindeberg \cite[Equations~(10) and~(12)]{Lin23-BICY}),
\begin{equation}
  \label{eq-distr-tau-values}
  \tau_k = \sum_{j = 1}^k \mu_j^2
             = c^{2(k-K)} \tau_{\max} \quad\quad (1 \leq k \leq K),
\end{equation}
and with the finer scale time constants according to
(see Lindeberg \cite[Equations~(13)--(14)]{Lin23-BICY})
\begin{align}
  \begin{split}
     \label{eq-mu1-log-distr}
     \mu_1 & = c^{1-K} \sqrt{\tau_{\ max}}
  \end{split}\\
  \begin{split}
     \label{eq-muk-log-distr}
     \mu_k & = \sqrt{\tau_k - \tau_{k-1}} = c^{k-K-1} \sqrt{c^2-1} \sqrt{\tau_{\ max}} \quad (2 \leq k \leq K),
  \end{split}
\end{align}
where, thus, the finer scale
temporal scale levels $\tau_k$
will cluster infinitely dense near scale
$t \downarrow 0^+$ as $K \rightarrow \infty$.

As described in Lindeberg \cite[Section~3.1.3]{Lin23-BICY}, the
resulting temporal limit kernel, {\em the time-causal limit kernel\/}, defined as the
infinite convolution of truncated exponential kernels with time
constants $\mu_k$, is covariant under temporal scaling transformations
of the form
\begin{equation}
  t' = c^j t
\end{equation}
for integer values of $j$
\begin{equation}
  \label{eq-temp-sc-cov-time-caus-limit-kern}
  \Psi(t';\; \tau', c) = \frac{1}{c^j} \, \Psi(t;\; \tau, c),
\end{equation}
provided that the temporal scale parameters are matched according to
\begin{equation}
  \tau' = c^{2j} \tau.
\end{equation}
The time-causal limit kernel $\Psi(t;\; \tau, c)$ also obeys the following cascade
smoothing property over temporal scales
(see Lindeberg \cite[Equation~(28)]{Lin23-BICY})
\begin{equation}
  \label{eq-recur-rel-limit-kernel}
   \Psi(\cdot;\; \tau, c) = h_{\exp}(\cdot;\; \tfrac{\sqrt{c^2-1}}{c} \sqrt{\tau}) * \Psi(\cdot;\; \tfrac{\tau}{c^2}, c),
\end{equation}
here, for simplicity, expressed only as a relation between adjacent
levels of temporal scale ($\tau$ and $\tau/c^2$).

The middle and bottom rows in Figure~\ref{fig-gauss-limit-kern-graphs}
show examples of time-causal limit kernels for a few
combinations of the temporal scale values $\sigma = \sqrt{\tau}$ and the
distribution parameter $c$.

\subsection{Time-causal analogue of the Gabor filter: The complex-valued extension of the time-causal limit kernel}

If we multiply the time-causal limit kernel with a complex sine wave $e^{i \omega t}$,
then we obtain the following complex-valued extension of the time-causal
limit kernel
\begin{equation}
  \label{eq-compl-time-caus-limit-kern}
  \chi(t, \omega;\; \tau, c) = \Psi(t;\; \tau, c) \, e^{-i \omega t},
\end{equation}
which can be seen as a time-causal analogue of the Gabor function.

The middle and bottom rows in
Figure~\ref{fig-gabor-compl-limit-kern-graphs}
show examples of such time-frequency kernels for a few combinations of
temporal scale values $\sigma = \sqrt{\tau}$ and angular frequencies
$\omega$.

\subsection{Time-causal analogue of the Gabor transform}

Motivated by the above construction, we can also define a time-causal
analogue of the Gabor transform (\ref{eq-gab-transf}) according to
\begin{equation}
  \label{eq-time-caus-anal-gab-transf}
  ({\cal H} f)(t, \omega;\; \tau, c) =
  \int_{u=-\infty}^{t}
     f(u) \,
    \Psi(t-u;\; \tau, c) \, e^{-i \omega  u} \, du.
\end{equation}
Note specifically, that, by the definition of the temporal smoothing kernel $\Psi(t;\; \tau, c)$
in terms of a set of truncated exponential kernels coupled in cascade,
in turn equivalent to a set of first-order integrators in cascade,
the temporal smoothing process, needed to implement this operation in
practice, can be performed in a fully time-recursive manner,
in the sense than no other temporal memory of the past is needed than
the information contained in the multi-scale representation of the
primitive temporal smoothing stages over multiple temporal scales itself.
In this respect, the time-causal analogue of the Gabor transform lends
itself to real-time modelling of physical or biological processes, as
well as to real-time processing of measurement signals, by computations
that are inherently local over time.

In analogy with the relationship (\ref{eq-rel-gabor-filt-gabor-transf}) between Gabor filtering
and the Gabor transform, the time-causal analogue of the Gabor
transform is related to filtering with the complex-valued extension of
the time-causal limit kernel (\ref{eq-compl-time-caus-limit-kern}) according to
\begin{align}
  \begin{split}
     &(\chi(\cdot, \omega;\; \tau, c) * f(\cdot))(t, \omega;\; \tau, c)
  \end{split}\nonumber\\
  \begin{split}
     & = \int_{u=-\infty}^{t} \chi(t-u, \omega;\; \tau, c) \, f(u) \, du
    = e^{i \omega t} \,  ({\cal H} f)(t, \omega;\; \tau, c).
  \end{split}
\end{align}
Thus, we can interpret the kernels shown in the bottom four rows in
Figure~\ref{fig-gabor-compl-limit-kern-graphs} as describing the
essential effect of the time-frequency filtering operations that take
place in the time-causal analogue of the Gabor transform.

A derivation of an inverse transform for the time-causal analogue
of the Gabor transform is given in
Appendix~\ref{sec-app-inv-transf-time-caus-gabor-transf}, as well as
an outline of how to define other possible inverse transforms of this
overcomplete time-frequency transform.

\section{Theoretical properties of the time-causal analogue of Gabor
  filtering}
\label{sec-theor-prop-time-caus-gabor-filt}

In this section, we will describe a set of theoretical properties that
the complex-valued extension of the time-causal limit kernel and the
time-causal analogue of the Gabor transform obey, with regard to
defining a time-frequency analysis over multiple levels of temporal
scales.

\subsection{Theoretical symmetry properties that lead to the time-causal limit kernel as a canonical
  temporal smoothing kernel over a time-causal temporal domain}

A complementary motivation for why the proposed 
time-frequency analysis concept can be regarded as a time-causal
analogue of the Gabor transform is given in
Appendix~\ref{sec-arg–why-time-caus-anal}.

In that appendix, first of all, a set of principled theoretical symmetry
arguments is presented, regarding theoretically well-founded
processing of signals over different temporal scales, that uniquely single out
the choice of the continuous Gaussian kernel for defining a
time-frequency transform over a non-causal temporal domain, where the
relative future in relation to any pre-recorded time moment can indeed
be accessed. These principled arguments, thus, uniquely lead to the
Gabor transform as a canonical time-frequency analysis concept over a
non-causal temporal domain.

Then, it is shown how the formulation of as similar principled
theoretical symmetry arguments as possible, given the additional
contraints due to the restriction to time-causal window functions
only, for which the future cannot be accessed,
lead to the choice of the time-causal limit kernel as a
canonical temporal window function for formulating a multi-scale
time-frequency analysis concept over a time-causal temporal domain.
These theoretical arguments do, thus, show how the proposed time-causal
time-frequency analysis concept can be determined in a principled way,
based on desirable symmetry properties with respect to processing over
multiple temporal scales.

Over purely either non-causal or time-causal temporal domains, as
well as over joint either non-causal or time-causal spatio-temporal
domains, corresponding principled ways of reasoning, based on related 
theoretical symmetry properties, can be shown to single out canonical
classes of temporal smoothing kernels in models for either purely
temporal or joint spatio-temporal models of receptive fields,
see Lindeberg \cite{Lin16-JMIV,Lin17-JMIV, Lin21-Heliyon,Lin23-BICY}.
Specifically, the Gaussian kernel is in these ways singled out as the
canonical temporal smoothing kernel over a non-causal temporal domain,
whereas the time-causal limit kernel arises as a canonical choice of
temporal smoothing kernel over a time-causal temporal domain.

In these respects, there are conceptual similarities between the proposed
way of defining a time-causal time-frequency analysis, and previously
developed approaches for processing either purely temporal signals or
joint spatio-temporal video in a strictly time-causal manner.

\subsection{Temporal scale covariance}

Under a scaling transformation of the temporal domain,
\begin{equation}
  t' = S \, t
\end{equation}
for some temporal scaling factor $S = c^j$,
for some $c > 1$ and any $j \in \bbbz$, the time-causal analogue of the
Gabor function, {\em i.e.\/}, the complex-valued extension of the
time-causal limit kernel (\ref{eq-compl-time-caus-limit-kern}), transforms according to
\begin{equation}
  \chi(t', \omega';\; \tau', c) = \frac{1}{S} \, \chi(t, \omega;\; \tau, c),
\end{equation}
provided that the angular frequency is transformed according to
\begin{equation}
  \omega' = \frac{\omega}{S}.
\end{equation}
and the temporal scale parameter according to
\begin{equation}
  \tau' = S^2 \, \tau.
\end{equation}
This property follows from the scaling property (\ref{eq-temp-sc-cov-time-caus-limit-kern})
of the time-causal limit kernel under temporal scaling transformations for $S = c^j$, where $j$ must be an integer.

Similarly, the time-causal analogue of the Gabor transform
(\ref{eq-time-caus-anal-gab-transf}) applied to a rescaled temporal input signal
\begin{equation}
  f'(t') = f(t) \quad\quad \mbox{for} \quad\quad t' = S \, t
\end{equation}
is related to the time-causal analogue of the Gabor transform applied
to the original signal according to
\begin{equation}
  ({\cal H} f')(t', \omega';\; \tau') = ({\cal H} f)(t, \omega;\, \tau), 
\end{equation}
with
\begin{multline}
  ({\cal H} f')(t', \omega';\; \tau', c) = \\ =
  \int_{u'=-\infty}^{t'}
     f'(u') \,
    \Psi(t'-u';\; \tau', c) \, e^{-i \omega'  u'} \, du'.
\end{multline}
In these ways, both the complex-valued extension of the time-causal
limit kernel and the time-causal analogue of the Gabor transform obey
similar transformation properties under temporal scaling
transformations of a time-causal temporal domain as the regular Gabor function
and the regular Gabor transform obey over a non-causal temporal domain.

\begin{figure}[hbt]
  \[
    \begin{CD}
       ({\cal H} f)(t, \omega;\; \tau, c)
       @>{\scriptsize\begin{array}{c} \scriptsize t' = S t  \\
                         \scriptsize \omega' = \omega/S \\
                         \scriptsize \tau' = S^2 \tau \end{array}}>> ({\cal H} f')(t', \omega';\; \tau', c) \\
       \Big\uparrow\vcenter{\rlap{$\scriptstyle{{*T(t;\; 
               \tau)}}$}} & &
       \Big\uparrow\vcenter{\rlap{$\scriptstyle{{*T(t';\; 
               \tau')}}$}} \\
       f(t) @>{\scriptsize t' = S t}>> f'(t')
    \end{CD}
 \]
\caption{Commutative diagram for the time-causal analogue of the Gabor transform under
  temporal scaling transformations for scaling factors $S$ that are
  integer powers of the distribution parameter $c$ of the time-causal
  limit kernel, {\em i.e.\/}, $S = c^j$ for integer $j$ for some $c > 1$.
  (This commutative diagram should be
   read from the lower left corner to the upper right corner, and means
   that irrespective of whether we first rescale the input signal $f(t)$
   to a rescaled signal $f'(t')$ and then compute the time-causal
   analogue of the Gabor transform, or
   first compute the time-causal analogue of the Gabor transform
   and then rescale it, we get the same
   result, provided that the values of the angular frequency parameter
   and the temporal scale parameter are matched according to
$\omega' = \omega/S$ and $\tau' = S^2 \tau$.)}
\label{fig-comm-diag-gabor-transf}
\end{figure}

\subsection{Cascade smoothing property with a time-causal simplifying
  temporal kernel over temporal scales}

Based on the cascade smoothing property over temporal scales
(\ref{eq-recur-rel-limit-kernel}) for the time-causal limit kernel
(\ref{eq-FT-comp-kern-log-distr-limit}),
it follows that the time-causal analogue of the Gabor transform will obey a similar cascade
smoothing property
\begin{equation}
  \label{eq-recur-rel-limit-kernel}
  ({\cal H} f)(t, \omega;\; \tau, c)
  = h_{\exp}(\cdot;\; \tfrac{\sqrt{c^2-1}}{c} \sqrt{\tau})
  * ({\cal H} f)(t, \omega;\; \tfrac{\tau}{c^2}, c),
\end{equation}
which in turn can be applied recursively, to express a cascade
smoothing property from any finer to any coarser (discrete) levels of
temporal scale $\tau_k  = c^{2k}$ in the time-causal analogue of the Gabor transform.

In this respect, the time-causal analogue of the Gabor transform obeys
a structurally similar cascade smoothing property over temporal scales
over a time-causal temporal domain
as the regular Gabor transform (\ref{eq-casc-smooth-prop-gab-transf})
obeys over a non-causal temporal domain,
with the minor differences that (i)~the non-causal Gaussian kernel is
replaced%
\footnote{Note in this context that both the non-causal Gaussian
  kernel and the time-causal truncated exponential kernels coupled in
  cascade are continuous scale-space kernels, in the sense that they
  guarantee a simplification property from finer to coarser levels of
  temporal scale in any temporal signal
  (see Lindeberg \cite{Lin23-BICY} Section~2.1).}
by a set of truncated exponential kernels coupled in cascade,
and (ii)~the temporal scale levels are discrete in the time-causal
analogue of the Gabor transform, whereas the temporal scale levels are
defined over a continuum for the regular Gabor transform.

\subsection{Temporal shifting property}

Under a temporal shift
\begin{equation}
  f'(t') = f(t) \quad\quad \mbox{for} \quad\quad t' = t + \Delta t,
\end{equation}
the time-causal analogue of the Gabor transform according to
\begin{equation}
  ({\cal H} f')(t', \omega;\; \tau, c)
  =  ({\cal H} f)(t, \omega;\; \tau, c) \, e^{-i \omega \Delta t}.
\end{equation}
In other words, the time-causal analogue of the Gabor transform is covariant under shifts of the
temporal axis.

\subsection{Relation to the Heisenberg group}

These properties together imply that the time-frequency representation
obtained from the time-causal analogue of the Gabor transform has the
theoretically attractive properties that it is closed under
(i)~temporal shifts and (ii)~uniform rescalings of the temporal axis
for temporal scaling factors that are integer powers of the
distribution parameter $c$ for the time-causal limit kernel.

In these respects, the presented method for time-causal frequency analysis obeys essentially similar
transformation properties as the regular Gabor transform over a
non-causal temporal domain, and as can
be described by the Heisenberg group
(see Feichtinger and Gr{\"o}chenig \cite{FeiGro92-Wavelets}).

\section{Discrete implementation}
\label{sec-disc-impl}

When to implement the above operations for processing discrete signals
in practice, there does indeed also exist a
corresponding discrete theory of temporal scale-space kernels to build
upon.

\subsection{First-order recursive filters coupled in cascade}

Following the results of an axiomatic classification of discrete temporal scale-space
kernels, characterized by the property that they are guaranteed to not
increase the number of local extrema or zero-crossings in a discrete
signal, in Lindeberg \cite{Lin23-BICY} Section~4.1, in turn based on
classical results by Schoenberg \cite{Sch48}, with these arguments
summarized in Appendix~\ref{app-class-temp-scsp-kern},
we do for discrete signals replace the temporal smoothing operation in
the time-causal limit kernel with a
cascade of first-order recursive filters
normalized to the form
\begin{equation}
  \label{eq-norm-update}
  f_{\outout}(t) - f_{\outout}(t-1)
  = \frac{1}{1 + \mu_k} \,
    (f_{\inin}(t) - f_{\outout}(t-1)),
\end{equation}
which constitutes both the discrete analogue of the first-order continuous integrators
(\ref{eq-first-ord-int}), and which is also maximally
well-conditioned with respect to possible numerical
errors.

Note in particular in this context that, since the time
constant $\mu_k$ is positive, the feedback factor in the recursive
update is guaranteed to always be less than 1, thus guaranteeing that
spurious perturbations in the discrete measurement signal or due to 
numerical errors in the computations are guaranteed to be reduced over
time. This damping effect is, of course, stronger for larger values of
the time constant $\mu_k$.

With respect to the numerical stability of this implementation method, it
does, furthermore specifically follow, from the concept of temporal
scale-space kernel itself,  that the number of local extrema
(or equivalently the number of zero-crossings) is guaranteed to not
increase by these discrete filtering operations. In this way, the
resulting discrete implementation is guaranteed to serve as
a formal smoothing transformation over time, to gradually suppress the
influence of spurious fine-scale structures in the signal, such as
local perturbations or noise.

The resulting discrete approximation can also be formally be shown to
correspond to a true numerical approximation of the corresponding
continuous theory, see Appendix~\ref{sec-proof-1st-order-int-nu-approx}.

\subsection{Determination of the discrete time constants $\mu_k$}
Then, to obtain the same amount of temporal smoothing as in the
continuous case, we make use of
the fact that the variance of a single first-order integrator is
\begin{equation}
   \Delta \tau_k = \mu_k^2 + \mu_k
\end{equation}
and compute $\mu_k$ from the desired scale increment%
\footnote{For the modelling of multiple temporal scale levels over a
  set of temporal window functions coupled in cascade, we make use of
  the fact that the temporal variances of the convolution kernels are
  additive, provided that the convolution kernels are non-negative.}
$\Delta \tau_k$
between adjacent levels of temporal scale levels
\begin{equation}
   \Delta \tau_k = \tau_k - \tau_{k-1} =c^{2k} - c^{2(k-1)}
\end{equation}
according to Lindeberg \cite{Lin23-BICY} Equation~(55)
\begin{equation}
  \label{eq-disc-time-constant}
  \mu_k = \frac{\sqrt{1 + 4 \Delta \tau_k}-1}{2},
\end{equation}
see Appendix~\ref{app-discr-impl} for a detailed
outline of how to implement these computations in practice.

By this way of determining the time constants $\mu_k$ for the discrete
temporal smoothing operations, the discrete variance of the discrete temporal
smoothing kernels, that define the temporal window functions in the resulting
discrete analogue of the time-causal analogue of the Gabor transform
\begin{multline}
  V(h(\cdot;\; \tau)) = \\
  = \frac{\sum_{n\in \bbbz} n^2 \, h(n;\; \tau)}{\sum_{n \in \bbbz} h(n;\; \tau)}
  - \left(
    \frac{\sum_{n \in \bbbz} n \, h(n;\; \tau)}{\sum_{n \in \bbbz} h(n;\; \tau)}
  \right)^2 = \tau,
\end{multline}
will be equal to the continuous variance of the corresponding
continuous smoothing kernels
\begin{multline}
  V(h(\cdot;\; \tau)) = \\
  = \frac{\int_{t \in \bbbr} x^2 \, h(t;\; \tau) \, dt}{\int_{x \in \bbbr} h(t;\; \tau) \, dt}
  - \left(
        \frac{\int_{t \in \bbbr} x \, h(t;\; \tau) \, dt}{\int_{x \in \bbbr} h(t;\; \tau) \, dt}
     \right)^2 = \tau,
\end{multline}
that define the temporal window functions in the true continuous
time-causal analogue of the Gabor transform. In this respect, the
notion of temporal scale, as reflected by the temporal duration
of the temporal window function, is preserved between the
continuous theory and the discrete implementation.

\subsection{Strictly time-recursive implementation for real-time implementation}

Specifically, all the temporal smoothing computations needed to
implement the time-causal analogue of the Gabor transform will
therefore be local over time, and do not require any complementary
memory of the past beyond the last previous time frame,
since all the necessary information of the past is
stored in the temporal multi-scale representation itself.
In this way, the time-causal analogue of the Gabor transform is fully
time recursive, and lends itself to real-time processing of signals
using a compact signal processing architecture.

Appendix~\ref{sec-app-real-time-alg} gives a detailed outline of
how to implement the temporal smoothing operations in terms
of strictly time-recursive computations for real-time applications.

\subsection{Default parameter settings}

For practical implementations, we usually choose the distribution parameter
$c$ of the time-causal limit kernel as $c = \sqrt{2}$ or $2$, and approximate the time-causal limit
kernel at the finest level of scale in a multi-scale analysis using
the first $4$ to $8$ first-order recursive filters having the longest
time constants.

Due to the exponential decrease of the time constants
as function of the scale depth, the numerical error caused by such a
truncation will for many purposes be quite moderate. If a more
accurate numerical implementation is needed for special purposes, then
the truncation error can be reduced by increasing the number of temporal
scale levels, however, at the cost of more computational work.

\section{Properties of the time-causal analogue of the Gabor
  transform}
\label{sed-theor-props-time-caus-gabor}

In this section, we will describe properties of the proposed
time-causal analogue of the Gabor transform, based on a theoretical
analysis of the temporal delays and the frequency selectivity
characteristics of the proposed methodology for
time-causal time-frequency analysis.

\begin{figure*}[hbtp]
  \begin{center}
    \begin{tabular}{ccc}
      {\footnotesize\em Gabor ($N = 4$)}
      & {\footnotesize\em time-causal ($N = 4$, $c = \sqrt{2}$)}
      & {\footnotesize\em time-causal ($N = 4$, $c = 2$)}  \\
      \includegraphics[width=0.30\textwidth]{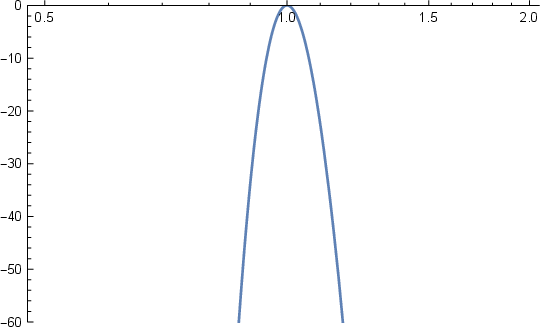}
      & \includegraphics[width=0.30\textwidth]{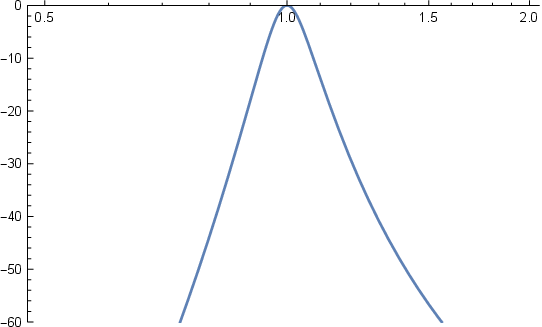}
      & \includegraphics[width=0.30\textwidth]{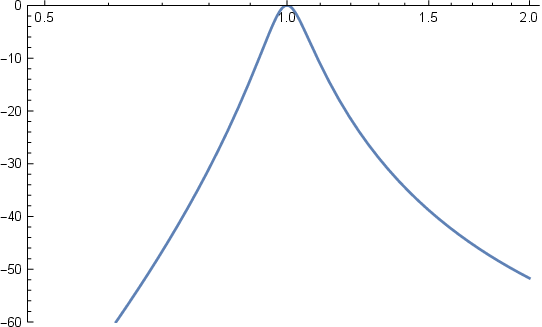} \\
      {\footnotesize\em Gabor ($N = 8$)}
      & {\footnotesize\em time-causal ($N = 8$, $c = \sqrt{2}$)}
      & {\footnotesize\em time-causal ($N = 8$, $c = 2$)}  \\
      \includegraphics[width=0.30\textwidth]{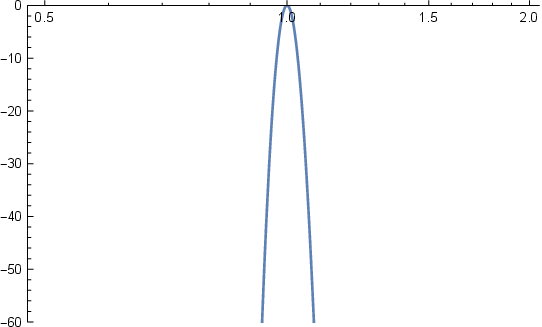}
      & \includegraphics[width=0.30\textwidth]{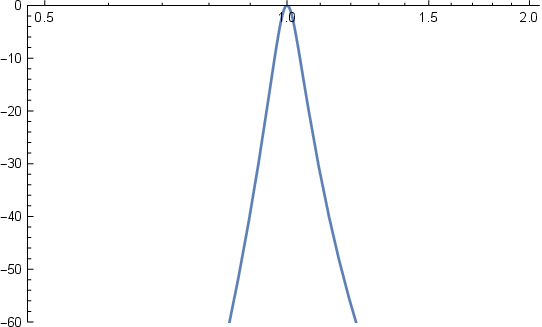}
      & \includegraphics[width=0.30\textwidth]{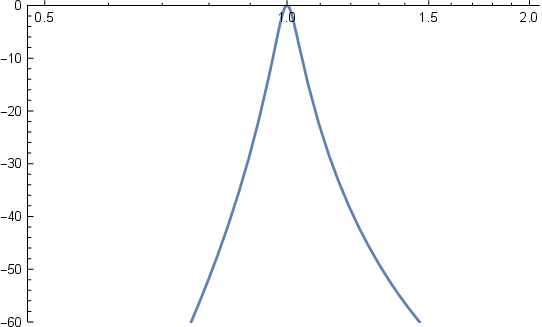} 
    \end{tabular}
  \end{center}
  \caption{Frequency selectivity properties of the Gabor transform and the
    time-causal analogue of the Gabor transform, based on the entity
    $R(\omega) = |\hat{h}(\omega - \omega_0;\; \tau(\omega))|$
    in (\ref{eq-freq-sel-entity}), for different values of the
    wavelength proportionality factor $N$ in
    (\ref{eq-def-tau-from-omega-N}) and different values of the
    distribution parameter $c$ in the time-causal limit kernel.
    As can be seen from these graphs, the frequency selectivity
    becomes more narrow for larger values of $N$, which in turn correspond to
    temporal window functions at coarser temporal scales.
    The frequency selectivity of the time-causal analogue of the Gabor
    transform also becomes more narrow when the distribution parameter $c$
    is decreased, however, then at the cost of longer temporal delays.
    Choosing appropriate parameter settings does in this respect
    correspond to a trade-off, which should be balanced for any given application.
    (Horizontal axis: angular frequency in the time-frequency
    transform, on a logarithmic scale, and
    relative to a reference angular frequency of $\omega_0 = 1$.
    Vertical axis: Value of
    $R(\omega;\ \tau) = |\hat{h}(\omega - \omega_0;\; \tau(\omega))|$ in dB.)}
  \label{fig-freq-selectivity}
\end{figure*}

\subsection{Temporal delays}
\label{sec-temp-delay}

Due to the temporal causality of the temporal window function in the
time-causal analogue of the Gabor transform, all such measurements
from a signal will be associated with non-zero temporal delays.

To estimate the temporal delay of the time-causal limit kernel, we can
estimate the temporal mean of the continuous time-causal limit kernel
as
\begin{equation}
  \label{eq-approx-temp-mean-time-caus-log-distr}
  m
  = \sum_{k=1}^{\infty} \mu_k
  = \sum_{k=1}^{\infty} c^{-k} \sqrt{c^2-1} \sqrt{\tau}
  = \sqrt{\frac{c+1}{c-1}} \sqrt{\tau}.
\end{equation}
Based on an approximation of the time-causal limit kernel in
terms of Koenderink's scale-time kernel \cite{Koe88-BC},
the temporal location of the temporal maximum of the time-causal limit
kernel can also be estimated as
(see Lindeberg \cite{Lin23-BICY} Equation~(39))
\begin{equation}
  \label{eq-approx-temp-pos-time-caus-log-distr}
  t_{\max} \approx
  \frac{(c+1)^2 \, \sqrt{\tau}}{2 \sqrt{2} \sqrt{(c-1) \, c^3}}.
\end{equation}
The latter estimate corresponds to a slight overestimate, while constituting
a better estimate of the temporal delay than the temporal mean.

From these expressions, we can clearly see that the amount of
temporal delay is proportional to the temporal scale of the
temporal window function in units of the standard deviation
of the temporal smoothing kernel $\sigma = \sqrt{\tau}$.
We can also see that the temporal delay becomes shorter, when
increasing the value of the distribution parameter $c$ in the
time-causal limit kernel.

For the two default values of the distribution parameter
$c = \sqrt{2}$ and $c = 2$ considered in this work, these temporal
delay estimates do specifically assume the values according to
Table~\ref{tab-est-temp-delay}, based on estimates in terms of the mean value
$\hat{\delta} = m$ or the temporal maximum $\hat{\delta} = t_{\max}$.

\begin{table}[hbtp]
  \begin{center}
    \begin{tabular}{ccc}
      \hline
      \multicolumn{3}{c}{Estimated temporal delay}\\
      \hline
      $c$ & $m$ & $t_{\max}$ \\
      \hline
      $\sqrt{2}$ & 2.414 & 1.904 \\
      $2$           & 1.732 & 1.125 \\
      \hline
    \end{tabular}
  \end{center}
  \caption{Numerical estimates of the temporal delay for the
    time-causal and time-recursive analogue of the Gabor transform,
    in units of $\sigma = \sqrt{\tau}$ and based
    on either the temporal mean $m$ or the temporal maximum point
    $t_{\max}$ for the time-causal limit kernel, according to
    (\ref{eq-approx-temp-mean-time-caus-log-distr}) and
    (\ref{eq-approx-temp-pos-time-caus-log-distr}).}
  \label{tab-est-temp-delay}
\end{table}

Specifically, the ratio between the temporal delays, based on the
position of the temporal maximum of the time-causal limit kernel
for the two default values $c = \sqrt{2}$ or $c = 2$, is given by
\begin{equation}
  \frac{t_{\max} |_{c=\sqrt{2}}}{t_{\max} |_{c=2}}
  \approx \frac{\frac{(\sqrt{2} + 1)^2}{\sqrt{(\sqrt{2} - 1) \sqrt{2}^3}}}
  {\frac{(2 + 1)^2}{\sqrt{(2 - 1) \, 2^3}}}
  \approx 1.692.
\end{equation}
In other words, the estimate of the temporal delay based on the
temporal position of the temporal maximum point of the temporal
smoothing kernel is for the same value of the temporal duration
of the temporal smoothing kernel in terms of its temporal variance $\tau$
about 70~\% longer when using $c = \sqrt{2}$
compared to using $c =2$.%
\footnote{When estimating the temporal delay from the temporal mean of
  the time-causal limit kernel, the ratio between the temporal delays
  for the two default values $c = \sqrt{2}$ or $c = 2$ is instead
  given by
  $(m |_{c=\sqrt{2}})/(m |_{c=2})
  = \sqrt{(\sqrt{2} + 1)/(\sqrt{2} - 1)}/\sqrt{(2 + 1)/(2 - 1)}
  \approx 1.394$. In other words, when measured in this way,
  the estimated temporal delay is about
40~\% longer when when using $c = \sqrt{2}$
compared to using $c =2$.}
Thus, if aiming at reducing the temporal
delays for real-time processing,
then using $c =2$ is very much preferably compared to using $c =
\sqrt{2}$.

Furthermore, since these estimates of the temporal delay are
proportional to the temporal standard deviation $\sigma = \sqrt{\tau}$ of the
time-causal limit kernel, it follows that, when choosing the temporal
standard deviation of the time-causal limit kernel proportional to the
wavelength $\lambda$ corresponding to the angular frequency $\omega$
in the time-frequency transform, as can be motivated from requirement
of temporal scale covariance, the temporal delays will be
different and specifically inversely proportional to each angular frequency. In
this respect, the time-causal analogue of the Gabor transform differs
fundamentally from the regular non-causal Gabor transform, in that special
attention has to be put on handling the different temporal delays for the
different angular frequencies $\omega$, that will result in a time-causal time-frequency
transform defined from the principled requirement of temporal scale covariance.

\subsection{Frequency selectivity properties of the resulting
  time-causal spectrograms}
\label{sec-freq-sel}

When using the time-causal analogue of the Gabor transform for
estimating local frequencies based on spectrograms, it is as
previously mentioned natural to
let the temporal scale parameter in units of the temporal standard
deviation $\sigma = \sqrt{\tau}$ be proportional to the wavelength
$\lambda = 2\pi/\omega$ of the angular frequency.
Denoting this proportionality factor by $N$, we should thus have:
\begin{equation}
  \label{eq-def-tau-from-omega-N}
  \tau(\omega) = (\sigma(\omega))^2
  = \left( \frac{2 \pi N}{\omega} \right)^2.
\end{equation}
If we use a temporal-scale-dependent temporal window function $h(t;\; \tau)$
for defining a time-frequency analysis
\begin{equation}
  ({\cal H} f)(t, \omega;\; \tau) =
  \int_{u=-\infty}^{\infty}
     f(u) \,
     h(t -u;\; \tau) \, e^{-i \omega  u} \, du,
\end{equation}
from which we then define the spectrogram from the absolute value
\begin{equation}
  |({\cal H} f)(t, \omega;\; \tau)|,
\end{equation}
and then want to measure the response to a sine wave of a given
angular frequency
\begin{equation}
  f(t) = \sin(\omega_0 \, t),
\end{equation}
then it can be shown
(see Lindeberg and Friberg \cite{LinFri15-PONE} Equation~(114),
as well as Section~\ref{sec-theor-anal-freq-sel-sine-wave} in the present
paper, specifically the extended derivation leading to
Equations~(\ref{eq-spectrogram-resp-single-sine-wave})
and~(\ref{eq-approx-spectr-R-in-deriv}))
that the dominant component in the frequency response, that describes
this dependency for values of $\omega$ near $\omega_0$, is given by the entity
\begin{equation}
  \label{eq-freq-sel-entity}
  R(\omega;\; \tau) = \left|\hat{h}(\omega - \omega_0;\; \tau(\omega))\right|,
\end{equation}
which on a logarithmic dB scale assumes the form
\begin{equation}
  R_{\dB}(\omega;\; \tau) = 20 \log_{10} |\hat{h}(\omega - \omega_0;\; \tau(\omega))|.
\end{equation}
The variability in this entity, as function of the free angular
frequency $\omega$, does, hence, describe the frequency
selectivity of the spectrogram, based on the time-frequency analysis
concept for the temporal window function $h(\omega;\; t)$.

For the regular non-causal Gabor transform, we have
\begin{equation}
  R_{\Gabor}(\omega;\; N) = e^{- \frac{2\pi^2 N^2  (\omega-\omega_0)^2}{\omega^2}} 
\end{equation}
or in dB
\begin{align}
  \begin{split}
  R_{\dB,\Gabor}(\omega;\; N) 
  & = - \frac{40 \pi^2 \, N^2 \, (\omega-\omega_0)^2}{\log 10 \, \omega^2}
   \end{split}\nonumber\\
  \begin{split}
  & = - \frac{40 \pi^2 \, N^2 }{\log 10} \left(1 - \frac{\omega_0}{\omega}\right)^2,
  \end{split}
\end{align}
whereas for the time-causal analogue of the Gabor transform, we obtain
\begin{equation}
  \label{eq-R-time-caus-tau-prop-N}
  R_{\timecaus}(\omega;\; N, c) =
  \frac{1}
          {\prod_{k=1}^{\infty}
                  \sqrt{
                  1 
                  + \frac{4 \pi^2 \, c^{-2k} \, (c^2-1) N^2 (\omega-\omega_0)^2}{\omega^2}}}
\end{equation}
or in dB
\begin{multline}
  R_{\dB,\timecaus}(\omega;\; N, c) =
- \frac{10}{\log 10} \times \\
              \sum_{k=1}^{\infty}
                 \log 
                    \left(
                            1 
                            + 4 \pi^2 \, c^{-2k} \, (c^2-1) \, N^2 
                               \left(1 - \frac{\omega_0}{\omega}\right)^2
                    \right).
\end{multline}
Figure~\ref{fig-freq-selectivity} shows the result of plotting these entities for
characterizing the frequency selectivity for the two 
values of the wavelength proportionality constant $N = 4$ and $N = 8$,
and also using the two different values of the distribution parameter
$c \in \{ \sqrt{2}, 2 \}$ for the time-causal limit kernel.

First of all, an immediate consequence of choosing the temporal
standard deviation $\sigma = \sqrt{\tau}$ of the temporal window
function proportional to the wavelength $\lambda$ corresponding to the
angular frequency $\omega$ is that the relative width of any spectral
band, as measured in terms of logarithmic angular frequencies
$\log(\omega/\omega_0)$,
will be independent of the angular frequency $\omega_0$ of any probing sine wave signal.
Again, this property reflects the basic requirement of temporal scale
covariance, so as to be able to handle input signals with different
frequency characteristics in a similar manner.

Furthermore, as can be seen from the graphs, the
frequency selectivity for the time-causal spectrogram based on the
time-causal analogue of the Gabor transform is less narrow than for the
spectrogram based on the regular Gabor transform.

The frequency selectivity also becomes less narrow when increasing the
distribution parameter $c$ in the time-causal limit kernel, used as the
temporal window function for defining the time-causal spectrograms.
In this respect, there is a trade-off issue in that a higher value
of the distribution parameter $c$ leads to shorter temporal delays,
while then also making the frequency selectivity less narrow.

The difference widths of the spectral bands obtained in these ways do,
in turn, affect the ability of the corresponding time-frequency
transforms to resolve nearby frequencies, in that
more narrow spectral bands decrease the interference effects between
nearby frequencies in the time-frequency transform, compared to
corresponding interference effects between nearby frequencies
caused by wider spectral bands.

Fundamentally, we cannot expect the time-causal analogue of the Gabor
transform to have as narrow frequency selectivity properties as the
regular Gabor transform, since the time-causal analogue of the Gabor
transform can only make use of information about what has occurred in
the past, whereas the Gabor transform also grabs information from
the relative future in relation to any pre-recorded time moment. Additionally, the
Gaussian temporal window function in the Gabor transform corresponds
to an infinitely divisible distribution over a continuum of temporal
scale increments, whereas the time-causal limit kernel used as
temporal window function in the time-causal limit kernel is based
on macroscopic non-infinitesimal scale steps.

The frequency
selectivity of the time-causal limit kernel can, however, be steered
to become more narrow, by decreasing the distribution parameter
$c$ down towards 1. Then, however, the temporal delay will increase,
and additionally more computational work will also be needed to
implement the corresponding time-frequency analysis method,
since a larger number of layers will be needed to sufficiently well
approximate the time-causal limit kernel with a truncated finite
number of layers, when the distribution parameter $c$ is closer to 1.

Additionally, the frequency selectivity does also, for spectrograms
based on both the non-causal Gabor transform and the time-causal
analogue of the Gabor transform, become sharper when increasing the
proportionality constant $N$ relative to the wavelength. Variations of
this parameter do thus also lead to a similar trade-off issue, in that a
larger value of $N$ will lead to sharper frequency selectivity
properties, while simultaneously increasing the temporal delay, since
the temporal scale of the temporal window function will increase.

From a requirement of being able to form the basis for making
decisions about possible actions in a real-time scenario, these
properties thus imply that in real-time scenarios it could be very
valuable to simultaneously perform multiple time-frequency analysis
stages over multiple time scales, which the proposed time-causal
analogue of the Gabor transform is highly suitable for.

\begin{figure*}[hbtp]
  \begin{center}
    {\bf Time-causal spectrograms}

    \bigskip
    
    \begin{tabular}{cc}
      {\footnotesize\em Spectrogram over shorter temporal scales for $c = \sqrt{2}$}
      &  {\footnotesize\em Spectrogram over shorter temporal scales for $c = 2$}\\
      \includegraphics[width=0.48\textwidth]{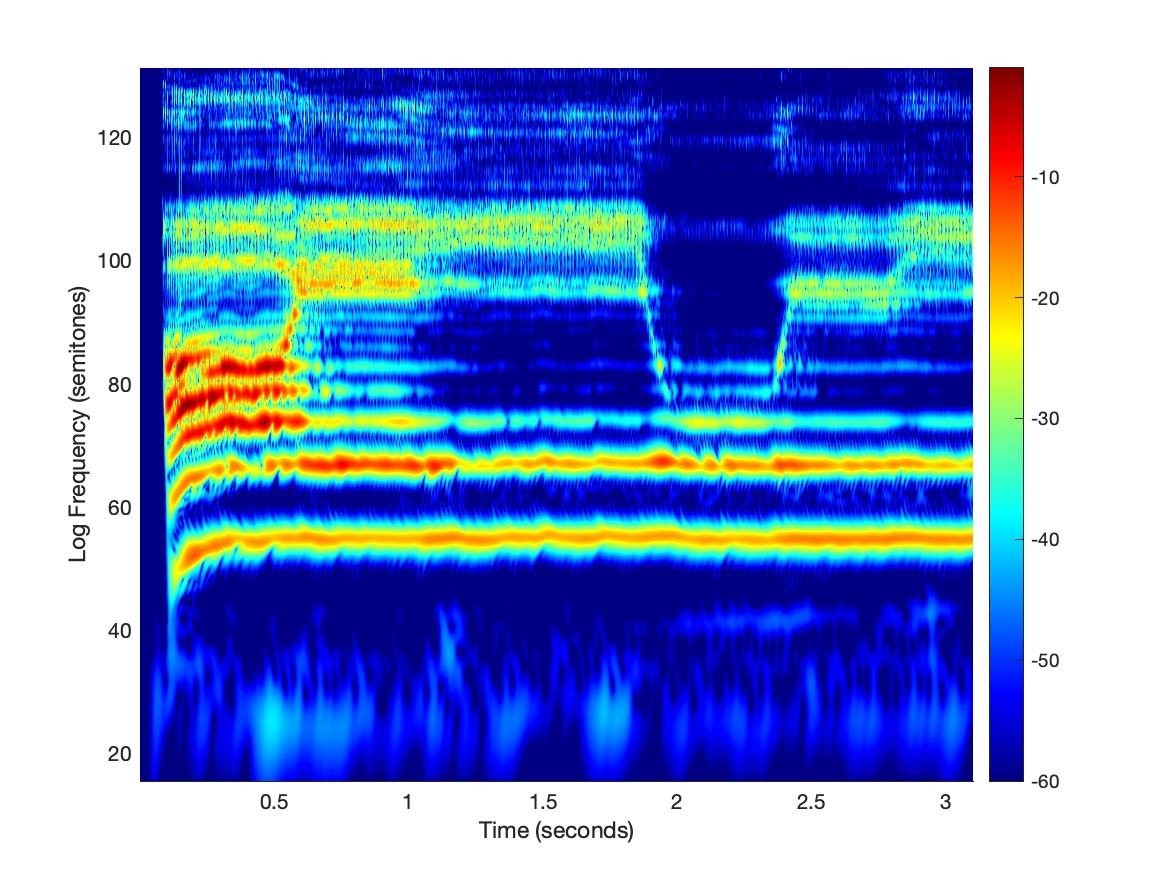}
      & \includegraphics[width=0.48\textwidth]{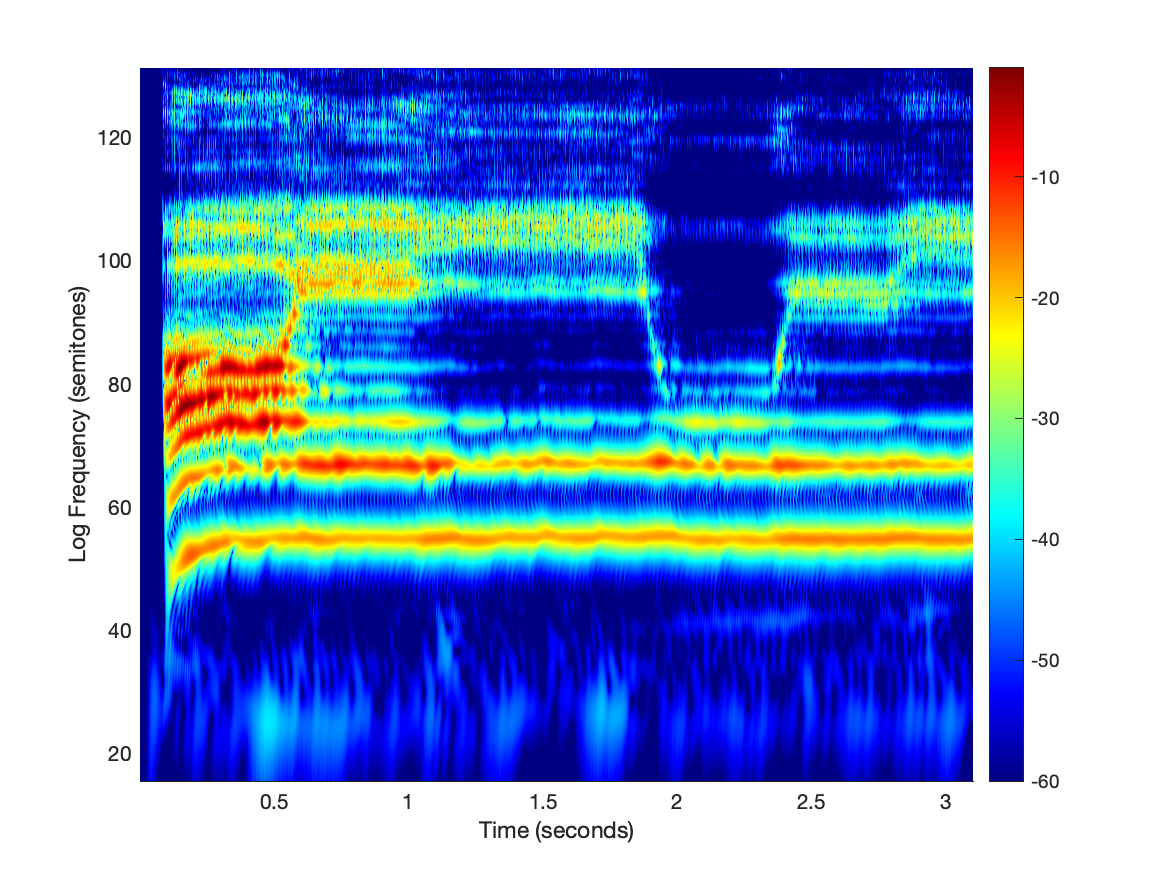}\\
      {\footnotesize\em Spectrogram over longer temporal scales for $c = \sqrt{2}$}
      &  {\footnotesize\em Spectrogram over longer temporal scales for $c = 2$} \\
      \includegraphics[width=0.48\textwidth]{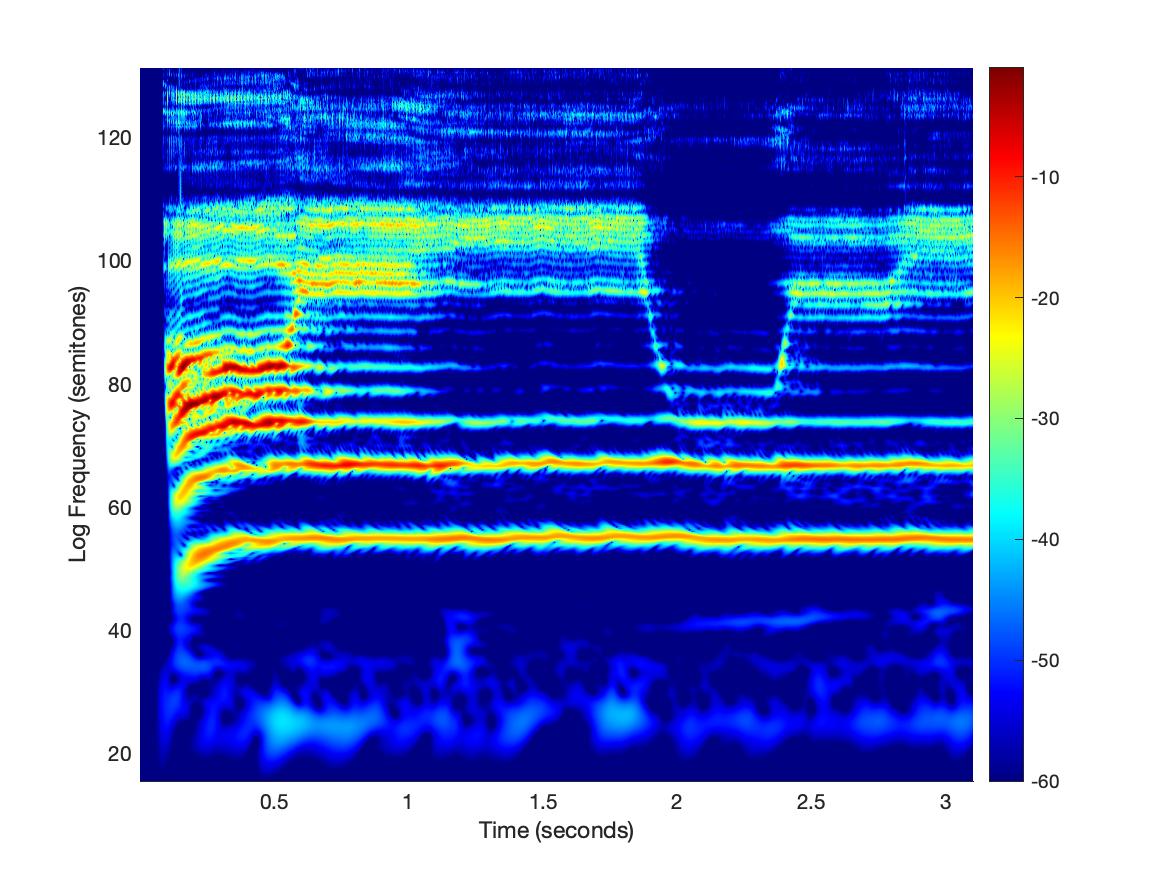}
      & \includegraphics[width=0.48\textwidth]{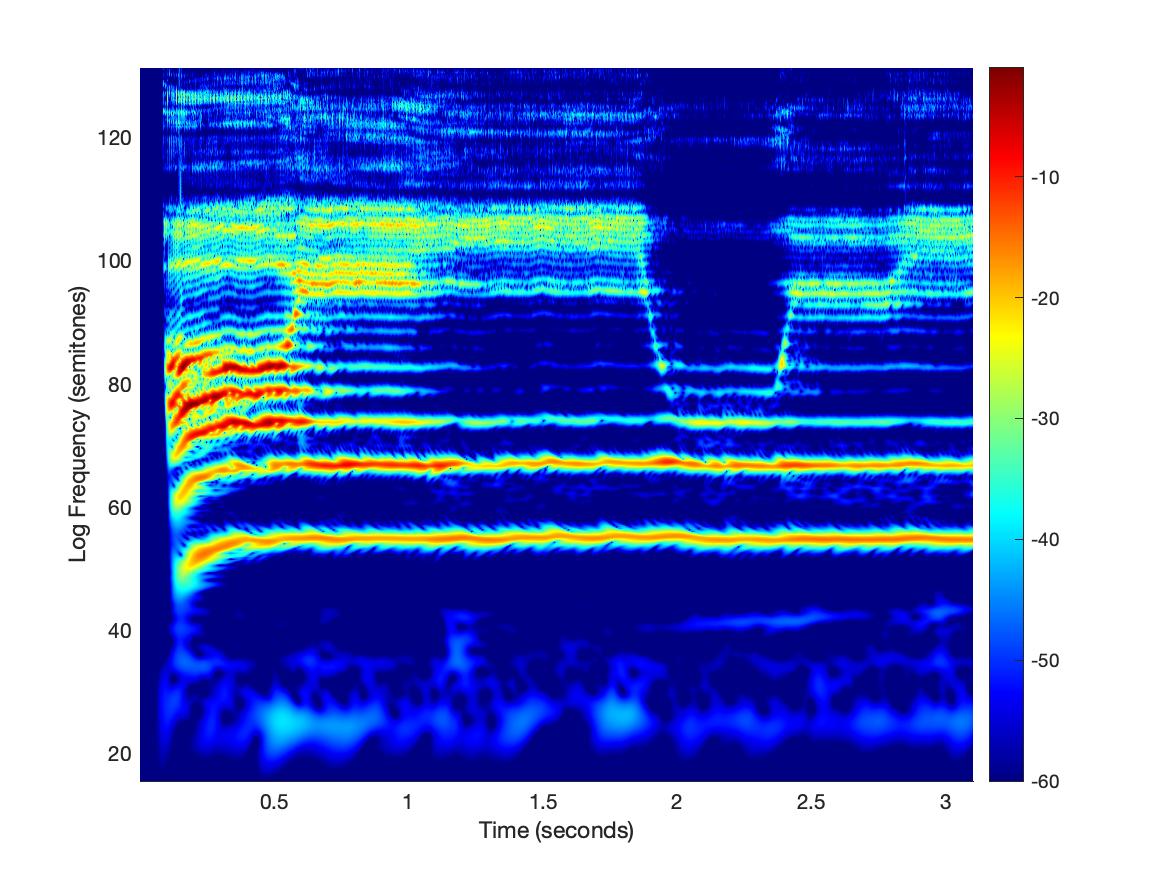}  
     \end{tabular}
     \end{center}
   \caption{Multi-scale spectrograms of an audio signal with a human speaker
     saying a series of vowels, computed with the proposed discrete
     analogue of the time-causal analogue of the Gabor transform, using
     discrete approximations 
     of the time-causal limit kernel for $c = \sqrt{2}$ and $c = 2$
     as the temporal window function in the
     time-causal time-frequency analysis, in terms of a formally infinite
     convolution of first-order integrators coupled in cascade
     truncated after the first 8 temporal filtering stages. In these spectrograms,
     the temporal scale in units of $\sigma = \sqrt{\tau}$
     is chosen proportional to the wavelength of each frequency, 
     for two different proportionality factors (and using
     soft thresholding of the temporal scales to prevent too
     long integration times for the lowest frequencies or too short
     integration times for the highest frequencies).
     (top row) Shorter temporal scales for proportionality factor 4.
     (bottom row) Longer temporal scales for proportionality factor 8.
     (Along the left vertical axis, the frequencies are expressed according
     to the MIDI standard
     $\nu = \nu_0 + C \, \log \left( \frac{\omega}{\omega_0} \right)$
     for $\nu_0 = 69$,  $C = 12/\log 2$ and $\omega_0 = 2 \pi \cdot
     440$. The magnitude of the spectrogram is, in turn, expressed in
     dB ($S_{\dB} = 20 \log_{10} \left( |({\cal H} f)(t, \omega;\;
       \tau, c)|/(\max_{t,\omega} |({\cal H} f)(t, \omega;\; \tau, c)| \right)$),
     according to the colour scale bar to the right, and clipped at
     -60~dB.)
      As can be seen from the spectrograms, a smaller value of the
      distribution parameter $c$ leads to more narrow frequency
      selectivity properties, while a larger value of $c$ leads to
      shorter temporal delays, in agreement with the results of the
      theoretical analysis in
      Section~\ref{sed-theor-props-time-caus-gabor}.
      Similarly, longer temporal scales lead to more narrow frequency
      selectivity, while shorter temporal scales lead to shorter
      temporal delays.}
   \label{fig-spectrograms}
\end{figure*}

\begin{figure*}[hbtp]
  \begin{center}
    {\bf Truncated time-shifted Gabor spectrograms}

    \bigskip
    
    \begin{tabular}{cc}
      {\footnotesize\em Spectrogram over shorter temporal scales for $c = \sqrt{2}$}
      &  {\footnotesize\em Spectrogram over shorter temporal scales for $c = 2$}\\
      \includegraphics[width=0.48\textwidth]{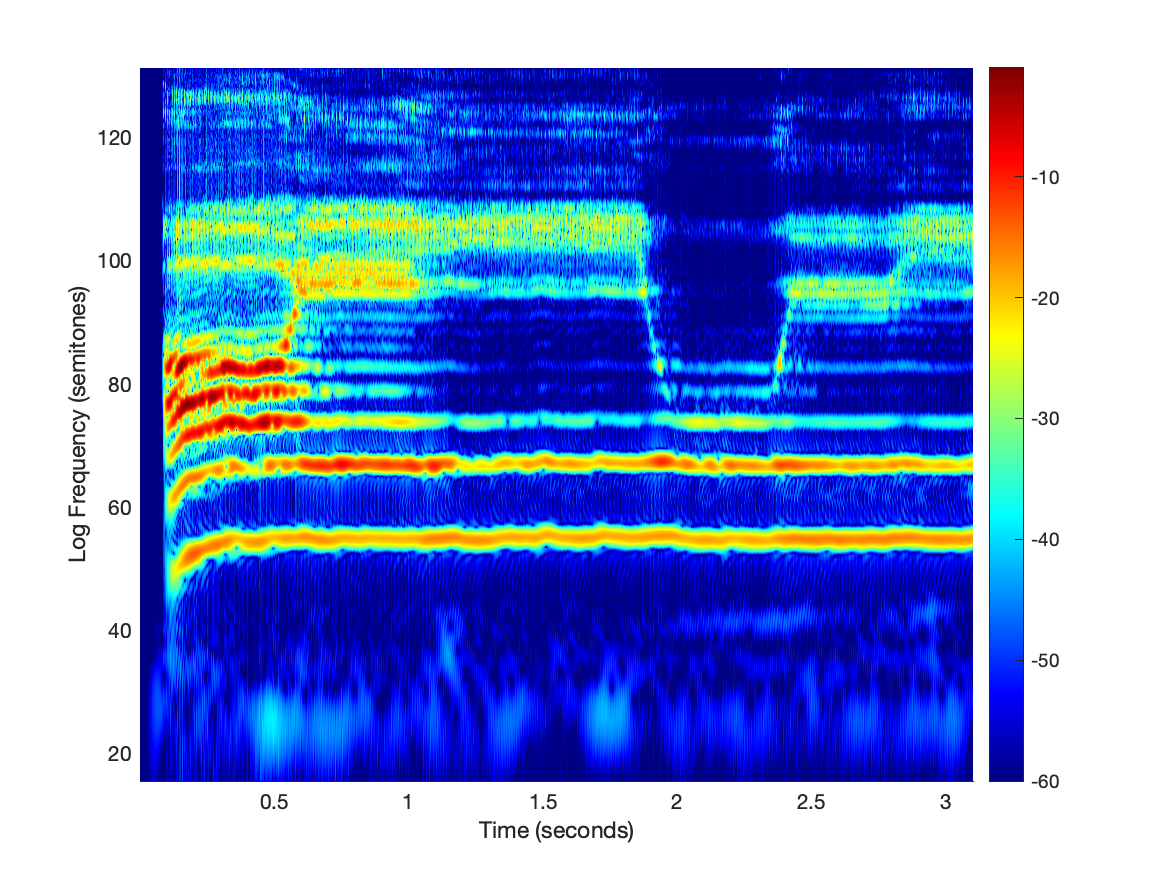}
      & \includegraphics[width=0.48\textwidth]{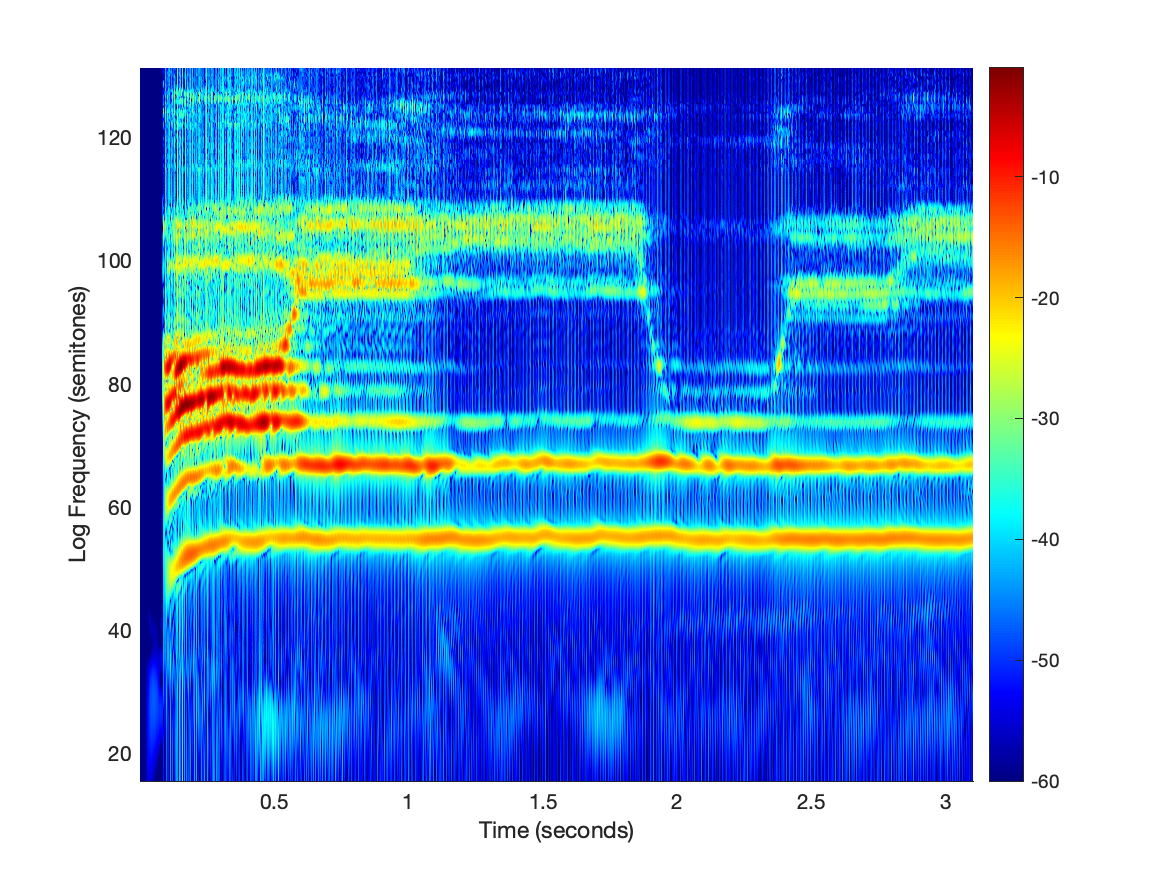}\\
      {\footnotesize\em Spectrogram over longer temporal scales for $c = \sqrt{2}$}
      &  {\footnotesize\em Spectrogram over longer temporal scales for $c = 2$} \\
      \includegraphics[width=0.48\textwidth]{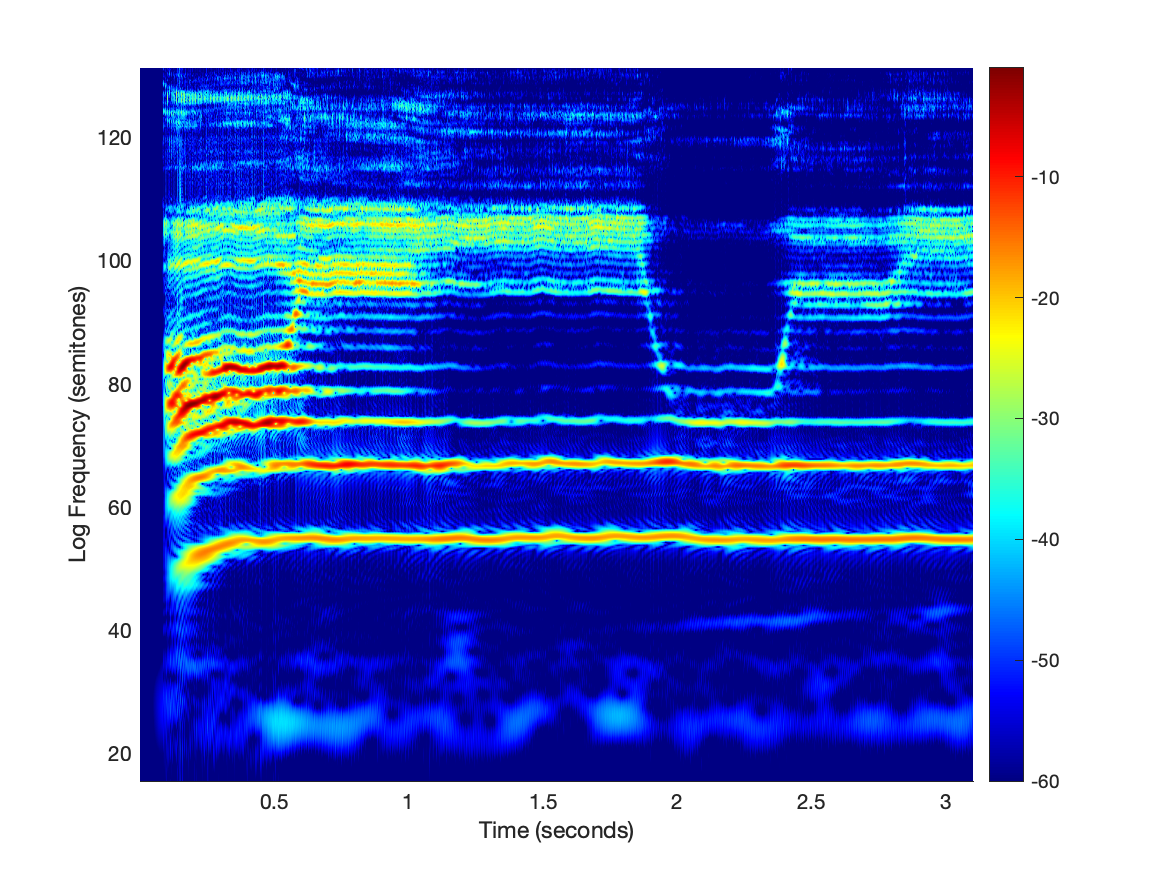}
      & \includegraphics[width=0.48\textwidth]{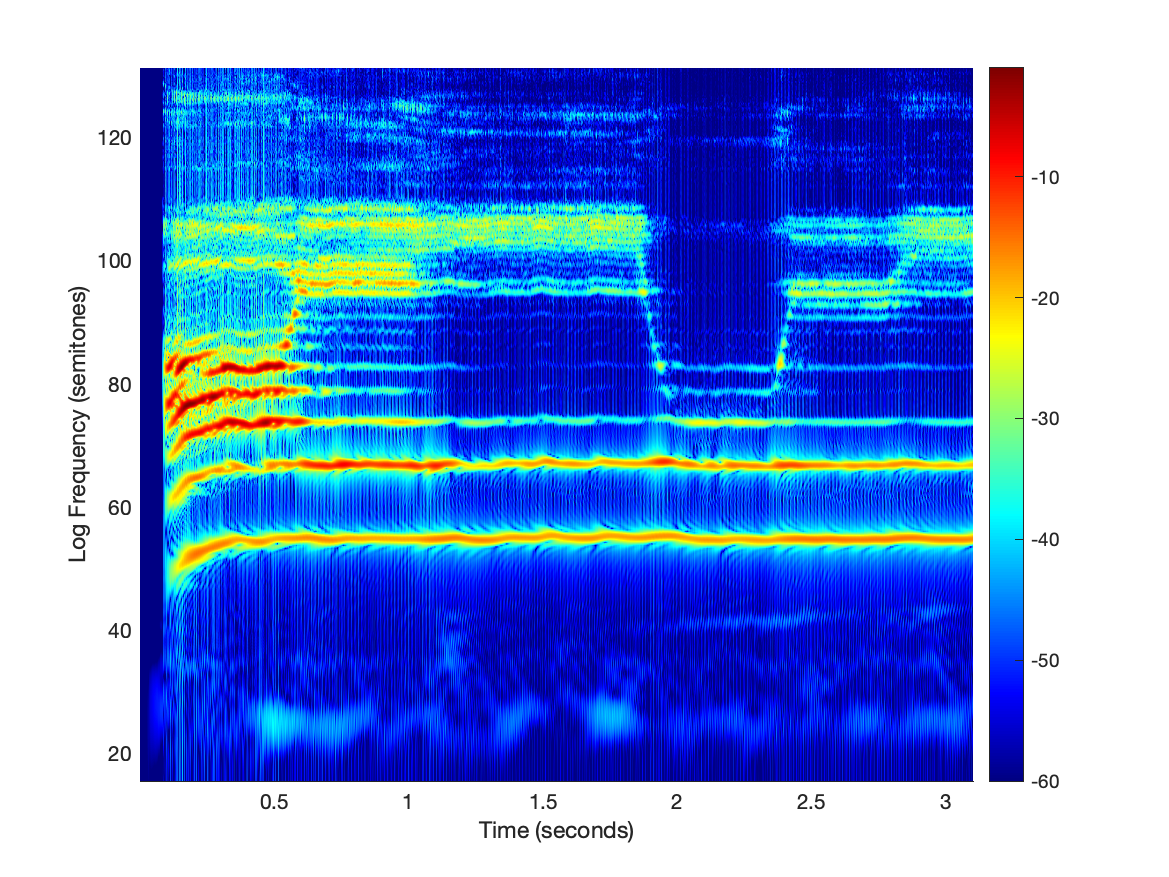}  
     \end{tabular}
     \end{center}
   \caption{Multi-scale spectrograms of an audio signal with a human speaker
     saying a series of vowels, computed with the truncated time-shifted
     discrete analogue of the Gabor transform, using similar temporal
     delays according to  (\ref{eq-approx-temp-pos-time-caus-log-distr}) as would
     be obtained when using the continuous time-causal limit kernel as the
     temporal window function for $c = \sqrt{2}$ and $c = 2$,
     and in this way serving as an {\em ad hoc\/} time-causal time-frequency analysis.
     In a similar way as for the previous principled time-causal
     time-frequency analysis in Figure~\ref{fig-spectrograms},
     the temporal scale in units of $\sigma = \sqrt{\tau}$
     is chosen proportional to the wavelength of each frequency, 
     for two different proportionality factors.
     (top row) Shorter temporal scales for proportionality factor 4.
     (bottom row) Longer temporal scales for proportionality factor 8.
     (Along the left vertical axis, the frequencies are expressed according
     to the MIDI standard
     $\nu = \nu_0 + C \, \log \left( \frac{\omega}{\omega_0} \right)$
     for $\nu_0 = 69$,  $C = 12/\log 2$ and $\omega_0 = 2 \pi \cdot
     440$. The magnitude of the spectrogram is, in turn, expressed in
     dB ($S_{\dB} = 20 \log_{10} \left( |({\cal H} f)(t, \omega;\;
       \tau, c)|/(\max_{t,\omega} |({\cal H} f)(t, \omega;\; \tau, c)| \right)$),
     according to the colour scale bar to the right, and clipped at
     -60~dB.)
      As can be seen from the spectrograms, the truncation of the
      temporal window function in the time-delayed Gabor transform,
      leads to substantial distortions in the spectrograms for $c = 2$
      ,
      demonstrating that the possible {\em ad hoc\/} solution of
      just shifting and truncating the Gabor transform, to enforce
      temporal causality for real-time applications, is not a
      suitable approach compared to using the proposed truly time-causal and
      time-recursive analogue of the Gabor transform proposed in this
      work. Additionally, the truncation of the Gabor transform 
      breaks the other desirable properties of the
      Gabor transform, in terms of its cascade smoothing property over
      temporal scales and
      temporal scale covariance, which, however, will hold for the
      truly time-causal analogue of the Gabor transform.}
   \label{fig-spectrograms-truncgabor}
\end{figure*}

\section{Experiments}
\label{sec-exp}

Figure~\ref{fig-spectrograms} shows an example of a straightforward
application of the presented theory for computing time-causal
spectrograms of an audio signal. In this application, we have let the
temporal scale $\sigma$ of the time-causal window function in the
time-causal frequency transform be proportional to the wavelength
$\lambda = 2\pi/\omega$ for each angular frequency $\omega$,
for two values of the distribution parameter $c = \sqrt{2}$ and $c = 2$
in the discrete approximation of the time-causal limit kernel,
as well as for the two different proportionality constants $N = 4$ and
$N = 8$.

As can be seen from these figures, the
frequency selectivity becomes more narrow when decreasing the
distribution parameter $c$, whereas the temporal delay on the other hand
becomes shorter when increasing this parameter, in agreement with
the results of the above theoretical analysis in
Section~\ref{sec-temp-delay} and Section~\ref{sec-freq-sel}.

We can also see that the fine-scale spectrograms in the
top row have better ability to resolve temporal transients, however, at
the cost of a coarser frequency resolution. For the coarser-scale
spectrogram in the bottom row, the temporal scale is longer, leading
to a
sharper resolution of the frequencies, however, at the cost of a lower
temporal resolution regarding transients. These conflicting
requirements, thus, strongly motivate a multi-scale time-frequency
analysis, so that different trade-offs between the relative advantages
of using either shorter or longer temporal scales for the temporal
window function can be obtained, or at best even be obtained
simultaneously, by combining information from multi-scale spectrograms
over different temporal scales.

As a result of the theoretical
properties of the presented time-causal frequency transform, the
coarser-scale spectrogram can be seen as simplifications of the
corresponding finer-scale spectrograms, due to the cascade smoothing property over
temporal scales. In this example, the only essential difference
between these spectrograms at the different scales is that the coarse scale spectrogram
is computed with {\em one\/} additional layer of first-order temporal
integration relative to the fine scale spectrogram, which thus also saves
substantial computational work, if several multi-scale spectrograms are to be
computed in parallel in a real-time time-frequency analysis.

If attempting to compare the results from a truly time-causal analysis
to a non-causal analysis in an offline scenario, where the non-causal
analysis has access to the relative future in relation to any time
moment, while the time-causal analysis has not, we could, of course,
in general assume that the non-causal analysis should be able to
perform better, since it has access to more information.
To compare a time-causal analysis to other approaches in a more
realistic manner, it is therefore
more appropriate to also prevent access to the relative future in
relation to any time moment, for the other non-causal approaches that the
time-causal analysis is compared to.

Figure~\ref{fig-spectrograms-truncgabor} shows the result of such a
comparison between the proposed time-causal analogue of the Gabor
transform with the regular Gabor transform, with the substantial
modifications of the original non-causal approach in that (i)~a
temporal delay has been added, corresponding to the temporal delay of
the corresponding (continuous)%
\footnote{To simplify the implementation, in our software with
  distinct pipelines for recursive filters {\em vs.\/} explicit
  temporal convolutions, we do here estimate the temporal
  delay directly from the continuous model of the continuous time-causal
  analogue of the Gabor transform. For the discrete implementation,
  the temporal delays are, however, shorter, because the method for
  discrete implementation has been chosen so as to mimic the temporal
  variances of the temporal window functions, which in turn determine the
  temporal scales. If the temporal delays for this comparison would
  instead have been chosen from the actual temporal delays of the
  discrete kernels in the implementation, then the temporal delays
  would have become shorter, with more visual similarity to the
  spectrograms in Figure~\ref{fig-spectrograms}. If the temporal delays
  for the truncated time-shifted Gabor transforms would have been
  determined to be shorter, then the distortions in the spectrograms
  would be expected to be even higher, since harsher truncation operations
  lead to stronger distortion effects.}
time-causal limit kernel for the same value of the
distribution parameter $c$ according to
(\ref{eq-approx-temp-pos-time-caus-log-distr}), and (ii)~with the Gaussian
window function then truncated for the time moments when it would have
implied access to the relative future in relation to any pre-recorded
time moment.

As can be seen from the resulting truncated time-shifted Gabor
spectrograms, the resulting
truncation of the Gaussian window function in the Gabor transform
corresponding to the faster temporal response properties for $c = 2$ then
leads to substantial distortions in the spectrograms. Specifically,
the otherwise sharper frequency selectivity properties of the regular
Gabor transform are by the temporal truncation operation
replaced by a substantial widening of the spectral
bands, notably in relation to transients in the signal.
In this way, the resulting spectrograms become more noisy,
which could be expected to lead to problems for later processing
stages, that are designed to handle such spectrograms as their input.

Furthermore, the desirable theoretical properties of the regular
Gabor transform, in terms of a cascade smoothing property over
temporal scales and temporal scale covariance, will no longer hold after
the temporal truncation of the time-delayed Gabor transform.

Additionally, the true time-causal analogue of the Gabor transform can also
be computed much more efficiently than the truncated time-delayed Gabor
transform, because of the implementation of the time-causal analogue
of the Gabor transform in terms of a few layers of computationally
highly efficient first-order recursive filters coupled in cascade,
whereas the implementation of the truncated time-delayed Gabor
transform is based on explicit convolutions over support intervals of
longer temporal duration.

In these ways, the proposed time-causal analogue of the Gabor
transform should constitute a much better choice for performing a
time-causal time-frequency analysis than {\em e.g.\/} performing an
{\em ad hoc\/} truncation of a time-shifted Gabor transform, to
enforce temporal causality.

\section{Theoretical estimates of the accuracy of time-causal frequency estimates}
\label{sec-theor-est-accuracy-freq-est}

Due to the requirement of temporal causality, we cannot,
as previously described, be able to expect fully similar
properties of a time-frequency analysis based on the time-causal
and time-recursive analogue of the Gabor transform as for
the regular non-causal Gabor transform, since a time-frequency
analysis based on the regular Gabor transform will for every time
moment also have access to information from future, which the time-causal
and time-recursive analogue of the Gabor transform does not have
access to. A non-causal time-frequency analysis will therefore be
based on more information than a time-causal time-frequency
analysis.

For this reason, it is of interest to characterize how the results
from a time-causal time-frequency analysis may differ from the results
of a non-causal time-frequency analysis, in terms of quantitative
performance measures.

In this section, we will perform such an analysis, based on
theoretical estimates regarding the the abilities of the time-causal
vs. the non-causal Gabor transforms to estimate the frequency of a
single sine wave. We will also provide explicit estimates of the width
of a spectral band for the time-causal {\em vs.\/}\ the non-causal
time-frequency analysis concepts, which constitutes a characteristing
property regarding the ability to separate nearby frequencies.

Then, in the following Section~\ref{sec-exp-robustness-to-noise}, we
will perform a complementary experimental analysis regarding the
robustness of local frequency estimates to noise, for actual
discrete implementations of the time-causal and time-recursive
analogue of the Gabor transform.

\subsection{Theoretical analysis for a single sine wave in the ideal
  noise free case}
\label{sec-theor-anal-freq-sel-sine-wave}

Following Lindeberg and Friberg \cite{LinFri15-PONE} Appendix
``Frequency selectivity of the spectrograms'' on page~45 in that
paper, let us consider
the response of the time-causal time-frequency analysis
to a sine wave signal with angular frequency $\omega_0$:
\begin{equation}
  f(t) = \sin \omega_0 t.
\end{equation}
When computing the time-causal analogue of the Gabor transform,
we first compute the real and imaginary components $c(t;\; \tau)$
and $s(t;\; \tau)$, respectively, according to
\begin{align}
  \begin{split}
     c(t, \omega;\; \tau, c) = \Psi(t;\; \tau, c) * \left( f(t) \cos \omega t) \right),
  \end{split}\\
  \begin{split}
     s(t, \omega;\; \tau, c) = -\Psi(t;\; \tau, c) * \left( f(t) \sin \omega t) \right).
  \end{split}
\end{align}
From basic rules for trigonometric functions, we have that:
\begin{multline}
     f(t) \cos \omega t 
     = \sin \omega_0 t \, \cos \omega t = \\
     = \frac{1}{2} 
             \left( 
              -\sin ((\omega - \omega_0) \, t) + \sin ((\omega + \omega_0) \, t)
             \right),
\end{multline}
\begin{multline}
     f(t) \sin \omega t 
     = \sin \omega_0 t \, \sin \omega t = \\
      = \frac{1}{2} 
            \left( 
               \cos ((\omega - \omega_0) \, t) - \cos ((\omega + \omega_0) \, t)
            \right).
\end{multline}
The result of convolving these components with the time-causal limit
kernel $\Psi(t;\; \tau, c)$ can then be expressed as a multiplication with
the magnitude of the the Fourier transform
$|\hat{\Psi}(\omega;\; \tau, c)|$ and adding the phase
$\theta(\omega;\; \tau, c) = \arg \hat{\Psi}(\omega;\; \tau, c)$ of the
time-causal limit kernel.

With additionally the notation
$\theta_+(\omega, \omega_0;\; \tau, c)
= \arg \hat{\Psi}(\omega + \omega_0;\; \tau, c)$
and
$\theta_-(\omega, \omega_0;\; \tau, c)
= \arg \hat{\Psi}(\omega - \omega_0;\; \tau, c)$,
however, with the arguments
of $\theta_+(\omega, \omega_0;\; \tau, c)$ and
$\theta_-(\omega, \omega_0;\; \tau, c)$
henceforth dropped to save space, we then get
\begin{multline}
     c(t, \omega;\; \tau, c) 
     = \frac{1}{2} 
         \left( 
           - \left|\hat{\Psi}(\omega - \omega_0;\; \tau, c)\right| \,
              \sin \left( (\omega - \omega_0) \, t + \theta_- \right)
          \right. \\
          \left.
            + \, \left|\hat{\Psi}(\omega + \omega_0;\; \tau, c)\right| \,
               \sin \left( (\omega + \omega_0) \, t + \theta_+ \right)
         \right),
\end{multline}
\begin{multline}
     s(t, \omega;\; \tau, c) 
    = - \frac{1}{2} 
         \left( 
           \left|\hat{\Psi}(\omega - \omega_0;\; \tau, c)\right| \,
           \cos ((\omega - \omega_0) \, t + \theta_-)
         \right. \\
          \left.
            - \, \left|\hat{\Psi}(\omega + \omega_0;\; \tau, c)\right|  \,
            \cos ((\omega + \omega_0) \, t + \theta_+)
         \right).
\end{multline}
Thereby, after simplification, the magnitude of the spectrogram
\begin{equation}
   |({\cal H} f)(t, \omega;\; \tau, c)| = \sqrt{c(t, \omega;\; \tau, c)^2 + s(t, \omega;\; \tau, c)^2}
\end{equation}
is given by
\begin{align}
  \begin{split}
    \label{eq-spectrogram-resp-single-sine-wave}
    & \left|({\cal H} f)(t, \omega;\; \tau, c)\right|^2 =
  \end{split}\nonumber\\
  \begin{split}  
    & =
    \frac{1}{4} 
     \left(
       \left|\hat{\Psi}(\omega - \omega_0;\; \tau, c)\right|^2
       + \left|\hat{\Psi}(\omega + \omega_0;\; \tau, c)\right|^2
     \right.
   \end{split}\nonumber\\
  \begin{split}
    & \hphantom{= \frac{1}{4} \left( \right.}
     \left.
        - 2 \, 
       \left|\hat{\Psi}(\omega - \omega_0;\; \tau, c)\right| \times
       \left|\hat{\Psi}(\omega + \omega_0;\; \tau, c)\right| \times
     \right.
   \end{split}\nonumber\\
  \begin{split}
    & \hphantom{= \frac{1}{4} \left( \right. - 2}
     \left.    
       \times \cos \left(
                               2 \, \omega_0 \, t \, +
                               \vphantom{\arg \hat{\Psi}(\omega + \omega_0;\; \tau, c}
                           \right.
    \right.
   \end{split}\nonumber\\
  \begin{split}
    & \hphantom{= \frac{1}{4} \left( \right. - 2 \times \cos}
    \left.
                           \left.
                             + \arg \hat{\Psi}(\omega + \omega_0;\; \tau, c)
                           \right.
    \right.
   \end{split}\nonumber\\
  \begin{split}
    & \hphantom{= \frac{1}{4} \left( \right. - 2 \times \cos}
    \left.
                           \left.
                             - \arg \hat{\Psi}(\omega - \omega_0;\; \tau, c)
                           \right) \,      
     \right),
   \end{split}
\end{align}
where $|\hat{\Psi}(\omega;\; \tau, c)|$ denotes the magnitude of the Fourier transform of
the time-causal limit kernel according to
(\ref{eq-FT-comp-kern-log-distr-limit})
and
$\arg \hat{\Psi}(\omega;\; \tau, c)$
the argument of the same Fourier transform.

\subsubsection{Approximations of the time-causal spectrogram for an ideal sine
  wave}

Since the time-causal limit kernel $\Psi(\omega;\; \tau, c)$
is a low-pass filter, it follows that for angular frequencies $\omega$
in the spectrogram near the angular frequency $\omega_0$ of the
input sine wave $f(t)$, the entity
\begin{equation}
  \label{eq-def-r2-spectral-anal}
  R^2(\omega;\; \tau, c)
  = \left|\hat{\Psi}(\omega - \omega_0;\; \tau, c)\right|^2
\end{equation}
will, except for the non-essential scaling factor $\tfrac{1}{2}$,
be the dominant term in the above expression for the spectral response,
and will thus for $\omega \approx \omega_0$ serve as
a reasonable approximation of the frequency selectivity properties%
\footnote{It is for this reason that we have studied the frequency selectivity
properties of this entity in Section~\ref{sec-freq-sel}.}
for the time-causal and time-recursive analogue of the Gabor transform
near the peak over the angular frequencies in the spectral band.

A specific consequence of the form for the expression
(\ref{eq-spectrogram-resp-single-sine-wave}) is that, if we
disregard the influence of the smaller terms
\begin{equation}
  \label{eq-def-t2-spectral-anal}  
  T^2(\omega;\; \tau, c) = \left|\hat{\Psi}(\omega + \omega_0;\; \tau, c)\right|^2
\end{equation}
and
\begin{align}
  \begin{split}
    \label{eq-osc-term}
    & O(t, \omega;\; \tau, c) =
  \end{split}\nonumber\\
  \begin{split}
    & = - 2 \,
           \left|\hat{\Psi}(\omega - \omega_0;\; \tau, c)\right| \times
           \left|\hat{\Psi}(\omega + \omega_0;\; \tau, c)\right| \times
 \end{split}\nonumber\\
  \begin{split}
    & \phantom{ = 2 \,}
           \times
           \cos(2 \, \omega_0 \, t
                   + \Delta \arg \hat{\Psi}(\omega;\; \tau, c))
  \end{split}\nonumber\\
  \begin{split}
    & = - 2 \, R(\omega;\; \tau, c) \, T(\omega;\; \tau, c) \times
\end{split}\nonumber\\
  \begin{split}
    & \phantom{ = 2 \,}
            \times
            \cos \left(
                        2 \, \omega_0 \, t
                        + \Delta \arg \hat{\Psi}(\omega;\; \tau, c)
                     \right) \, 
  \end{split}
\end{align}
in (\ref{eq-spectrogram-resp-single-sine-wave}),
where
\begin{multline}
  \Delta \arg \hat{\Psi}(\omega;\; \tau, c) = \\
  = \arg \hat{\Psi}(\omega + \omega_0;\; \tau, c)
      - \arg \hat{\Psi}(\omega - \omega_0;\; \tau, c),
\end{multline}
then in the ideal
noise free case, the maximum value over the angular 
frequencies $\omega$ in the spectrogram, as modelled according to the
approximation
\begin{equation}
  \label{eq-approx-spectr-R-in-deriv}
  |({\cal H} f)(t, \omega;\; \tau, c)| \approx \frac{1}{2} \, R(\omega;\; \tau, c),
\end{equation}
will be assumed for the angular frequency
\begin{equation}
  \label{eq-ideal-freq-est-coarse-approx}
  \hat{\omega} = \omega_0,
\end{equation}
since the magnitude of the Fourier transform $|\hat{\Psi}(\omega;\; \tau, c)|$
in Equation~(\ref{eq-def-r2-spectral-anal})
assumes its maximum value over the angular frequencies $\omega$
for $\omega = 0$. To that order of approximation, the estimate of
the angular frequency could therefore be regarded as bias free.

Next, one may ask: Could we also estimate the influence on the
frequency estimates, obtained from local extrema over the angular
frequency in the spectrogram, caused by the lower order terms
$O(t, \omega;\; \tau, c)$ and $T^2(\omega;\; \tau, c)$
in the reformulation of the expression
(\ref{eq-spectrogram-resp-single-sine-wave}) into
\begin{align}
  \begin{split}
    \label{eq-spectrogram-resp-single-sine-wave-ROT-descr}
    & |({\cal H} f)(t, \omega;\; \tau, c)| =
  \end{split}\nonumber\\
  \begin{split}      
    & = \frac{1}{2}
           \sqrt{R^2(\omega;\; \tau, c)
                    + O(t, \omega;\; \tau, c)
                    + T^2(\omega;\; \tau, c)} \\
  \end{split}\nonumber\\
  \begin{split}      
     & = \frac{1}{2} \, R(\omega;\; \tau, c) \,
            \sqrt{1 
                     + \frac{O(t, \omega;\; \tau, c)}{R^2(\omega;\; \tau, c)}
                     + \frac{T^2(\omega;\; \tau, c)}{R^2(\omega;\; \tau, c)}},
  \end{split}
\end{align}
and also simultaneously study the influence of local perturbations in
the signal, in case there would be
interfering structures in the signal for other frequencies,
such as additional sine waves with nearby frequencies, or noise added to the
idealized model signal?

\subsubsection{Estimates of the relative influence of the oscillatory 
  term $O(t, \omega;\; \tau, c)$}
\label{sec-est-rel-infl-osc-term}

For angular frequencies $\omega$ near the angular frequency $\omega_0$
of the input sine wave, it will hold that the absolute value of
the oscillatory term $O(t, \omega;\; \tau, c)$ will on average be
larger than the value of $T^2(\omega;\; \tau, c)$
\begin{align}
  \begin{split}
    & E(\left| O(t, \omega;\; \tau, c) \right|)
  \end{split}\nonumber\\
  \begin{split}
    & = \frac{1}{\lambda_0}
           \int_{t = 0}^{\lambda_0}
              2 \,
              \left|\hat{\Psi}(\omega - \omega_0;\; \tau, c)\right| \times
              \left|\hat{\Psi}(\omega + \omega_0;\; \tau, c)\right| \times
 \end{split}\nonumber\\
  \begin{split}
   & \hphantom{= \frac{1}{\lambda_0} \int_{t = 0}^{\lambda_0} \,}
              \times
              \left|
                 \cos \left(
                             2 \, \omega_0 \, t
                             + \Delta \arg \hat{\Psi}(\omega;\; \tau, c)
                          \right)
               \right|
          \, dt
  \end{split}\nonumber\\
  \begin{split}
    & \gg
       \left|
         \hat{\Psi}(\omega + \omega_0;\; \tau, c)
       \right|^2
      = T^2(\omega;\; \tau, c),
  \end{split}
\end{align}
where $\lambda_0 = 2\pi/\omega_0$ is the wavelength corresponding the angular
frequency $\omega_0$ of the input sine wave. This relationship follows
from the fact that for $\omega$ near $\omega_0$, we have that
\begin{equation}
  \label{eq-omegadiff-gg-omegasum}
  \left| \hat{\Psi}(\omega - \omega_0;\; \tau, c) \right|
  \gg
  \left| \hat{\Psi}(\omega + \omega_0;\; \tau, c) \right|,
\end{equation}
because of the low-pass nature of the time-causal limit kernel.

Thus, to the next order of approximation, if we disregard the
influence on the spectrogram $|({\cal H} f)(t, \omega;\; \tau, c)|^2$
in (\ref{eq-spectrogram-resp-single-sine-wave}) due to
the term $T^2(\omega;\; \tau, c)|$, then the oscillatory factor
$2 \cos \left( 2 \, \omega_0 \, t + \Delta \arg \hat{\Psi}(\omega;\; \tau, c) \right)$
in the oscillatory term $O(t, \omega;\; \tau, c)$
in Equation~(\ref{eq-osc-term}) will after a Tayor expansion
approximation of the form
\begin{multline}
  \hspace{-4mm}
  \left|({\cal H} f)(t, \omega;\; \tau, c)\right|^2
  = \frac{1}{2} \, R(\omega;\; \tau, c)
  \sqrt{1 + \frac{O(t, \omega;\; \tau, c)}{R^2(\omega;\; \tau, c)} +
    {\cal O}(\cdot)} \\
  \approx
  \frac{1}{2} \, R(\omega;\; \tau, c)
  \left(
    1 + \frac{1}{2} \frac{O(t, \omega;\; \tau, c)}{R^2(\omega;\; \tau, c)}
    +  {\cal O}(\cdot)
   \right),
\end{multline}
with the informal ordo notation ``${\cal O}(\cdot)$'' here used as a place holder for
terms that are on average of smaller magnitude than the preceding terms,
have a relative influence on the
spectral band in the spectrogram $|({\cal H} f)(t, \omega;\; \tau, c)|$, in relation to
the dominating term $R^2(\omega;\; \tau, c)$,  of order
\begin{multline}
  \varepsilon(\omega;\; \tau, c) =
  \frac{1}{2} \, \frac{\sup_t |O(t, \omega;\; \tau, c)|}
                              {R^2(\omega;\; \tau, c)} = \\
  = \frac{T(\omega;\; \tau, c)}{R(\omega;\; \tau, c)}
  = \frac{\left| \hat{\Psi}(\omega + \omega_0;\; \tau, c) \right|}
                 {\left| \hat{\Psi}(\omega - \omega_0;\; \tau, c) \right|},
\end{multline}
which when choosing $\omega = \omega_0$ reduces to
\begin{equation}
  \varepsilon_{\omega_0}(\tau, c)
  = \varepsilon(\omega;\; \tau, c) |_{\omega = \omega_0}
  = \left| \hat{\Psi}(2 \, \omega_0;\; \tau, c) \right|.
\end{equation}
Furthermore, if we choose the temporal duration
$\sigma = \sqrt{\tau}$ of the time-causal limit kernel proportional to
the wavelength $\lambda$ corresponding to the angular frequency
$\omega$ in the time-frequency transform according to
(\ref{eq-def-tau-from-omega-N}), then it follows after inserting
$\tau$ according to (\ref{eq-def-tau-from-omega-N})
and $\omega = \omega_0$ into the Fourier transform
(\ref{eq-FT-comp-kern-log-distr-limit}) of
the time-causal limit kernel, that the resulting
perturbation measure
\begin{multline}
  \label{eq-def-perturb-meas-eps}
  \tilde{\varepsilon}(N;\; c) 
  = \left|
         \hat{\Psi}(2 \, \omega_0;\; (\tfrac{2 \pi N}{\omega_0})^2, c)  
      \right| = \\
  = \prod_{k=1}^{\infty}
       \left|
         \frac{1}{1 + i \, 4 \pi \, N  \, c^{-k} \sqrt{c^2-1} }
       \right|,
\end{multline}
or in dB
\begin{multline}
  \tilde{\varepsilon}_{\dB} (N;\; c)
    = 20 \log_{10} \tilde{\varepsilon}(N;\; c)  \\
    = 
    - \frac{10}{\log 10} \times 
              \sum_{k=1}^{\infty}
                 \log 
                    \left(
                            1 
                            + 4 \pi^2 \, N^2 \, c^{-2k} \, (c^2-1) 
                    \right),
\end{multline}
will be independent of the angular frequency $\omega_0$, in accordance with
required property of temporal scale covariance.

Hence, we can simply read off an estimate of the order of magnitude of
the relative influence of the oscillatory term
$O(t, \omega;\; \tau, c)$ in relation to the magnitude of the
dominant term $R(\omega;\; \tau, c)$ in the approximation of
the spectrogram $|({\cal H} f)(t, \omega;\; \tau, c)|$ near the center
of a spectral band from the values of the graphs in
Figure~\ref{fig-freq-selectivity} for $\omega = 2 \, \omega_0 = 2 \times 1$,
for each one of the four different use cases
regarding the default choices of the wavelength proportionality factor $N$ 
and the distribution parameter $c$ for the time-causal limit kernel.

Computed numerically, we do, in turn, obtain the values according
to Table~\ref{tab-eps-omega0-use-cases-time-caus} for the four main
use cases obtained by choosing the wavelength proportionality factor
$N \in \{ 4, 8 \}$ and the distribution parameter of the time-causal
limit kernel $c \in \{ \sqrt{2}, 2 \}$.
For comparison, Table~\ref{tab-eps-omega0-use-cases-non-caus} shows
corresponding perturbation measures computed for the regular
non-causal Gabor transform for $N \in \{ 4, 8 \}$.

From these estimates, we can conclude that, although the regular Gabor
non-causal Gabor transform leads to substantially smaller values of
these perturbation measures, the values of the perturbation measures for
the time-causal and time-recursive analogue of the Gabor transform
are quite low for the four main use cases studied here, with the largest magnitude
$\varepsilon_{\omega_0} \approx -51.7$~dB assumed for the use
case with the wavelength proportionality factor $N = 4$ and the
distribution parameter $c = 2$ for the time-causal limit kernel.
That use case does, in turn, correspond to the shortest temporal delay out
of the here considered four use cases.

\begin{table}[hbtp]
  \begin{center}
    \begin{tabular}{lr}
      \hline
      \multicolumn{2}{c}{Time-causal spectrograms}\\
      \hline
      Use case \vphantom{$\hat{X}_X$} & $\tilde{\varepsilon}_{\dB}(N;\; c)$ \\
      \hline
      $N = 4$, $c = \sqrt{2}$ & -77.3 \\
      $N = 4$, $c = 2$ & -51.7 \\
      $N = 8$, $c = \sqrt{2}$ & -118.8 \\
      $N = 8$, $c = 2$ & -78.4 \\      
      \hline
    \end{tabular}
  \end{center}
  \caption{Numerical values of the frequency-independent perturbation measure
    $\tilde{\varepsilon}_{\dB}(N;\; c)$
    according to (\ref{eq-def-perturb-meas-eps}) regarding
                the relative order of magnitude of the oscillatory term
                $O(t, \omega;\; \tau, c)$ according to (\ref{eq-osc-term})
                in relation to the dominant term
                $R^2(\omega;\; \tau, c)$ according to (\ref{eq-def-r2-spectral-anal})
                in the spectrogram $|({\cal H} f)(t, \omega;\; \tau, c)|$
                according to (\ref{eq-spectrogram-resp-single-sine-wave})
                of an ideal sine wave, computed using the time-causal
                analogue of the Gabor transform, and as function of
                the wavelength proportionality factor $N$ and
                the distribution parameter $c$ for the time-causal limit kernel.}
  \label{tab-eps-omega0-use-cases-time-caus}
\end{table}

\begin{table}[hbtp]
  \begin{center}
    \begin{tabular}{lr}
      \hline
      \multicolumn{2}{c}{Non-causal spectrograms}\\
      \hline
      Use case \vphantom{$\hat{X}_X$} & $\tilde{\varepsilon}_{\dB}(N)$ \\
      \hline
      $N = 4$ & -685.8  \vphantom{$\sqrt{2}$} \\
      $N = 8$ & -2743.2 \\  
      \hline
    \end{tabular}
  \end{center}
  \caption{Numerical values of the frequency-independent perturbation measure
    $\tilde{\varepsilon}_{\dB}(N)$
    according to (\ref{eq-def-perturb-meas-eps}) regarding
                the relative order of magnitude of the oscillatory term
                $O(t, \omega;\; \tau, c)$ according to (\ref{eq-osc-term})
                in relation to the dominant term
                $R^2(\omega;\; \tau, c)$ according to (\ref{eq-def-r2-spectral-anal})
                in the spectrogram $|({\cal H} f)(t, \omega;\; \tau, c)|$
                according to (\ref{eq-spectrogram-resp-single-sine-wave})
                of an ideal sine wave, computed using the regular
                non-causal Gabor transform, and as function of
                the wavelength proportionality factor $N$.}
  \label{tab-eps-omega0-use-cases-non-caus}
\end{table}

\subsubsection{Estimates of the relative influence of the remaining 
  term $T^2(t, \omega;\; \tau, c)$}

Concerning the remaining term $T^2(\omega;\; \tau, c)$
according to (\ref{eq-def-t2-spectral-anal}) in the
spectrogram $|({\cal H} f)(t, \omega;\; \tau, c)|$ according to
(\ref{eq-spectrogram-resp-single-sine-wave}),
the influence of that term differs from the influence of the
oscillatory term $O(t, \omega;\; \tau, c)$, in that it only affects
the variability in one direction of the angular frequency,
and may thus lead to a systematic bias.
In terms of magnitude, the relative influence of this term in
relation to the dominant term $R^2(\omega;\; \tau, c)$
is, however, of the order of
\begin{equation}
  b(\omega;\; \tau, c)
  =   \frac{T^2(\omega;\; \tau, c)}
                 {R^2(\omega;\; \tau, c)}
 = \varepsilon^2(\omega;\; \tau, c)
\end{equation}
with the corresponding relationship for $\omega = \omega_0$ when the
temporal duration $\sigma = \sqrt{\tau}$ is proportional to the
wavelength $\lambda$ corresponding to the angular frequency
$\omega = \omega_0$ in the spectrogram
\begin{equation}
  \tilde{b}(N;\; c)
  = \frac{T^2(\omega_0;\; (\tfrac{2 \pi N}{\omega_0})^2, c)}
                 {R^2(\omega_0;\; (\tfrac{2 \pi N}{\omega_0})^2, c)}
 = \tilde{\varepsilon}^2(N;\; c),
\end{equation}
or in dB
\begin{equation}
  \tilde{b}_{\dB}(N;\; c)
  = 20 \log_{10} \tilde{b}(N;\; c)
  = 2 \, \tilde{\varepsilon}_{\dB}(N;\; c),
\end{equation}
where this entity is also independent of the angular frequency
$\omega_0$, because of the temporal scale covariance property.

In terms of numerical values, the above relation implies that we
obtain the dB value for the perturbation measure
$\tilde{b}(N;\; c)$ by multiplying
the dB value for the perturbation measure $\tilde{\epsilon}(N;\; c)$ by 2.
In view of the dB values, listed for $\tilde{\epsilon}(N;\; c)$
in Table~\ref{tab-eps-omega0-use-cases-time-caus},
the influence of the term $\tilde{b}(N;\; c)$ on the frequency
estimate from a single sine wave ought therefore to be very marginal
for the four main use cases considered in this work.

\subsection{Theoretical estimates of the accuracy of frequency
  estimates in the ideal noise free case}
\label{sec-theor-estim-acc-freq-est-noise-free}

To estimate how much the frequency estimate of an ideal sine wave can
be expected to vary from the ideal estimate $\hat{\omega} = \omega_0$
according to (\ref{eq-ideal-freq-est-coarse-approx}), let us in the
following approximate the closed-form expression
(\ref{eq-spectrogram-resp-single-sine-wave}) for the square of the
spectrogram $|({\cal H} f)(t, \omega;\; \tau, c)|^2$ by a
second-order Taylor expansion around the angular frequency
$\omega = \omega_0$ of the input sine wave.

By combining
Equations~(\ref{eq-spectrogram-resp-single-sine-wave-ROT-descr})
and~(\ref{eq-osc-term}), let us write the square of the spectrogram in
Equation~(\ref{eq-spectrogram-resp-single-sine-wave}) as
\begin{align}
  \begin{split}
    \label{eq-spectr-sine-wave-V-components}
    & \left|({\cal H} f)(t, \omega;\; \tau, c)\right|^2 =
  \end{split}\nonumber\\
  \begin{split}
    = \frac{1}{4}
    \left(
      R^2(\omega;\; \tau, c)
      + T^2(\omega;\; \tau, c)
      \right.
  \end{split}\nonumber\\
  \begin{split}
    \phantom{= \frac{1}{4} \left( \right.} 
    \left. \vphantom{R^2}
      - 2 R(\omega;\; \tau, c) \times
            T(\omega;\; \tau, c) \times
      C(t, \omega;\; \tau, c)
    \right),
  \end{split}
\end{align}
where the only explicitly time-dependent entity
\begin{multline}
  \label{eq-def-C}
  C (t, \omega;\; \tau, c) = \\
   = \cos \left(
                   2 \, \omega_0 \, t
                   + \arg \hat{\Psi}(\omega + \omega_0;\; \tau, c)
                   - \arg \hat{\Psi}(\omega - \omega_0;\; \tau, c)
                \right)
\end{multline}
assumes values in $[-1, 1]$.

Let us next, again, let the temporal scale $\sigma = \sqrt{\tau}$ be proportional to
the angular frequency $\omega$ according to
(\ref{eq-def-tau-from-omega-N}), and additionally reparameterize the variations
in the angular frequency $\omega$ around the angular frequency
$\omega_0$ of the sine wave according to
\begin{equation}
   \omega = \omega_0 \, e^{\gamma},
\end{equation}
where the relative frequency variability variable $\gamma$ should then
assume values near $\gamma = 0$.

By calculating the second-order Taylor expansion 
of the logarithm for
the resulting expression for the square of the spectrogram
$ |({\cal H} f)(t, \omega;\; \tau, c)|^2$, truncated to the first $K = 8$
factors%
\footnote{The reason why we truncate the closed-form expression for
  the spectrogram of an ideal sine wave for a fixed number of $K = 8$
convolution operations in cascade, as an approximation of the
time-causal limit kernel, which comprises an infinite number
when $K \rightarrow \infty$, in this way, is that this makes the
calculations much easier to handle in Mathematica, also compared to instead
performing the calculations for a general value of $K$.},
with the calculations performed in Mathematica,
we then obtain a resulting expression of the form
\begin{multline}
  \left|({\cal H} f)(t, \omega_0 \, e^{\gamma};\;
  \left(\tfrac{2 \pi N}{\omega_0 \, e^{\gamma}}\right)^2, c)\right|^2 = \\
  = A_0(N, c, C)
  + A_1(N, c, C) \, \gamma
  + A_2(N, c, C) \, \frac{\gamma^2}{2} 
  + {\cal O}(\gamma^3),
\end{multline}
where the explicit expressions for $A_0(N, c, C)$,
$A_1(N, c, C)$ and $A_2(N, c, C)$ are, however, 
too complex to be reproduced here.

Truncating this Taylor expansion after the second-order term, and then
differentiating the resulting expression with respect to $\gamma$,
thereby gives that the estimate of the relative frequency variability
variable $\gamma$, given by
\begin{equation}
  \label{eq-gamma-estim}
  \hat{\gamma} = - \frac{A_1(N, c, C)}{A_2(N, c, C)}.
\end{equation}
This result does, in turn, mean that the frequency estimate $\hat{\omega}$
will be off from the ideal value $\hat{\omega} = \omega_0$
by a factor of
\begin{equation}
  \frac{\hat{\omega}}{\omega_0} = e^{\hat{\gamma}}.
\end{equation}
By plotting the graphs of the estimated value $\hat{\gamma}$ of the
relative frequency variability variable according to
(\ref{eq-gamma-estim}) as function of the time-dependent factor
$C \in [-1, 1]$ according to (\ref{eq-def-C}) for each one of the main four
use cases, thus for the different values of the distribution wavelength proportionality
factor $N \in \{4, 8\}$ and the distribution parameter of the
time-causal limit kernel $c \in \{\sqrt{2}, 2\}$
studied in this article,
we find that the graphs decrease monotonically as function of $C$,
and that the graphs are also approximately symmetric around the origin,
with the extreme values, thus, assumed at either of the ends of the
interval $C \in [-1, 1]$.

\begin{table}[hbtp]
  \begin{center}
    \begin{tabular}{llll}
      \hline
      \multicolumn{4}{c}{Time-causal spectrograms}\\
      \hline
      Use case \vphantom{$\frac{\hat{X}_x}{\hat{X}_x}$} 
      & $\max_C |\hat{\gamma}|$
      & $\min_C \frac{\hat{\omega}}{\omega_0} - 1$
      & $\max_C \frac{\hat{\omega}}{\omega_0} - 1$ \\
      \hline
      $N = 4$, $c = \sqrt{2}$
        & $3.6 \times 10^{-11}$
        & $-3.6 \times 10^{-11}$
        & $+3.6 \times 10^{-11}$ \\
      $N = 4$, $c = 2$           
        & $1.3 \times 10^{-8}$
        & $-1.3 \times 10^{-8}$
        & $+1.3 \times 10^{-8}$ \\
      $N = 8$, $c = \sqrt{2}$ 
        & $3.9 \times 10^{-14}$
        & $-3.9 \times 10^{-14}$
        & $+3.9 \times 10^{-14}$ \\
      $N = 8$, $c = 2$           
        & $4.6 \times 10^{-11}$
        & $-4.6 \times 10^{-11}$
        & $+4.6 \times 10^{-11}$ \\
      \hline
    \end{tabular}
  \end{center}
  \caption{Theoretical estimates of the maximum relative offset from the ideal value
    $\hat{\omega} = \omega_0$ for local frequency estimation based on
    the maximum value of the magnitude of the time-causal spectrogram
    computed for a single sine wave with angular frequency $\omega_0$,
  computed from an approximation of the ideal spectrogram by
  truncating the infinite convolution operation in the time-causal
  limit kernel after the first $K = 8$ components.}
  \label{tab-freq-offset}
\end{table}

Table~\ref{tab-freq-offset} summarizes the extreme values
obtained in this way for each one of the main use cases for the
time-causal and time-recursive analogue of the Gabor transform.
As can be seen from this table, the resulting frequency estimates
become very close to the ideal value for the four considered main use
cases, in the ideal noise free continuous case considered here.

Specifically, for a discrete implementation of the time-causal and
time-recursive analogue of the Gabor transform, we could then expect this
source of error in the resulting frequency estimates to be far below the
errors caused by quantizing the frequencies in a discrete
approximation of the continuous time-frequency transform.

\subsection{Theoretical estimates of the width of a spectral band}
\label{sec-theor-estim-width-spectr-band}

Beyond estimating the variability in the frequency estimates, it is
also of interest to estimate how wide the spectral bands will be for
the different types of time-frequency analysis methods.
Starting from the approximation of the spectrogram
$|({\cal H} f)(t, \omega;\; \tau, c)|$ for an ideal sine wave
in terms of the dominant component $R(\omega;\; \tau, c)$
according to (\ref{eq-approx-spectr-R-in-deriv}) and
(\ref{eq-def-r2-spectral-anal}), let us therefore estimate
the width of a spectral band from the two values of
$\omega = \omega_-$ and $\omega = \omega_+$
for which
\begin{equation}
  R(\omega;\; \tau, c) = \frac{1}{2},
\end{equation}
where $\omega_-$ represents the lower bound and $\omega_+$
represents the upper bound, with $\omega_- < \omega_+$.

Again, letting the temporal scale $\sigma = \sqrt{\tau}$ be proportional to
the angular frequency $\omega$ according to
(\ref{eq-def-tau-from-omega-N}),
and reparameterizing the angular frequency $\omega$ as
\begin{equation}
   \omega = \omega_0 \, e^{\gamma},
\end{equation}
we are therefore, after using the explicit expression for $R(\omega;\; \tau, c)$
according to Equation~(\ref{eq-R-time-caus-tau-prop-N}), to solve the
equation%
\footnote{It should be remarked that measuring the width of a spectral
band from the neighbouring frequencies, where the absolute value of the
spectrogram has decreased to a factor of $1/2$ of its peak
value, constitutes a rather conservative way of measuring the width of
a the spectral band, in the sense that the resulting estimates will be rather
narrow. Of course, it is also possible to estimate the width of a
spectral band for other fractions of the peak value, which may then
lead to different estimates, and then also different ratios
between the estimated widths of the spectral bands computed for different types of
time-frequency analysis concepts used for defining the spectrograms.}
\begin{multline}
  R(\omega_0 \, e^{\gamma};\;
      \left(\tfrac{2 \pi N}{\omega_0 \, e^{\gamma}}\right)^2, c)
      = \left|\hat{\Psi}(\omega_0 \, e^{\gamma}- \omega_0;\;
                           \left(\tfrac{2 \pi N}{\omega_0 \, e^{\gamma}}\right)^2, c)\right| = \\
  = \frac{1}
          {\prod_{k=1}^{\infty}
                  \sqrt{
                  1 
                  + \frac{4 \pi^2 \, c^{-2k} \, (c^2-1) \, N^2 \, (e^{\gamma}-1)^2}{e^{2\gamma}}}}
    = \frac{1}{2}
\end{multline}
in terms of the variable $\gamma$,
to give the two roots $\gamma_-$ and $\gamma_+$ with $\gamma_- < \gamma_+$.

Table~\ref{tab-width-spectr-bands-time-caus}
shows numerical values concerning the
estimates of the widths of the spectral bands obtained in this way,
for the four main use cases considered in the this paper, regarding the different values of the
wavelength proportionality factor $N \in \{ 4, 8 \}$ and the distribution parameter
$c \in \{ \sqrt{2}, 2 \}$ of the time-causal limit kernel, based on approximations using
the first $K = 8$ factors in Mathematica.

In this table, we have also
defined the following compact measure of the width of the spectral band
\begin{equation}
  \Delta \gamma = \gamma_+ - \gamma_-,
\end{equation}
as well as computed the relative lower and upper bands of the spectral
bands according to
\begin{equation}
  \frac{\omega_-}{\omega_0} = e^{\gamma_-},
  \quad\quad
  \frac{\omega_+}{\omega_0} = e^{\gamma_+},
\end{equation}
which are notably independent of the angular frequency $\omega_0$ of the input
sine wave, due to the temporal scale covariance property for the
time-causal and time-recursive analogue of the Gabor transform.


\begin{table}[hbtp]
   \begin{center}
    \begin{tabular}{lccccc}
      \hline
      \multicolumn{6}{c}{Time-causal spectrograms}\\
      \hline
      Use case \vphantom{$\frac{\hat{X}_x}{\hat{X}_x}$}
      & $\gamma_-$
      & $\gamma_+$
      & $\Delta \gamma$
      & $\tfrac{\omega_-}{\omega_0}$
      & $\tfrac{\omega_+}{\omega_0}$ \\
      \hline
      $N = 4$, $c = \sqrt{2}$
        & -0.0511
        & +0.0539
        & 0.105
        & 0.950
        & 1.055 \\
      $N = 4$, $c = 2$           
         & -0.0556
         & +0.0588
         & 0.114
         & 0.946
         & 1.061 \\
      $N = 8$, $c = \sqrt{2}$ 
        & -0.0259
        & +0.0266
        & 0.0525
        & 0.974
        & 1.027  \\
      $N = 8$, $c = 2$           
        & -0.0282
        & +0.0290
        & 0.0572
        & 0.972
        & 1.029  \\
      \hline
    \end{tabular}
  \end{center}
  \caption{Measures, that estimate the widths of the spectral bands for the
    time-causal and time-recursive analogue of the Gabor transform
    applied to an ideal sine wave, for the four main use cases
    considered in this article, regarding the wavelength
    proportionality factor $N$ and the distribution parameter
    $c$ of the time-causal limit kernel.}
  \label{tab-width-spectr-bands-time-caus}
\end{table}

\begin{table}[hbtp]
  \begin{center}
    \begin{tabular}{lccccc}
      \hline
      \multicolumn{6}{c}{Non-causal spectrograms}\\
      \hline
      Use case \vphantom{$\frac{\hat{X}_x}{\hat{X}_x}$}
      & $\gamma_-$
      & $\gamma_+$
      & $\Delta \gamma$
      & $\tfrac{\omega_-}{\omega_0}$
      & $\tfrac{\omega_+}{\omega_0}$ \\
      \hline
      $N = 4$ \vphantom{$\sqrt{2}$}
        & -0.0458
        & +0.0480
        & 0.0938
        & 0.955
        & 1.049  \\
       $N = 8$ 
        & -0.0231
        & +0.0237
        & 0.0469
        & 0.977
        & 1.024  \\
     \hline
    \end{tabular}
  \end{center}
  \caption{Measures, that estimate the widths of the spectral bands for the
    regular non-causal Gabor transform
    applied to an ideal sine wave, for different values of the wavelength
    proportionality factor $N$.}
  \label{tab-width-spectr-bands-non-caus}
\end{table}

Table~\ref{tab-width-spectr-bands-non-caus} shows numerical values of
corresponding descriptors obtained for the regular non-causal Gabor
transform, obtained by solving the equation,
\begin{equation}
  \left|\hat{g}( \omega_0 \, e^{\gamma}- \omega_0;\;
               \left(\tfrac{2 \pi N}{\omega_0 \, e^{\gamma}}\right)^2)\right|
  = e^{- \frac{2\pi^2 N^2  (e^{\gamma} -1)^2}{e^{2\gamma}}} 
  = \frac{1}{2}.
\end{equation}
As can be seen from comparing the time-causal results in
Table~\ref{tab-width-spectr-bands-time-caus} to the non-causal results
in Table~\ref{tab-width-spectr-bands-non-caus}, the spectral bands are
somewhat wider for the time-causal and time-recursive analogue of the
Gabor transform than for the regular non-causal Gabor transform,
with the values of the compact bandwidth descriptor
$\Delta \gamma$ having the following ratios
\begin{align}
  \begin{split}
    \frac{\Delta \gamma |_{\timecaus, N = 4, c = \sqrt{2}}}
            {\Delta \gamma |_{\Gabor, N=4}}
    & \approx 1.120,
  \end{split}\\
  \begin{split}
    \frac{\Delta \gamma |_{\timecaus, N = 4, c = 2}}
            {\Delta \gamma |_{\Gabor, N=4}}
    & \approx 1.220,
  \end{split}\\
  \begin{split}
    \frac{\Delta \gamma |_{\timecaus, N =8, c = \sqrt{2}}}
            {\Delta \gamma |_{\Gabor, N=8}}
    & \approx 1.120,
  \end{split}\\
  \begin{split}
    \frac{\Delta \gamma |_{\timecaus, N =8, c = 2}}
            {\Delta \gamma |_{\Gabor, N=8}}
    & \approx 1.220,
  \end{split}
\end{align}
between the time-causal {\em vs.\/}\ the non-causal time-frequency
analysis concepts for the four different main uses.

Thus, the measure $\Delta \gamma$ of the width of a spectral band, in
terms of logarithmic frequencies, is about
12~\% or 22~\% larger for the time-causal and time-recursive analogue
of the Gabor transform compared to the regular non-causal Gabor
transform, depending on whether the value of the distribution
parameter $c$ for the time-causal limit kernel is chosen as either
$c = \sqrt{2}$ or $c = 2$.

\section{Experimental characterization of the robustness to noise for time-causal frequency estimates}
\label{sec-exp-robustness-to-noise}

Beyond the above characterizations of the proposed time-causal
time-frequency analysis method in the ideal noise free case, it is
also of interest to investigate how sensitive the resulting frequency
estimates will be due to noise. First of all, we can note that the
frequency sensitivity curve for a spectral band, as approximated
by $|({\cal H} f)(t, \omega;\; \tau, c)| \approx R(\omega;\; \tau, c)$
according to (\ref{eq-approx-spectr-R-in-deriv}) for the time-causal
and time-recursive analogue of the Gabor transform, is, for values of the
angular frequency rather near the peak angular frequency
at $\omega = \omega_0$, rather symmetric around $\omega = \omega_0$,
as can seen in Figure~\ref{fig-freq-selectivity} and from the
approximate symmetry of the relative delimiters $\gamma_-$ and
$\gamma_+$ of a spectral band in Table~\ref{tab-width-spectr-bands-time-caus}.

Thus, if a signal contains superimposed added noise with rather uniform
spectral properties over the different angular frequencies $\omega$,
then we could expect the contributions from the noise to the variations
over the angular frequency $\omega$ across a spectral band to
comparably well balance each other, such that the angular frequency
estimate $\hat{\omega}$ obtained by applying a peak detector over the
variations over the angular frequency in the spectrogram ought to not
deliver excessively biased frequency estimates
in either direction of the angular frequency.

In this section, we will perform a characterization of this property
experimentally, and also in the discrete case, which then,
beyond the idealized continuous theoretical analysis
in Section~\ref{sec-theor-est-accuracy-freq-est}, thereby also
characterizes the influence on the accuracy of local frequency
estimates due to the discretization steps, needed to
transfer the continuous formulation of the time-causal and
time-recursive analogue to a discrete implementation,
according to Section~\ref{sec-disc-impl}.

\subsection{Experimental analysis of the accuracy of local frequency
  estimates for a single sine wave with different amounts of added
  white Gaussian noise}

To analyze the frequency selectivity properties for a discrete
implementation of the proposed time-causal and time-recursive analogue
of the Gabor transform, we proceeded as follows:
\begin{itemize}
\item
  A set of 5 logarithmically spaced frequency intervals,
  $[240~\mbox{Hz}, 480~\mbox{Hz}]$,
  $[480~\mbox{Hz}, 960~\mbox{Hz}]$,
  $[960~\mbox{Hz}, 1920~\mbox{Hz}]$,
  $[1920~\mbox{Hz}, 3840~\mbox{Hz}]$ and
  $[3840~\mbox{Hz}, 7680~\mbox{Hz}]$,
  was selected. Within each such frequency interval, 10 random
  frequencies were chosen, drawn%
\footnote{The motivation for choosing the frequencies randomly in this
  way, is first of all to ensure that they will be located without any systematic
  relations to the discrete sampling frequencies in the discretized time-frequency
  transforms. Specifically, by later
  finally pooling the results over 5 frequency intervals with 10
  random samples within each such interval, we both ensure that the
  frequencies should sufficiently well span the entire frequency range
  of the union $[240~\mbox{Hz}, 7680~\mbox{Hz}]$ of the 
  frequency intervals, and also with a total number of 50 pooled samples
  per noise level sufficiently well represent each resulting statistical
  measure, computed as either a logarithmic mean
  value or a logarithmic standard deviation.}
  from a uniform random distribution
  over a logarithmic frequency domain.
\item
  For each such frequency, a sine wave of amplitude 1 and duration
  3.0~seconds%
\footnote{This temporal duration of the signals was chosen, to make it possible
  to meaningfully listen to the generated audio files.}
  was
  generated, with sampling frequency 44.1~kHz, corresponding
  to CD quality. Furthermore, different amounts of white Gaussian
  noise was added to each signal, with the following 6 standard deviations
  for the noise $\nu \in \{ 0, 0.01, 0.0316, 0.10, 0.316, 1.0\}$.
\item
  For each such noisy signal, 6 different types of auditory spectrograms were
  computed, with 4 of these auditory spectrograms being time-causal,
  based on the time-causal and time-recursive analogue of the Gabor
  transform for wavelength proportionality factor $N \in \{ 4, 8 \}$
  and distribution parameter $c \in \{ \sqrt{2}, 2 \}$ for the
  time-causal limit kernel, and 2 of these auditory spectrograms being
  non-causal, based on the regular regular non-causal Gabor transform
  with wavelength proportionality factor $N \in \{ 4, 8 \}$.
\item
  The discrete implementation of the time-causal and time-recursive
  analogue of the Gabor transform was according to the treatment in
  Section~\ref{sec-disc-impl}, with the temporal smoothing operation
  performed in terms of 8 layers of first-order recursive filters
  coupled in cascade. The discrete implementation of the Gabor
  transform was based on temporal smoothing with the sampled%
\footnote{For this comparison to the regular non-causal Gabor
  transform, the discrete implementation of the Gabor transform was
  chosen to be based on using the sampled Gaussian kernel instead of the
  discrete analogue of the Gaussian kernel, since using sampled
  Gaussian kernel may constitute the otherwise most common way of implementing
  Gaussian convolution.}
  Gaussian kernel truncated at the tails for a truncation error $\epsilon$
  below $10^{-8}$.
\item
  The auditory spectrograms were computed over the frequency interval
  $[20~\mbox{Hz}, 16000~\mbox{Hz}]$ with 48 uniformly sampled frequencies per
  octave over a logarithmic frequency scale, and with the standard deviations
  $\sigma$ for the temporal window functions in the time-frequency
  transforms proportional to the frequency in an interior part of the
  frequency interval. In the transitions between the interior part of
  the frequency and the exterior parts, we did, however, here use
  hard thresholding of the temporal scale values of the temporal
  window functions (as opposed to the soft thresholding approach according to
  Appendix~\ref{app-adapt-auditory-spectrograms}, that we otherwise use
  for computing auditory spectrograms), to guarantee true
  proportionality with respect to the interior frequencies in the frequency
  range. The reason for using this thresholding operation on the temporal scale
  values is to prevent too long temporal
  scale values for lower frequencies or too short temporal scales for
  higher frequencies.
\item
  For the wavelength proportionality factor $N = 4$, the bounds on the
  linear range of the temporal scale values were
  $\sigma \in [0.5~\mbox{ms}, 20~\mbox{ms}]$, while for
  the wavelength proportionality factor $N = 8$, the bounds on the
  linear range of the temporal scale vales were
  $\sigma \in [1~\mbox{ms}, 40~\mbox{ms}]$.
\item
  For both of the cases $N = 4$ and $N = 8$, the temporal scales of
  the temporal window functions were therefore proportional to the
  frequency in the spectrogram for frequencies in the rangle
  $[200~\mbox{Hz}, 8000~\mbox{Hz}]$, that is for a frequency interval
  that with a reasonable margin clearly contains the range of the randomly generated
  frequencies for the signals for which the spectrograms are computed.
\end{itemize}

\begin{table*}[hbtp]
  \begin{center}
    \begin{tabular}{lcccccc}
      \hline
      \multicolumn{7}{c}{Accuracy of local frequency estimates for synthetic
      sine waves with added white Gaussian noise}\\
      \hline
      Use case
      & $\nu = 0$ & $\nu = 1~\%$ & $\nu \approx 3.16~\%$ 
      & $\nu = 10~\%$ & $\nu \approx 31.6~\%$ & $\nu = 100~\%$ \\
      \hline
      Time-causal $N = 4$, $c = \sqrt{2}$
        & 1.0001 */ 1.0001 & 1.0001 */ 1.0001 &  1.0001 */ 1.0001 
        & 1.0001 */ 1.0003 & 1.0001 */ 1.0008 &  1.0001 */ 1.0076 \\
      Time-causal $N = 4$, $c = 2$
        & 1.0001 */ 1.0001 & 1.0001 */ 1.0001 &  1.0001 */ 1.0002
        & 1.0001 */ 1.0003 & 1.0001 */ 1.0008 &  1.0003 */  1.0153 \\
      Time-causal $N = 8$, $c = \sqrt{2}$
        & 1.0000 */ 1.0004 & 1.0000 */ 1.0004 &  1.0000 */  1.0004
        & 1.0000 */ 1.0004 & 1.0000 */ 1.0005 &  1.0000 */ 1.0011 \\
      Time-causal $N = 8$, $c = 2$
        & 1.0000 */ 1.0004 & 1.0000 */ 1.0004 &  1.0000 */ 1.0004
        & 1.0000 */ 1.0004 & 1.0000 */ 1.0005 &  1.0000 */ 1.0012 \\
      Non-causal $N = 4$
         & 1.0001 */ 1.0001 & 1.0001 */ 1.0001 &  1.0001 */ 1.0001
         & 1.0001 */ 1.0002 & 1.0001 */ 1.0007 &  1.0001 */ 1.0055 \\
      Non-causal $N = 8$
         & 1.0001 */ 1.0002 & 1.0001 */ 1.0002 &  1.0001 */ 1.0002
         & 1.0001 */ 1.0003 & 1.0001 */ 1.0004 &  1.0001 */ 1.0009 \\
      \hline
    \end{tabular}
  \end{center}
  \caption{Experimental characterization of the accuracy of local
    frequency estimates obtained with the proposed time-causal and
    time-recursive analogue of the
    Gabor transform as well as with the regular non-causal Gabor
    transform, for synthetic sine waves with amplitude 1 and
    different amounts of added white Gaussian noise with standard
    deviations $\nu \in \{ 0, 0.01, 0.0316, 0.10, 0.316, 1.0\}$.
    The local frequency estimates have been computed by for
    each time moment $t$ by detecting the
    global maximum over the frequencies in the absolute value of the
    spectrogram, and then interpolating that discrete estimate to
    higher resolution, by local parabolic interpolation over the
    nearest neighbours in the frequency direction.
    The results are shown in terms of (i) a bias factor $b$, according to
    (\ref{eq-def-bias-freq-est}), obtained by computing the
    mean of the logarithm of the ratio
    between the estimated frequency and the reference frequency and
    then exponentiating this result, and (ii) a spread factor $s$,
    according to (\ref{eq-def-spread-freq-est}),
    obtained by computing the standard deviation of the logarithm of the ratio
    between the estimated frequency and the reference frequency and
    then exponentiating that result. As can be seen from the data, the
  local frequency estimates computed in this way are very close to the
true values (with the ideal values being that both the bias factor $b$
and
the spread factor $s$ should be equal to 1). Notably, the deviations from the
ideal values are essentially always below or far below the quantization
error caused by sampling the frequency domain with 48
logarithmically distributed quantized frequencies per octave.}
  \label{tab-accuracy-freq-est-sines-with-noise}
\end{table*}

\noindent
Then, to obtain local frequency estimates and to quantify the accuracy
of these estimates, we proceeded as follows:
\begin{itemize}
\item
  To avoid possible boundary effects, the first 10~\% and the last
  10~\% of the temporal points were discarded. Thus, the following
  temporal operations were only performed within the central 80~\%
  of the temporal sample points.
\item
  For each time moment $t$, the discrete frequency $\hat{f}_{\disc}(t)$
  for the global maximum in the absolute value of the spectrogram over the
  logarithmic frequencies was determined:
  \begin{equation}
    \hat{f}_{\disc}(t)
    = \frac{1}{2\pi}
    \argmax_{\omega_{\disc}} |({\cal H} f)(t, \omega_{\disc};\;
                                                             \left( \frac{2 \pi N}{\omega_{\disc}} \right)^2, c)|.
  \end{equation}
\item
  Each such discrete frequency estimate $\hat{f}_{\disc}(t)$ was then
  interpolated to a subresolution frequency estimate $\hat{f}(t)$ of
  higher resolution by parabolic interpolation, that is by fitting a
  second-order polynomial to the variation over the discrete
  logarithmic frequencies over the nearest neighbours just below and
  just below the determined discrete frequency estimate, and then
  determining the peak position of that local second-order polynomial:
  \begin{equation}
    \hat{f}(t)
    = \operatorname{parabolicinterpol}(\operatorname{neighbours}(\hat{f}_{\disc}(t))).
  \end{equation}
\item
  Measures of the temporal mean and the temporal standard deviations of
  these frequency estimates were computed, by first computing the
  logarithm%
\footnote{The motivation for computing the statistical mean values and
  standard deviations after a logarithmic transformation of the
  frequency ratios is that that frequency ratio can be expected to be
  more symmetric with regard to the reference value over a logarithmic
  scale than over a linear scale.}
  of the ratio between the frequency estimate $\hat{f}(t)$ and the
  reference value $f_{\refref}$ of the true frequency, and then exponentiating
  the statistical mean and standard deviation
  measures computed over the logarithmic frequency
  domain, thus leading to a multiplicative bias measure $b$
  and a multiplicative spread measure $s$ of the forms
  \begin{align}
    \begin{split}
      \label{eq-def-bias-freq-est}
      b = \exp(\operatorname{mean}_t(\log\left(\frac{\hat{f}(t)}{f_{\refref}}\right))),
    \end{split}\\
    \begin{split}
      \label{eq-def-spread-freq-est}
     s = \exp(\operatorname{sdev}_t(\log\left(\frac{\hat{f}(t)}{f_{\refref}}\right))).
    \end{split}
  \end{align}
\end{itemize}

\noindent
Table~\ref{tab-accuracy-freq-est-sines-with-noise} shows the results
of performing these operations for the 4 types of time-causal
spectrograms and the 2 types of non-causal spectrograms, with the
results for all the 5 frequency intervals with 10 samples in each interval
pooled into the reported values for the multiplicative bias values $b$
and the multiplicative spread values $s$, and with the
results for each use case and each noise level compactly shown on the form
\begin{equation}
   b */ s.
\end{equation}
As can be seen from these results,  the multiplicative bias values $b$ are very close
to the ideal value $b = 1$, with the deviations being of the order of
$10^{-4}$ or lower for all the noise levels.
The multiplicative spread values $s$ are somewhat further away from their ideal
values $s = 1$, with increasing deviations from the ideal value $s = 1$ for
increasing noise levels, but still for noise levels
up to $31.6~\%$ the deviations are below $10^{-3}$. Only when the noise level is as high
as $100~\%$, the deviations of the multiplicative spread values from
the ideal value of 1 reach values of the order of $1~\%$.

Notably also, comparing to the effects of the quantization error,
which with the chosen parameter settings of using 48 logarithmically
distributed frequency levels per octave should correspond to a
relative quantization error of the order of $2^{1/48} \approx 1.0145$,
that is to a quantization error of the order of $1.5~\%$, we can note
that the shapes of the frequency selectivity curves for the
time-causal and time-recursive analogue of the Gabor transform are
very compatible with the chosen method for parabolic interpolation, to
obtain frequency estimates of much higher resolution than the raw
discretization of the frequency values in the discrete implementation.

In these respects, the frequency estimates obtained from the time-causal
and time-recursive analogue must be regarded as very accurate, and
specifically also consistent with the results from the previous theoretical analysis
of the frequency selectivity properties for the fully continuous spectrograms, as
derived in Section~\ref{sec-theor-estim-acc-freq-est-noise-free} and as
summarized in Table~\ref{tab-freq-offset}.

A possible explanation of why these error measures are very low,
also in the presence of substantial amounts of noise,
is that the temporal filtering operations in the proposed discrete
implementation of the new time-causal time-frequency transform
are based on provably variation-diminishing filtering operations
(see Appendices~\ref{app-class-temp-scsp-kern}--\ref{app-discr-impl}),
which should then strongly reduce the influence of local perturbations and noise.
  
{\em Remarks:\/}
Finally, it should be remarked that we have in this experiment
deliberately chosen a very much simplified peak detection algorithm,
that for at each temporal moment
detects just the main peak (the global maximum) over the logarithmic frequencies in
the spectrogram. More realistically, in an actual algorithm for
performing local frequency estimation in real-world signals, it is
more appropriate to detect multiple peaks (several local maxima) over the
logarithmic frequencies, to be able to simultaneously handle
responses to multiple local frequencies in the input signal.

Additionally, to prevent spurious responses to noise or other interfering
structures in the data, it can also,  for auditory signals recorded from sound
sources with more complex spectral characteristics than a pure sine
wave, be more appropriate to precede the above peak
detection step with a local bandpass filtering stage, {\em e.g.\/} by
filtering the absolute value of the spectrogram with a negative second-order derivative
of a Gaussian kernel in the logarithmic frequency direction, as done
in Lindeberg and Friberg \cite{LinFri15-PONE} Section
``Auditory features from second layer receptive fields'', see
specifically the subparagraph on ``Spectral sharpening'' on page~31 in
that paper.

If additional temporal delays can furthermore be acceptable, it is also
possible to combine such a spectral sharpening operation with
complementary filtering in the temporal
direction, to additionally reduce the spread of the local frequency
estimates.

In this treatment, we have on the other hand deliberately chosen an as
much simplified peak detection method over the logarithmic frequencies
as possible, in order to as far as possible reveal the properties of
the pure time-frequency transforms, without having the results being
further influenced by the properties of more refined methods for
spectral sharpening.

\section{Covariance properties of the time-causal and time-recursive
  analogue of the Gabor transform}
\label{sec-summ-cov-props}

To summarize, similarly to the regular Gabor transform,
the proposed time-causal and time-recursive analogue of the Gabor
transform is covariant under the
following transformations of the input signal $f(t)$:
\begin{itemize}
\item
  Temporal shift:
  \begin{equation}
     f'(t') = f(t) \quad\quad \mbox{for} \quad\quad t' = t + \Delta t.
  \end{equation}
\item
  Temporal rescaling:
  \begin{equation}
    f'(t') = f(t) \quad\quad \mbox{for}%
\footnote{With the additional restriction that temporal scaling factor $S_t$ must
  be of the form $S_t = c^j$ for the time-causal and time-recursive
  analogue of the Gabor transform, with $c$ being the distribution
  parameter of the time-causal limit kernel and $j$ being an integer.} \quad\quad t' = S_t\, t.
  \end{equation}
\item
  Frequency shift:
  \begin{equation}
    \hat{f'}(\omega') = \hat{f}(\omega) \quad\quad \mbox{for} \quad\quad \omega' = \frac{\omega}{S_{\omega}}.
  \end{equation}
\item
  Amplitude rescaling:
 \begin{equation}
    f(t') = A \, f(t) \quad\quad \mbox{for} \quad\quad t' = t.
  \end{equation}
\end{itemize}
In this respect, the proposed time-causal and time-recursive analogue of
the Gabor transform can be expected to have a robust behaviour for
signals that may undergo these basic types of transformations over the
signal domain.

In the area of computer vision, the formulation of
provable covariance properties for mathematically based image
operations has been established as an effective way of
substantially increasing the robustness of
visual operations to the influence of the variabilities generated by
natural image transformations,
see Lindeberg \cite{Lin13-BICY,Lin24-arXiv-UnifiedJointCovProps} and the
references therein. In a similar way, we propose that the formulation
of provable covariance properties for other classes of signal domains,
such as the temporal and auditory signals considered in this paper,
can be essential for increasing the robustness of signal processing
operations, that are to operate on real-world data generated from
physical or biological processes with substantial variabilities.

\section{Summary and discussion}
\label{sec-summ-disc}

We have presented a framework for performing time-frequency analysis
over multiple temporal scales in a way that is fully time-causal and
time-recursive. In this way, the presented theory makes it possible to
perform time-frequency analysis for both modelling and processing
real-world physical or biological signals, as well as expressing
corresponding signal processing operations in real time.

The time-causal analogues to the Gabor function and the Gabor
transform, that we have presented, obey true temporal scale covariance
under uniform rescalings of the temporal domain, for rescaling factors
that are integer powers of the distribution parameter $c$ for the
time-causal limit kernel, that represents the temporal smoothing kernel
in the time-causal analogue of the Gabor transform. The time-causal
analogue of the Gabor transform, that we have proposed in this work,
does also obey a cascade smoothing property over temporal scales, which
ensures that a time-frequency transform at a coarse temporal scale can
be treated as a simplification of the corresponding time-frequency
transform at any finer temporal scale. In these ways, the proposed
representations generalize unique properties of the Gabor functions
from a non-causal temporal domain to a truly time-causal temporal
domain.

In essence, the Gaussian kernel, which is used as the temporal window function in
the regular non-causal Gabor transform, can, based on the arguments developed in more
detail in Appendix~\ref{sec-arg–why-time-caus-anal},
be regarded as the canonical temporal
smoothing kernel over a non-causal temporal domain.
The time-causal limit kernel, used as temporal window function in the
proposed new time-causal frequency analysis, can based on as close as
possible arguments over a time-causal temporal domain, be regarded as the
canonical temporal smoothing kernel over a time-causal temporal
domain. For these conceptual reasons, we refer to the new time-causal
frequency analysis concept as a time-causal analogue of the Gabor transform.

Concerning the discrete implementation, we have also presented
an axiomatically based theoretical framework, outlined in
Appendix~\ref{app-class-temp-scsp-kern}, for defining a
discrete analogue of the proposed time-causal analogue of the Gabor
transform, based on a set of first-order recursive filters coupled in
cascade. Due to the positive time constants in these recursive
filters, the corresponding temporal filtering operations are guaranteed to obey
provably variation-diminishing properties, which is important for the numerical
stability, by provably reducing the influence of local perturbations
and noise.
This method for discrete implementation does also constitute a true
numerical approximation of the corresponding continuous theory.

An explicit algorithm for computing a discrete analogue of the
proposed time-causal analogue of the Gabor transform has been outlined
in Appendix~\ref{app-discr-impl}, with further details regarding a
strictly time-recursive implementation in
Appendix~\ref{sec-app-real-time-alg}.

In contrast to the regular non-causal Gabor transform, the proposed
time-causal time-frequency analysis concept is inherently associated
with temporal delays, for which we have presented closed-form
estimates in Section~\ref{sec-temp-delay}, showing that the temporal
delay is proportional to the temporal scale parameter measured in
dimension $[\mbox{time}]$, and that the temporal delay becomes shorter for
increasing values of the distribution parameter $c$ of the time-causal
limit kernel.
We have also in Sections~\ref{sec-freq-sel}
and~\ref{sec-theor-estim-width-spectr-band}
analyzed the frequency
selectivity properties of the proposed time-causal analogue of the
Gabor transform, with comparisons to the regular non-causal Gabor
transform, with specific emphasis on how different choices of the temporal
scale parameter $\tau$ and the distribution parameter $c$ of the time-causal
limit kernel affect the frequency selectivity properties, specifically
showing that the frequency selectivity becomes less narrow for
increasing values of the distribution parameter $c$ of the time-causal
limit kernel.

In these respects, requirements of short temporal delays
and narrow frequency selectivity properties constitute conflicting
requirements, that should be balanced for any specific application.
In this context, however, computing the time-frequency analysis over
multiple temporal scales, as the proposed time-causal analogue of the
Gabor transform is highly suitable for, should, however, also be
considered as an alternative, since then multiple trade-offs between
temporal delays and frequency selectivity properties can be obtained
simultaneously, over variations of the temporal scale in the temporal
window kernel of the time-causal time-frequency transform, and with a
very small amount of additional computations.

To quantify how accurate the resulting local frequency estimates can
be expected to be for the time-causal and time-recursive analogue of
the Gabor transform, as well as how the properties of the proposed time-causal and
time-recursive analogue of the Gabor transform differ from the
properties of the regular non-causal Gabor transform, we have in
Section~\ref{sec-theor-est-accuracy-freq-est} formulated theoretical
estimates based on closed-form expressions for the resulting
time-causal spectrograms defined from a single ideal sine wave. These estimates
indicate that the variations in local frequency estimates ought to,
in the continuous case, be
very low for the studied four main use cases, regarding the default
values of the wavelength proportionality factor $N$ and the distribution
parameter $c$ for the time-causal limit kernel. The spectral bands
obtained with the time-causal and time-recursive analogue of the Gabor
transform will, however, be somewhat wider
(of the order of 12~\% or 22~\% wider in terms of logarithmic frequencies when measuring
the width of the spectral band at half the peak value)
compared to the spectral bands obtained with the regular
non-causal Gabor transform, depending on whether choosing
the distribution parameter as $c = \sqrt{2}$ or $c = 2$.

In Section~\ref{sec-exp-robustness-to-noise}, we have then performed an
experimental characterization of the accuracy of the resulting local
frequency estimates, for the proposed and computationally very
efficient discrete implementations of the proposed
time-causal and time-recursive analogue of the Gabor transform, in
terms of a low number (4 to 8) of first-order recursive filters coupled in cascade.
For a set of synthetic sine waves with added white Gaussian noise,
we have shown that even for
noise levels up to 100~\%, the multiplicative bias values and the
multiplicative spread measures
(see Table~\ref{tab-accuracy-freq-est-sines-with-noise})
are very close to their ideal values to be equal to 1,
thus showing that the proposed time-causal and time-recursive
analogue of the Gabor transform should have the possibility to compute
very accurate local frequency estimates, if embedded within otherwise
well-designed algorithms for local frequency estimation.

We have also derived a family of inverse transforms of the proposed
time-causal analogue of the Gabor transform in
Appendix~\ref{sec-app-inv-transf-time-caus-gabor-transf}.
In contrast to the forward transform, those inverse transforms are,
however, not time-causal, and inevitably associated with additional
temporal delays. Thus, those inverse transforms may not be
directly applicable for time-critical real-time applications, why
this contribution should in that context mainly be regarded as of theoretical
interest, to clearly show that the proposed time-causal analogue of the Gabor
transform constitutes a true time-frequency transform, or intended for
applications that are not time-critical, such as offline analysis.

We propose that the theoretical constructions described in this
article could serve as a valuable tool for expressing the first layers
of time-frequency analysis when modelling physical or biological
processes in situations where non-causal access to the future is
simply not realistic, as well as for expressing real-time
time-frequency analysis methods for real-time processing, in
particular in situations where a multi-scale analysis is warranted to
capture different types of temporal structures at different temporal
scales, by using multiple temporal window functions of different
temporal duration.

Additionally, because of the computationally very efficient
discrete implementation, using the proposed time-causal and
time-recursive analogue of the Gabor transform could also be
considered as a computationally very efficient tool, when to perform
offline time-frequency analysis on larger datasets, especially if
the time-frequency analysis is to be performed over multiple temporal
scales.

\section*{Acknowledgements}

Python code, that implements a subset of the time-frequency analysis
methods described in this paper, is available
in the pygabor package, available at GitHub:
\begin{quote}
  \tt https://github.com/tonylindeberg/pygabor
\end{quote}
as well as through PyPi:
\begin{quote}
  \tt pip install pygabor
\end{quote}

\medskip
\noindent
I would like to thank the anonymous reviewers for valuable comments
that have improved the presentation.

\appendix

\subsection{In what respects the proposed time-causal time-frequency
  analysis concept can
  be regarded as a time-causal analogue of the Gabor transform}
\label{sec-arg–why-time-caus-anal}

In this appendix, we will present conceptual theoretical arguments
for time-frequency analysis that (uniquely) lead to using the Gaussian kernel as
a canonical temporal window function in a non-causal time-frequency transform,
and thus to the Gabor transform over a non-causal
temporal domain. Then, we will present as corresponding as possible
arguments for a time-frequency analysis over a time-causal temporal
domain, that then lead to instead using the time-causal limit kernel as the
canonical temporal window function, in the proposed time-causal
frequency analysis concept.

\subsubsection{Choosing the temporal window function in time-frequency
  analysis over multiple temporal scales}
Let us assume that we, for a non-causal temporal domain, are going to
define a time-frequency analysis over multiple temporal scales by
pointwise multiplication of any temporal signal $f(t)$ by a complex
sine wave $e^{-i \omega t}$ and then integrating the result using some
temporal window function $h(t;\; \tau)$, that depends on some temporal
scale parameter $\tau$:
\begin{equation}
  \label{eq-gab-transf}
  ({\cal T} f)(t, \omega;\; \tau) =
  \int_{u=-\infty}^{\infty}
     f(u) \,
    h(t - u;\; \tau) \, e^{-i \omega  u} \, du.
\end{equation}
A fundamental problem then concerns what temporal window functions
should be regarded as natural or desirable? Specifically, a crucial
problem concerns how to relate temporal window functions at different
scales, to ensure that a time-frequency representation at a coarse
scale can be regarded as some sort of simplification or abstraction of
the time-frequency representation at any finer scale.

If we choose address this problem by requiring that the convolutions of
the real and imaginary components of the product $f(u) \, e^{-i \omega  u}$
with the temporal smoothing function should be regarded as 
simplifications, in the sense that the number of local extrema or
the number of zero-crossings in the
temporally smoothed product must not increase from finer to coarser
temporal scales, then this problem can be addressed by the tools of
temporal scale-space theory, described in
Lindeberg \cite{Lin23-BICY} Section~2.1.

\subsubsection{Classification of temporal smoothing kernels}

According to this temporal scale-space theory, a temporal smoothing
kernel is referred to as a temporal scale-space kernel if for any input signal
it has the property that the number of local extrema (or equivalently the number of
zero-crossings) in the convolved signal cannot exceed the number of
local extrema in the input signal. Based on results by
Schoenberg \cite{Sch50}, it can furthermore be stated that a
continuous smoothing kernel is a scale-space kernel if and only if it has
a bilateral Laplace-Stieltjes transform
of the form
 \begin{equation}
    \label{eq-char-var-dim-kernels-cont-case-Laplace}
    \int_{\xi = - \infty}^{\infty} e^{-s \xi} \, h(\xi) \, d\xi =
    C \, e^{\gamma s^2 + \delta s}
    \prod_{i = 1}^{\infty} \frac{e^{a_i s}}{1 + a_i s}
      \quad
  \end{equation}
for $-c < \mbox{Re}(s) < c$ and some $c > 0$,
where $C \neq 0$, $\gamma \geq 0$, $\delta$ and $a_i$ are real and
$\sum_{i=1}^{\infty} a_i^2$ is convergent.

In this expression, the factor $e^{\gamma s^2 + \delta s}$ corresponds
to possibly time-delayed Gaussian kernels
\begin{equation}
  \label{eq-gauss-kern-argum-anal-Gabor}
   h(t;\; \tau, \Delta t) = \frac{1}{\sqrt{2 \pi}} \, e^{-(t- \Delta t)^2/2\tau},
\end{equation}
while the factors $1/(1 + a_i s)$ correspond to truncated exponential
functions
 \begin{equation}
    \label{eq-truncexp-polya-comp-I}
    h(\xi;\, \mu) =
    \left\{
      \begin{array}{lcl}
        e^{- |\mu| \xi} & & \xi \geq 0, \\
        0                 & & \xi < 0,
      \end{array}
    \right.
\end{equation}
or
\begin{equation}
    \label{eq-truncexp-polya-comp-II}  
    h(\xi;\; \mu) =
    \left\{
      \begin{array}{lcl}
        e^{|\mu| \xi} & & \xi \leq 0, \\
        0                 & & \xi > 0,
      \end{array}
    \right.
\end{equation}
for some strictly positive $|\mu|$.
The product form in (\ref{eq-char-var-dim-kernels-cont-case-Laplace})
does furthermore imply that all possible convolutions of such
primitive smoothing kernels constitute possible candidates to choose
from, when to define a transformation from any finer scale to any coarser
scale, in a multi-scale time-frequency analysis concept that is
guaranteed to ensure formal simplifications from any finer scale to
any coarser scale.

\subsubsection{Arguments that uniquely single out the Gaussian kernel}

A complementary question then concerns if it would be appropriate to
impose further restrictions to reduce the search space for possible
time-frequency analysis methods? A very desirable complementary
requirement to impose is to require the possibility of defining
a continuum of possible temporal scale parameters $\tau$. A formal way of
adding such a requirement is by imposing a semi-group condition
\begin{equation}
  h(t;\; \tau_1) * h(t;\; \tau_2) = h(t;\; \tau_1 + \tau_2) 
\end{equation}
on the family of temporal smoothing kernels. If we, in addition,
as is possible over a non-causal temporal domain,
require this semi-group property to be continuous with
respect to the temporal scale parameter, then the Gaussian kernel
(\ref{eq-gauss-kern-argum-anal-Gabor}) is singled out as the unique
choice among the otherwise possible primitive temporal primitives
(\ref{eq-gauss-kern-argum-anal-Gabor}),
(\ref{eq-truncexp-polya-comp-I}) and
(\ref{eq-truncexp-polya-comp-II}).

If we, in addition, require the temporal smoothing kernels to be
symmetric, so that they do not introduce any temporal delays, then we are
thus lead to the conclusion that the resulting multi-scale
time-frequency analysis must correspond to the Gabor transform.

\subsubsection{Arguments that lead to the time-causal limit kernel}

If we, on the other hand, consider temporal signals over a time-causal
temporal domain, then we cannot use the Gaussian kernel or those
truncated exponential kernels that would imply forbidden access to the future.
The only possible temporal smoothing kernel to choose from will then
be the time-causal truncated exponential kernels of the form
(\ref{eq-truncexp-polya-comp-I}).

In principle, all possible combinations of such kernels would guarantee a
simplification from finer to coarser levels of scale. Again, one may,
however, ask if it would be desirable to impose further conditions to
reduce the search space? Notably, we cannot impose a semi-group
structure with respect to a continuous scale parameter in the
time-causal case, since the convolutions with sets of truncated
exponential kernels in cascade by necessity implies that the temporal
scale levels have to be discrete. Additionally, theoretical arguments
in Lindeberg \cite{Lin17-JMIV} Appendix~1 show that a requirement of
semi-group structure over a time-causal temporal domain would lead to
unnecessarily long temporal delays.

For these reasons, we should instead seek other forms of complementary
requirements. A weaker extension of the semi-group property of the
Gaussian kernels is to instead focus on the cascade smoothing
property
\begin{equation}
  L(\cdot;\; \tau_2) = g(\cdot;\; \tau_2 - \tau_1) * L(\cdot;\; \tau_1),
\end{equation}
that applies to the non-causal temporal scale-space
representation $L(t;\; \tau)$ obtained by
smoothing any temporal signal $f(t)$ with a Gaussian kernel
\begin{equation}
  L(\cdot;\; \tau) = g(\cdot;\; \tau) * f(\cdot).
\end{equation}
Thus, for a time-causal temporal scale-space representation, obtained
by smoothing with a time-causal temporal scale-space representation
with a time-causal temporal smoothing kernel $h(t;\; \tau)$
\begin{equation}
  L(\cdot;\; \tau) = h(\cdot;\; \tau) * f(\cdot),
\end{equation}
we require that the time-causal temporal scale-space representation
should satisfy a cascade smoothing property of the form
\begin{equation}
  L(\cdot;\; \tau_2) = \kappa(\cdot;\; \tau_1 \mapsto \tau_2) * L(\cdot;\; \tau_1) 
\end{equation}
for some incremental convolution kernel $\kappa(\cdot;\; \tau_1 \mapsto \tau_2)$,
that transforms the temporal scale-space representation from the
temporal scale level $\tau_1$ to the temporal scale level $\tau_2$.

Specifically, to guarantee that the amount of information in the
signal, measured in terms of the number of local extrema, or
equivalently the number of zero-crossings, must not increase from the
temporal scale level $\tau_1$ to the temporal scale level $\tau_2$,
then that incremental convolution kernel
$\kappa(\cdot;\; \tau_1 \mapsto \tau_2)$
should be a temporal scale space
kernel. Specifically, from the classification of such temporal
scale-space kernels above, the incremental kernel must correspond to a
set of truncated exponential kernels coupled in cascade.

If we additionally restrict ourselves to the fact that the temporal
scale levels have to be discrete for a time-causal scale-space
representation, because of the discrete nature of the temporal scale
levels, as arising from the restriction to truncated exponential
kernels coupled in cascade, then the simplest choice will therefore be that
the transformation between adjacent temporal scale levels should be
given by convolution with a single truncated exponential kernel.

In this way, we have arrived at an overall architecture, where the
temporal scale-space representation should be constructed as a cascade
of successive convolutions with truncated exponential kernels with
possibly different time constants. What then remains, is to determine
how to choose these time constants $\mu_k$.

For the ability of the temporal smoothing
kernels to appropriately reflect a quantitative notion of temporal
scale, it is specifically desirable that the temporal smoothing
kernels should also transform
properly under temporal scaling transformations. In other words,
under a scaling transformation of the temporal domain of the form
\begin{equation}
  t' = S \, t
\end{equation}
for some temporal scaling factor $S > 0$, the temporal smoothing kernel
should transform according
\begin{equation}
  h(t';\; \tau') = \frac{1}{S} \, h(t;\; \tau).
\end{equation}
A deeper theoretical analysis in
Lindeberg \cite{Lin23-BICY} Section~3 of the property of temporal scale
covariance in relation to temporal smoothing kernels, that are
constructed from truncated exponential kernels coupled in cascade,
shows that the particular definition of the time-causal limit kernel,
from having a Fourier transform of the form
\begin{align}
  \begin{split}
     \hat{\Psi}(\omega;\; \tau, c) 
     & = \prod_{k=1}^{\infty} \frac{1}{1 + i \, c^{-k} \sqrt{c^2-1} \sqrt{\tau} \, \omega},
  \end{split}
\end{align}
ensures that the temporal smoothing kernels will then lead to temporal
scale covariance. This will hold in the sense that for temporal scaling factors $S$ of the
form
\begin{equation}
  S = c^j,
\end{equation}
with temporal scaling transformations of the form
\begin{equation}
  t' = S \, t,
\end{equation}
the temporal smoothing kernels will transform according to
\begin{equation}
  \Psi(t';\; \tau', c) = \frac{1}{S} \, \Psi(t;\; \tau, c),
\end{equation}
provided that the temporal scale parameters are matched according to
\begin{equation}
  \tau' = S^2 \, \tau.
\end{equation}

\subsubsection{Conceptual similarities between the Gaussian kernel and
  the time-causal limit kernel}

Over a non-causal temporal domain, the continuous Gaussian kernel also
obeys such a similar scaling transformation property
\begin{equation}
  g(t';\; \tau') = \frac{1}{S} \, g(t;\; \tau).
\end{equation}
as the time-causal limit kernel obeys.

The Gaussian kernel and the time-causal limit kernel do also
constitute the preferred choices among kernels that guarantee
non-creation of new local extrema (or equivalently zero-crossings)
from any finer to any coarser level of scale, over the respective
domains of either non-causal or time-causal kernels. In these
respects, these kernels constitute the canonical kernels with
strong information-reducing temporal smoothing properties,
over either non-causal or time-causal temporal domains.

In this way, the time-causal limit kernel carries as many as
possible of these information-reducing theoretical properties of the Gaussian kernel from a
non-causal temporal domain to a time-causal temporal domain.

Since these underlying properties do also carry over to time-frequency analysis
in the ways described in Section~\ref{sec-theor-prop-gab-filt} and
Section~\ref{sec-time-caus-anal-gab-filt}, we do therefore refer to
the proposed new time-causal time-frequency analysis concept as a
time-causal analogue of the Gabor transform.

Concerning treating the time-causal limit kernel as a time-causal
analogue of the Gaussian kernel, it has previously been shown that the
time-causal limit kernel can replace the role of the non-causal
temporal Gaussian kernel in time-causal models of spatio-temporal receptive fields
for video analysis (see Lindeberg
\cite{Lin16-JMIV,Lin17-JMIV,Lin18-JMIV} and
Jansson and Lindeberg \cite{JanLin18-JMIV}) as well as in time-causal models
of spatio-temporal receptive fields in biological vision
(see Lindeberg \cite{Lin21-Heliyon,Lin23-FrontCompNeuroSci}).

\subsubsection{Conceptual differences between the Gaussian kernel and
  the time-causal limit kernel}

Note that, due to the fundamental differences between a time-causal and
non-causal domain, we cannot, however, expect to be able to carry over all the
theoretical properties of the Gabor transform to a time-causal
frequency analysis.

First of all, by not having access to information from the future, an
analysis based on the time-causal limit kernel will not have access to
the same amount of information as an analysis based on the Gaussian
kernel, and can therefore not be expected to be able to compete with the Gaussian
kernel on a fair basis,
if evaluated in an offline scenario,
if comparisons would be made between an analysis based on the Gaussian
kernel, with an associated oracle that has non-causal access to
the future, and an analysis based on the time-causal limit kernel,
with restricted access to only what can be derived from information that has
occurred in the past.

Fundamentally, because of the temporal causality, the temporal
time-frequency analysis is associated with inescapable temporal
delays, caused by filtering information from the past only, and using
filters of non-infinitesimal size, as further described in
Section~\ref{sec-temp-delay}.

As described above, it is neither possible nor desirable to aim at
carrying over a semi-group property over temporal scales to a
time-causal temporal domain, because such a semi-group property would lead to
excessive temporal delays. Due to the semi-group property over a
continuous temporal scale parameter, the
Gaussian distribution is infinitely divisible, which among other
things implies special properties in terms of noise suppression for any
regular temporal signal and specifically more narrow frequency
selectivity in the context of a time-frequency analysis, as further
detailed in Section~\ref{sec-freq-sel}.

By decreasing the distribution parameter $c$ towards 1, the
time-causal limit kernel could, however, be made successively more
divisible, and then also making the frequency selectivity properties more
narrow. Such a decrease in the distribution parameter $c$ would,
however, also lead to longer temporal delays, as well as a need for a
larger amount of computations, and may therefore not be
desirable in time-critical situations.

\subsubsection{Underlying philosophy of the proposed time-causal
  time-frequency analysis concept}

What we have focused on in the development of this time-causal
time-frequency analysis concept is therefore instead the provably
variation-diminishing properties of the temporal window functions,
which also lead to excellent numerical properties for the resulting
discrete implementation of the forward transform.

\subsection{Determination of a discrete analogue of the time-causal
  analogue of the Gabor transform}
\label{app-class-temp-scsp-kern}

This appendix describes theoretical arguments, by which we can arrive
at the conclusion that a both natural, theoretically well-founded and
computationally very efficient way of implementing a discrete analogue
of the time-causal analogue of the Gabor transform is by filtering the
product between a discrete input signal $f(n \, \Delta t)$ and a sampled complex
sine wave $e^{-\omega \, n \, \Delta t}$
with a set of first-order recursive filters coupled in cascade.

\subsubsection{Classification of discrete temporal scale-space kernels}

Following the treatment in Lindeberg \cite{Lin23-BICY}
Section~4.1, a discrete kernel is referred to as a discrete temporal
scale-space kernel if it: (i)~obeys temporal causality in the sense
that it does not require access from the future, and
(ii)~guarantees that for any temporal signal the number of local extrema
(or equivalently the number of zero-crossings) in the convolved signal
does not exceed the number of local extrema in the original signal.

Let us initially, first consider the not necessarily time-causal case.
According to the classification of discrete (not necessarily
time-causal) scale-space
kernels in Lindeberg \cite{Lin23-BICY} Section~4.1, based on
theoretical results by Schoenberg \cite{Sch48}, a (not necessarily
time-causal) discrete kernel
guarantees non-creation of new local extrema (or equivalently
zero-crossings) if and only if it has a generating function of the form
  \begin{equation}
    \label{eq-char-pf}
    \varphi(z) = c \; z^k \; e^{(q_{-1}z^{-1} + q_1z)}
                 \prod_{i=1}^{\infty} \frac{(1+\alpha_i z)(1+\delta_i
    z^{-1})} {(1-\beta_i z)(1-\gamma_i z^{-1})}
  \end{equation}
where
$c > 0$, $k \in \bbbz$, 
$q_{-1}, q_1, \alpha_i, \beta_i, \gamma_i, \delta_i \geq 0$ and
$\sum_{i=1}^{\infty}(\alpha_i + \beta_i + \gamma_i + \delta_i) < \infty$.

To interpret this result, we can from the components in this
expression note that
\begin{itemize}
\item
  the factors $1+\alpha_i z$ and $1+\delta_i z^{-1}$
correspond to binomial smoothing of the form
   \begin{equation}
      \label{eq-gen-bin-filt}
      \begin{split}
        f_{\outout}(x)
        & =
          f_{\inin}(x) +
          \alpha_i \, f_{\inin}(x - 1)
          \quad (\alpha_i \geq 0),\\
        f_{\outout}(x)
        & =
          f_{\inin}(x) +
          \delta_i \, f_{\inin}(x + 1)
          \quad (\delta_i \geq 0),
      \end{split}
    \end{equation}
\item
  the factors $1-\beta_i z$ and $1-\gamma_i z^{-1}$ correspond to
  first-order recursive filters of the form
    \begin{equation}
      \label{eq-first-order-rec-filters}
      \begin{split}
        f_{\outout}(x)
        & =
        f_{\inin}(x) +
          \beta_i \, f_{\outout}(x - 1) \quad
          (0 \leq \beta_i < 1), \\
        f_{\outout}(x)
        & =
          f_{\inin}(x) +
            \gamma_i \, f_{\outout}(x + 1) \quad
            (0 \leq \gamma_i < 1),
      \end{split}
    \end{equation}
  \item
    the factor
    $ e^{(q_{-1}z^{-1} + q_1z)}$ corresponds to infinitely divisible
  distributions (see the monograph on this topic by Sato \cite{Sat99-Book}), where the 
  case specifically $q_{-1} = q_1$ corresponds to convolution with the
  non-causal discrete analogue of the Gaussian
kernel (see Lindeberg \cite{Lin90-PAMI}) and the case $q_{-1} = 0$
corresponds to convolution with
time-causal temporal Poisson kernel
(see Lindeberg and Fagerstr{\"o}m \cite{LF96-ECCV}), which obeys a
continuous semi-group property over temporal scales for discrete
temporal signals, however, at the cost of longer temporal delays.
\end{itemize}
The product form of this expression does, furthermore, mean that
(not necessarily time-causal) discrete scale-space kernels correspond
to convolutions of the above primitive kernels.

\subsubsection{Choice of discrete implementation method out of the
  theoretically possible candidates}

If we are to implement a discrete analogue of time-frequency analysis
based on variation-diminishing temporal kernels, then the above primitive
temporal kernels do therefore constitute a very natural set of primitive
temporal smoothing kernels to choose from. Due to the requirement of temporal
causality, we do, however, have to exclude those candidates that would
imply forbidden access to the future.

For our task of also approximating the properties of the time-causal
analogue of the Gabor transform numerically, the first-order recursive
filters of the form
\begin{equation}
        f_{\outout}(x)
        =
        f_{\inin}(x) +
          \beta_i \, f_{\outout}(x - 1) \quad
          (0 \leq \beta_i < 1)
\end{equation}
do in this context stand out as the natural candidate to select,
since such first-order filters do precisely constitute appropriate
numerical approximations of the first-order integrators
(\ref{eq-first-ord-int}), that represent the functional effect of
performing convolutions with truncated exponential kernels
of the form (\ref{eq-trunc-exp-kernel}) (see the treatment below for an explicit proof).

For convenience, we specifically choose to express the discrete
first-order integrators of the form
\begin{equation}
  \label{eq-first-order-rec-filt-app}
  f_{\outout}(t) - f_{\outout}(t-1)
  = \frac{1}{1 + \mu_k} \,
    (f_{\inin}(t) - f_{\outout}(t-1)),
\end{equation}
which is also maximally well-conditioned with respect to possible numerical
errors or perturbations in the input signals.

\subsubsection{Proof of numerical approximation property between the
  continuous and the discrete first-order integrators}
\label{sec-proof-1st-order-int-nu-approx}

To show that the discrete recursive filter
(\ref{eq-first-order-rec-filt-app}) can be regarded as a numerical
approximation of the continuous first-order integrator, let us first
introduce the following specialized notation for the time constant $\mu_{k,\cont}$ in the
continuous case
(\ref{eq-first-ord-int})
\begin{equation}
  \label{eq-first-ord-int}
   (\partial_t f_{\outout})(t) 
   = \frac{1}{\mu_{k,\cont}} \left( f_{\inin}(t) - f_{\outout}(t)\right).
\end{equation}
Let us then, given a temporal sampling distance $\Delta t > 0$,
approximate the temporal derivative
$(\partial_t f_{\outout})(t)$ using Euler's method \cite{DB74}
\begin{equation}
  (\partial_t f_{\outout})(t) = \frac{f_{\outout}(t) - f_{\outout}(t-1)}{\Delta t},
\end{equation}
which gives
\begin{equation}
  \frac{f_{\outout}(t) - f_{\outout}(t-1)}{\Delta t}
  = \frac{1}{\mu_{k,\cont}} \left( f_{\inin}(t) - f_{\outout}(t)\right),
\end{equation}
and which, in turn, can be rewritten as
\begin{equation}
  \mu_{k,\cont} \, f_{\outout}(t) - \mu_{k,\cont} \, f_{\outout}(t-1)
  = \Delta t \, f_{\inin}(t) - \Delta t \, f_{\outout}(t).
\end{equation}
Removing $\Delta t \, f_{\outout}(t-1)$ from each side and rearranging the
terms, then gives
\begin{multline}
  (\Delta t + \mu_{k,\cont}) (f_{out}(t) - \mu_{k,\cont} \, f_{\outout}(t-1))
  = \\ = \Delta t \, (f_{\inin}(t) - f_{\outout}(t-1)),
\end{multline}
which can be rewritten as
\begin{multline}
  f_{out}(t) - \mu_{k,\cont} \, f_{\outout}(t-1) = \\
  = \frac{\Delta t }{\Delta t + \mu_{k,\cont}} \, (f_{\inin}(t) - f_{\outout}(t-1)),
\end{multline}
and which clearly corresponds to the form
\begin{equation}
  \label{eq-first-order-rec-filt-app-again}
  f_{\outout}(t) - f_{\outout}(t-1)
  = \frac{1}{1 + \mu_{k,\disc}} \,
    (f_{\inin}(t) - f_{\outout}(t-1)),
\end{equation}
provided that the continuous time constant $\mu_{k,\cont}$ is related
to the discrete time constant $\mu_{k,\disc}$ in such a way that
\begin{equation}
  \frac{\Delta t }{\Delta t + \mu_{k,\cont}}
  = \frac{1}{1 + \mu_{k,\disc}},
\end{equation}
that is, if and only if
\begin{equation}
  \mu_{k,\disc} = \frac{\mu_{k,\cont}}{\Delta t}.
\end{equation}

\subsubsection{Summary of the derived theoretical results}

In this way way have shown that proposed discrete method for
implementing the discrete temporal smoothing stage in the
proposed discrete analogue of the time-causal analogue of the Gabor
transform can be regarded as a true numerical approximation of the
continuous temporal smoothing stage in the proposed continuous
time-causal analogue of the continuous Gabor transform.

Combined with the above theoretical result, showing that this form of
temporal smoothing is also guaranteed to have variation-diminishing
properties in the sense of guaranteeing that the number of local
extrema, or equivalently the number of zero-crossing in the filtered
signal cannot increase with the temporal scale, we propose to refer to
the resulting composed discrete implementation method as the discrete
analogue of the time-causal analogue of the Gabor transform.

\subsection{Explicit algorithm for discrete implementation of a
  discrete analogue of the time-causal analogue of the Gabor
  transform}
\label{app-discr-impl}

This appendix gives an explicit overview of how to implement the
computation of the discrete analogue of the time-causal analogue of the
Gabor transform for a digital signal $f(t)$. For simplicity, we first
describe the case for offline data, where the computations are not
performed in a real-time situation.

Let us assume that the signal has been sampled with a time increment
$\Delta t > 0$, and that we for some proportionality factor
$N \geq 1$ are going to compute the discrete
time-frequency transforms for the angular frequencies $\omega_j$
at the respective temporal scale levels, proportional to the
wavelengths corresponding the angular frequencies
\begin{equation}
  \sigma_{j,0} = \frac{2\pi \, N}{\omega_j}.
\end{equation} 
Then, we first define the temporal sampling rate $r$ as
\begin{equation}
  r = \frac{1}{\Delta t},
\end{equation}
and the temporal scales adjusted to the temporal sampling rate as
$\sigma_j = r \, \sigma_{j,0}$, leading to the following values, when
expressed in units of the temporal variances of the temporal window functions
\begin{equation}
  \label{eq-def-tau-j-ref}
  \tau_{j,\refref}
  = r^2 \, \sigma_{j,0}^2
  = r^2 \left( \frac{2\pi \, N}{\omega_j} \right)^2.
\end{equation}
Given a pre-selected value of the distribution parameter $c > 1$ of
the time-causal limit kernel, we should then perform the corresponding
operations:%
\footnote{This algorithmic outline constitutes an extension of a
  previously formulated algorithm in
  Lindeberg \cite{Lin23-BICY} Appendix~B for temporal filtering with
  the discrete analogue of the time-causal limit kernel.}
\begin{enumerate}
\item
  Given the set of angular frequencies $\omega_j$ and the set of temporal
  sampling indices $n \in \bbbz$,
  multiply the input signal $f(n \, \Delta t)$ by sine and cosine functions
  \begin{align}
    \begin{split}
      \label{eq-f-cos-alg}
      f_{\cos}(n, \omega_j) = f(n \, \Delta t) \, \cos(\omega_j \, n \, \Delta t),
     \end{split}\\
    \begin{split}
      \label{eq-f-sin-alg}      
      f_{\sin}(n, \omega_j) = - f(n \, \Delta t) \, \sin(\omega_j \, n \, \Delta t).
     \end{split}
  \end{align}
\item
  For each angular frequency $\omega_j$,
  compute a set of temporal scale levels $\tau_{j,k}$ according to a geometric
   distribution (\ref{eq-distr-tau-values}):
   \begin{equation}
     \tau_{j,k} = c^{2(k-K)} \tau_{j,ref} \quad\quad (1 \leq k \leq K).
   \end{equation}
   with $\tau_{j,\refref}$ according to (\ref{eq-def-tau-j-ref}).
\item
  For each angular frequency $\omega_j$,
  compute a corresponding set of scale increments:
   \begin{equation}
     \Delta \tau_{j,k} = \tau_{j,k} - \tau_{j,k-1} \quad\quad (1 \leq k \leq K)
   \end{equation}
   with the additional definition $\tau_{j,0} = 0$.
\item
  For each angular frequency $\omega_j$,
  compute the time constants $\mu_{j,k}$ for a set of temporal recursive
    filters of the form (\ref{eq-norm-update})
    according to (\ref{eq-disc-time-constant}):
    \begin{equation}
      \label{eq-def-mu-j-k-alg-outline}
      \mu_{j,k} = \frac{\sqrt{1 + 4 \Delta \tau_{j,k}}-1}{2} \quad\quad (1 \leq k \leq K).
    \end{equation}
  \item
    \label{item-time-rec-filt}
  For each angular frequency $\omega_j$ and for
  both the cosine and the sine parts of input signal multiplied by
  cosine and sine waves according to (\ref{eq-f-cos-alg}) and
  (\ref{eq-f-sin-alg}), couple the
  following sets of first-order recursive filters in
  cascade (\ref{eq-norm-update}) over increasing values
  of the temporal scale levels $k$ (and for a unit parameterization of
  time with time increments of the form $\Delta t = 1$)
  \begin{equation}
    \label{eq-app-rec-filt-casc}
       f_{\outout}(t) - f_{\outout}(t-1)
       = \frac{1}{1 + \mu_{j,k}} \,
            (f_{\inin}(t) - f_{\outout}(t-1))
   \end{equation}
   using the discrete signals $f_{\cos}(n, \omega_j)$ and $f_{\sin}(n, \omega_j)$,
   respectively, as the input data to the chain, and perform the explicit discrete
   filtering operations
   (see Appendix~\ref{sec-app-real-time-alg} for a more detailed
   description of how this can be done in a real-time situation).
 \item
   This results in the following real and imaginary components of the discrete analogue
   of the time-causal analogue of the Gabor transform:
   \begin{align}
     \begin{split}
        ({\cal H} f)_{\cos}(n \, \Delta t, \omega_j;\; \tau_{j,k}, c),
     \end{split}\\
     \begin{split}
        ({\cal H} f)_{\sin}(n \, \Delta t, \omega_j;\; \tau_{j,k}, c),
     \end{split}
   \end{align}
   over the temporal sampling indices $n$, the angular frequencies
   $\omega_j$ and the temporal scale levels $\tau_{j,k}$
 \item
   Compute the spectrogram components as
   \begin{multline}
     |({\cal H} f)(n \, \Delta t, \omega_j;\; \tau_k, c)| = \\
     \sqrt{({\cal H} f)^2_{\cos}(n \, \Delta t, \omega_j;\; \tau_k, c)
              + ({\cal H} f)^2_{\sin}(n \, \Delta t, \omega_j;\; \tau_k, c)}
   \end{multline}
 \end{enumerate}

\subsubsection{Specific adaptations when computing audio spectrograms}
\label{app-adapt-auditory-spectrograms}

When applying this method for computing an audio spectrogram, intended
to analyze audio data, as are to be perceived by a human, it can,
additionally, be valuable to perform soft thresholding on the temporal
scale levels at the lowest and the highest frequencies, to prevent the
temporal delays from becoming too long for the lowest frequencies and
preventing the temporal integration time to become too short for the
highest frequencies.

Therefore, instead of defining $\tau_{j,\refref}$ according to
(\ref{eq-def-tau-j-ref}), when computing audio spectrograms,
we define $\tau_{j,\refref}$ as
\begin{equation}
  \label{eq-tau-depend-on-omega-gauss-exp-spectr-soft-threshold}
  \tau_{j,\refref} =
  r^2\left( \tau_0 + \left( \frac{2 \pi \, N}{\omega} \right)^2 \right),
\end{equation}
where $\tau_0 = \sigma_0^2$ denotes a lower bound on the temporal
window scale, and where one may chose {\em e.g.\/}¶ $\sigma_0 = 1~\mbox{ms}$.

Correspondingly, to prevent the temporal delay from being too long for
low frequencies, a soft upper bound on the temporal scale is defined
as
\begin{equation}
  \tau_{j,\refref}'
  = \frac{\tau_{j,\refref}}
             {\left(
                1 + \left( \frac{\tau_{j,\refref}}{\tau_{\infty}} \right)^p
              \right)^{1/p}}
 \end{equation}
for suitable values of $\tau_{\infty}$ and $p$. As default values for
these parameters, we use $p = 2$ and $\tau_{\infty} =
\sigma_{\infty}^2$ with $\sigma_{\infty} = \mbox{40~ms}$.

By these adaptations, self-similarity over temporal scales will, as a
consequence, only hold in an intermediate range of the temporal
frequencies, consistent with previous
evidence that the resolution of
pitch perception is the highest in an intermediate range of frequencies and then
decreases for both lower and higher frequencies
(see Hartmann \cite{Har96-JASA} and Moore \cite{Moo73-JASA}).

Of course, other types of modifications to delimit the range of
temporal scales for low and high frequencies could also be considered.

\begin{algorithm*}

  \begin{algorithmic}[0]
    \Procedure{time-caus-time-freq-transform}{$f, \omega, \mu$}\Comment{$f$
      input stream, $\omega$ of size $J$, $\mu$ of size $J \times K$}
       \State $level \gets 0$ \Comment{of size $J \times K (\times 2)$}
       \State $level\_prev \gets 0$ \Comment{of size $J \times K
         (\times 2)$}
       \State $n \gets 0$ \Comment{time counter}

       \Repeat
         \State $signal \gets f(n \, \Delta t)$
            \Comment{read the input stream with time increment $\Delta t$}
          \For{$j \gets 1, J$}
              \State $input \gets signal \times \{ cos, -sin \} (\omega_j \, n \, \Delta t)$
                 \Comment{multiply input signal with complex
                   exponential at current time frame}
              \For{$k \gets 1, K$}
                \If{$k = 1$}
                   \State $level_{j, k} \gets level\_prev_{j, k} + (input - level\_prev_{j, k})/(1 + \mu_{j, k})$
                   \Comment{the first layer}
                \Else
                   \State $level_{j, k} \gets level\_prev_{j, k} +
                   (level_{j, k-1} - level\_prev_{j, k})/(1 + \mu_{j, k})$
                   \Comment{the higher layers}
                \EndIf
              \EndFor
            \EndFor
            \State $level\_prev \gets level$ \Comment{update the buffer for
              the previous time frame}
        \State $n \gets n+1$ \Comment{prepare for the next time frame}
     \Until{interrupt}
   \EndProcedure
\end{algorithmic}


%
%
%
%
  \caption{Pseudocode for the core time-causal temporal filtering module in the set of
    first-order recursive filters coupled in cascade
    (\ref{eq-app-rec-filt-casc}), that implements
    the convolution with the temporal window function in the discrete
    analogue of the time-causal analogue of the Gabor transform.
    Here, it is assumed that a set of angular frequencies $\omega_j$
    has been already defined and that the time constants
    $\mu_{j,k} > 0$
    have been already computed according to
    (\ref{eq-def-mu-j-k-alg-outline}). The variable $level\_prev$ 
    represents a memory from the previous frame, necessary to compute
    the temporal differences that drive the recursive filters. In this respect, the
    algorithm is strictly time-recursive, since it only makes use of
    information from the present moment and a very short-term
    memory from the previous frame. Note, specifically, that the outer for loop over
    the frequency index $j$
     can be computed in parallel on a
    multi-core architecture.
  }
  \label{fig-pseudo-code-rec-filt-casc}
\end{algorithm*}

\subsection{Strictly time-recursive online algorithm for computing the
  discrete analogue of the time-causal analogue of the Gabor transform
  in real-time situations}
\label{sec-app-real-time-alg}

This appendix describes how to implement the multi-scale temporal
filtering operations underlying the computation of the discrete analogue of
the time-causal analogue of the Gabor transform, in a fully time-causal
and time-recursive way suitable for real-time computations.

For this purpose, we focus on item~\ref{item-time-rec-filt} in the above outline in
Appendix~\ref{app-discr-impl}, concerning the overall implementation of
the discrete time-causal time-frequency transform.

For each angular frequency $\omega_j$, a cascade of recursive filter
is to be initiated, with the time-constants for the recursive filters
determined according to (\ref{eq-def-mu-j-k-alg-outline}).
Then, the actual procedure for coupling these recursive filters in
cascade can be expressed on the form outlined in the
pseudocode in Algorithm~\ref{fig-pseudo-code-rec-filt-casc}.

\subsubsection{Computational work}

The computational work required to implement this algorithm for a
single angular frequency $\omega_j$, besides the initial computation
of the sine and cosine functions, basically
corresponds to two additions and one multiplication for each of the
two cosine and sine channels, multiplied by the
$K$ number of layers, which by necessity have to be computed sequentially.

Then, when handling $J$ multiple angular frequencies, which for a
purely serial implementation requires computational work
proportional to the number $J$ of
angular frequencies, the computational work depends upon to what
extent the inherently parallel computations over the functionally independent
angular frequencies can be distributed over multiple cores.

Compared to an implementation of the regular Gabor transform, the
computational work for the discrete analogue of the Gabor transform
can for moderate values $K$ of the number of layers (usually 4 to 8 layers)
specifically be expected to be substantially lower than for an
implementation of the regular Gabor transform, based on explicit
temporal convolutions over temporal intervals of longer duration.

\subsection{Discrete analogue of the Gabor transform}
\label{app-disc-anal-gabor-transf}

In view of the classification of the discrete scale-space kernels in
Appendix~\ref{app-class-temp-scsp-kern} above, there is one family of
discrete smoothing kernels that stands out as a discrete analogue of
the Gaussian kernel over a non-causal domain
(see Lindeberg \cite{Lin90-PAMI}) of the form
\begin{equation}
  T(n;\; s) = e^{-s} I_n(s),
\end{equation}
where $I_n(s)$ denote the modified Bessel functions of integer order
(Abramowitz and Stegun \cite{AS64}). These kernels obey a semi-group
property over scales
\begin{equation}
  T(\cdot;\; s_1) * T(\cdot;\; s_2) = T(\cdot;\; s_1 + s_2) 
\end{equation}
and can specifically be shown to constitute a better discrete
approximation of the properties of the continuous Gaussian kernel than
the sampled Gaussian kernel (see Lindeberg \cite{Lin24-JMIV-discgaussders}).

Based on using this discrete analogue of the Gaussian kernel as the
temporal window function in a discrete time-frequency transform
\begin{equation}
  \label{eq-disc-gab-transf}
  ({\cal T} f)(t, \omega;\; \tau) =
  \sum_{n=-\infty}^{\infty}
     f(n) \,
    T(t - n;\; \tau) \, e^{-i \omega  n},
\end{equation}
for integer $t$, we can thus treat the definition
(\ref{eq-disc-gab-transf}) as a discrete analogue of the regular
non-causal Gabor transform.

The spectrograms in Figure~\ref{fig-spectrograms-truncgabor} have been
computed using this method for discrete implementation of the Gabor
transform, complemented with temporal shifts and truncations of the
filter for the input values that would have implied forbidden access
to the future in relation to any pre-recorded time moment.

\subsection{Inverse transforms of the time-causal analogue of the Gabor
  transform}
\label{sec-app-inv-transf-time-caus-gabor-transf}

Following the treatment in Teolis \cite{Teo95-book} Section~4.5.2, regarding the
inverse transform of a windowed Fourier transform, we can derive an
inverse transform of the time-causal analogue of the Gabor transform as
follows:

The time-causal analogue of the Gabor transform is defined according
to Equation~(\ref{eq-time-caus-anal-gab-transf}):
\begin{equation}
  ({\cal H} f)(t, \omega;\; \tau, c) =
  \int_{u=-\infty}^{\infty}
     f(u) \,
    \Psi(t-u;\; \tau, c) \, e^{-i \omega  u} \, du,
  \end{equation}
where $\Psi(t;\; \tau, c)$  denotes the time-causal limit kernel defined from having a
Fourier transform of the form (\ref{eq-FT-comp-kern-log-distr-limit}),
or equivalently defined from an infinite set of truncated exponential kernels
coupled in cascade, with the time constants $\mu_k$ defined from
the temporal scale parameter $\tau$ and the distribution parameter $c$ according to
(\ref{eq-mu1-log-distr}) and (\ref{eq-muk-log-distr}). The time-causal
limit kernel $\Psi(t;\; \tau, c)$ is specifically equal to
0 for $t < 0$, because of the temporal causality.
  
If we define the following combined translation and reversal operator
\begin{equation}
  (\delta_t h)(u) = h(t - u),
\end{equation}
then we see that the time-causal analogue of the Gabor transform
$({\cal H} f)(t, \omega;\; \tau, c)$ is the Fourier transform of the
function $f(u) \, (\delta_t \Psi)(u;\; \tau, c)$, {\em i.e.\/},
\begin{align}
  \begin{split}
    & ({\cal H} f)(t, \omega;\; \tau, c) =
  \end{split}\nonumber\\
  \begin{split}
    & = \int_{u=-\infty}^{\infty}
     f(u) \,
    (\delta_t\Psi)(u;\; \tau, c) \, e^{-i \omega  u} \, du
  \end{split}\nonumber\\
  \begin{split}
    & = {\cal F}(f(\cdot) \, (\delta_t \Psi)(\cdot;\; \tau, c))(\omega).
  \end{split}
\end{align}
According to the inverse of the Fourier transform, we then have
\begin{align}
  \begin{split}
      & f(u) \, (\delta_t \Psi)(u;\; \tau, c) =
  \end{split}\nonumber\\
  \begin{split}
     & = {\cal F}^{-1}(({\cal H} f)(t, \cdot;\; \tau, c))(u)
  \end{split}\nonumber\\
  \begin{split}
     \label{eq-inv-gabor-transf-basic-rel}
     & = \frac{1}{2 \pi}
             \int_{\omega = -\infty}^{\infty}
               ({\cal H} f)(t, \omega;\; \tau, c) \, e^{i \omega u} \, d\omega.
   \end{split}
\end{align}
Multiplying both sides of this equation by $(\delta_t \Psi)(u;\; \tau, c)$
and integrating over $t$, then gives
\begin{align}
  \begin{split}
    & f(u)
       \int_{t = -\infty}^{\infty}
           ((\delta_t \Psi)(u;\; \tau, c))^2 \, dt
  \end{split}\nonumber\\
  \begin{split}
    & = \frac{1}{2 \pi}
           \int_{t = -\infty}^{\infty}
               (\delta_t \Psi)(u;\; \tau, c)
  \end{split}\nonumber\\
  \begin{split}
    & \phantom{= \quad\quad} \quad\quad \int_{\omega = -\infty}^{\infty}
              ({\cal H} f)(t, \omega;\; \tau, c) \, e^{i \omega u} \, d\omega \, dt,
  \end{split}
\end{align}
from which we, noting that $(\delta_t \Psi)(u;\; \tau, c) = \Psi(t - u;\; \tau, c) = 0$ for
$t - u < 0$, get
\begin{align}
  \begin{split}
    & f(u)
    = \frac{1}{2 \pi}
        \left(
           \int_{t = u}^{\infty}
             \Psi(t - u;\; \tau, c)
      \right.
   \end{split}\nonumber\\
   \begin{split}
      & \phantom{f(u) = \frac{1}{2 \pi}}
         \quad\quad
          \left.
            \int_{\omega = -\infty}^{\infty}
               ({\cal H} f)(t, \omega;\; \tau, c) \, e^{i \omega u} \, d\omega \, dt
          \right)/
  \end{split}\nonumber\\
  \begin{split}
    & \phantom{f(u) = \frac{1}{2 \pi}}
    \int_{t = u}^{\infty}
           (\Psi(t-u;\; \tau, c))^2 \, dt,
   \end{split}
\end{align}
which, in turn, by the change of variables $v = t - u$, can be simplified to
\begin{align}
  \begin{split}
    & f(u)
    = \frac{1}{2 \pi}
        \left(
           \int_{v = 0}^{\infty}
             \Psi(v;\; \tau, c)
      \right.
   \end{split}\nonumber\\
   \begin{split}
      & \phantom{f(u) = \frac{1}{2 \pi}}
         \quad\quad
          \left.
            \int_{\omega = -\infty}^{\infty}
               ({\cal H} f)(u + v, \omega;\; \tau, c) \, e^{i \omega u} \, d\omega \, dv
          \right)/
  \end{split}\nonumber\\
  \begin{split}
    \label{eq-inv-transform-simplified}
    & \phantom{f(u) = \frac{1}{2 \pi}}
    \int_{v = 0}^{\infty}
           (\Psi(v;\; \tau, c))^2 \, dv.
   \end{split}
\end{align}
This is an explicit expression for an inverse of the time-causal
analogue of the Gabor transform.

Note, that this inverse transform is, however, not time-causal, since
for any given time moment, only values of the time-causal analogue of
the Gabor transform from the future will have an influence on the signal
at the present moment.

Note, furthermore, that there are also other possible
ways to define inverse transforms of the time-causal analogue of the
Gabor transform based on the relationship
(\ref{eq-inv-gabor-transf-basic-rel}), by estimating
the original signal $f(u)$ from a combination of the inverse Fourier
transform of the time-causal analogue of the Gabor transform
${\cal F}^{-1}(({\cal H} f)(t, \cdot;\; \tau, c))(u)$ with
the time-causal limit kernel $(\delta_t \Psi)(u;\; \tau, c) = \Psi(t - u;\; \tau, c)$
at either a single time moment $t$ or a set of time moments over a more
compact time interval than the entire temporal axis for $t \geq u$.

\subsubsection{Temporal delays associated with the inverse
  transform}

Note, specifically, that if we, with the aim of reducing the temporal
delays when computing the inverse transform, choose to instead use the
relationship (\ref{eq-inv-gabor-transf-basic-rel}) for gathering
information from the inverse Fourier transform of the time-causal
analogue of the Gabor transform ${\cal F}^{-1}(({\cal H} f)(t, \cdot;\; \tau, c))(u)$
over a shorter temporal interval $[u, u + \Delta u]$, such that the
resulting inverse transform corresponding to (\ref{eq-inv-transform-simplified})
will then be reduced to the form
\begin{align}
  \begin{split}
    & f(u)
    = \frac{1}{2 \pi}
        \left(
           \int_{v = 0}^{\Delta u}
             \Psi(v;\; \tau, c)
      \right.
   \end{split}\nonumber\\
   \begin{split}
      & \phantom{f(u) = \frac{1}{2 \pi}}
         \quad\quad
          \left.
            \int_{\omega = -\infty}^{\infty}
               ({\cal H} f)(u + v, \omega;\; \tau, c) \, e^{i \omega u} \, d\omega \, dv
          \right)/
  \end{split}\nonumber\\
  \begin{split}
    \label{eq-inv-transform-truncated}
    & \phantom{f(u) = \frac{1}{2 \pi}}
    \int_{v = 0}^{\Delta u}
           (\Psi(v;\; \tau, c))^2 \, dv.
   \end{split}
\end{align}
we cannot, however, reduce the duration $\Delta u$ of the temporal
support interval too much, and specifically not let the duration
$\Delta u$ tend to zero, since the time-causal limit kernel
$\Psi(v;\; \tau, c)$ tends to zero when the time variable $v$ tends to zero.

Thus, the need for accumulating information, about the values of
time-causal analogue of the Gabor transform over extended periods of time, will lead to
inevitable additional temporal delays when computing the inverse
transform. This points to a trade-off problem in that a larger
duration $\Delta u$ could be expected to enable more accurate
estimates of the inverse transform, while then also increasing the
temporal delay, which may, thus, strongly influence the potential applicability of using
the inverse transform in time-critical applications.

\bibliographystyle{SSBAtrans}
\bibliography{defs,tlmac}

\end{document}